\newif\ifcomments     
\commentsfalse        

\documentclass{article}

\usepackage{bigfoot}
\usepackage{amsmath,amssymb,amsthm}
\usepackage[dvipsnames]{xcolor}
\usepackage{verbatim}
\usepackage[utf8]{inputenc}
\usepackage[hyphens]{url}
\usepackage[hidelinks,bookmarksnumbered,pdfencoding=auto,psdextra]{hyperref}
\usepackage{supertabular}
\usepackage{listings}
\usepackage{textcomp}
\usepackage{xspace}
\usepackage{ottalt}
\usepackage[T1]{fontenc}

\usepackage{lstpi}
\usepackage{lsthaskell}
\newcommand\cd[1]{\lstinline[language=Haskell]{#1}}

\newcommand\pif{\texttt{pi-forall}\xspace}
\newcommand\unbound{\texttt{Unbound}\xspace}

\theoremstyle{definition}
\newtheorem{definition}{Definition}[section]
\newtheorem{lemma}{Lemma}[section]

\title{Implementing Dependent Types in \pif}
\author{Stephanie Weirich}

\begin{document}

\maketitle


\newcommand{\ottdrule}[4][]{{\displaystyle\frac{\begin{array}{l}#2\end{array}}{#3}\quad\ottdrulename{#4}}}
\newcommand{\ottusedrule}[1]{\[#1\]}
\newcommand{\ottpremise}[1]{ #1 \\}
\newenvironment{ottdefnblock}[3][]{ \framebox{\mbox{#2}} \quad #3 \\[0pt]}{}
\newenvironment{ottfundefnblock}[3][]{ \framebox{\mbox{#2}} \quad #3 \\[0pt]\begin{displaymath}\begin{array}{l}}{\end{array}\end{displaymath}}
\newcommand{\ottfunclause}[2]{ #1 \equiv #2 \\}
\newcommand{\ottnt}[1]{\mathit{#1}}
\newcommand{\ottmv}[1]{\mathit{#1}}
\newcommand{\ottkw}[1]{\mathbf{#1}}
\newcommand{\ottsym}[1]{#1}
\newcommand{\ottcom}[1]{\text{#1}}
\newcommand{\ottdrulename}[1]{\textsc{#1}}
\newcommand{\ottcomplu}[5]{\overline{#1}^{\,#2\in #3 #4 #5}}
\newcommand{\ottcompu}[3]{\overline{#1}^{\,#2<#3}}
\newcommand{\ottcomp}[2]{\overline{#1}^{\,#2}}
\newcommand{\ottgrammartabular}[1]{\begin{supertabular}{llcllllll}#1\end{supertabular}}
\newcommand{\ottmetavartabular}[1]{\begin{supertabular}{ll}#1\end{supertabular}}
\newcommand{\ottrulehead}[3]{$#1$ & & $#2$ & & & \multicolumn{2}{l}{#3}}
\newcommand{\ottprodline}[6]{& & $#1$ & $#2$ & $#3 #4$ & $#5$ & $#6$}
\newcommand{\ottfirstprodline}[6]{\ottprodline{#1}{#2}{#3}{#4}{#5}{#6}}
\newcommand{\ottlongprodline}[2]{& & $#1$ & \multicolumn{4}{l}{$#2$}}
\newcommand{\ottfirstlongprodline}[2]{\ottlongprodline{#1}{#2}}
\newcommand{\ottbindspecprodline}[6]{\ottprodline{#1}{#2}{#3}{#4}{#5}{#6}}
\newcommand{\ottprodnewline}{\\}
\newcommand{\ottinterrule}{\\[5.0mm]}
\newcommand{\ottafterlastrule}{\\}
\newcommand{\ottmetavars}{
\ottmetavartabular{
 $ \ottmv{tname} ,\, \ottmv{x} ,\, \ottmv{y} ,\, \ottmv{z} ,\, \ottmv{f} ,\, \ottmv{g} ,\, \ottmv{n} $ & \ottcom{variables} \\
 $ \ottmv{dcname} ,\, \ottmv{K} $ & \ottcom{data constructor names} \\
 $ \ottmv{tcname} ,\, \ottmv{T} $ & \ottcom{data type names} \\
 $ \ottmv{i} $ & \ottcom{indexvar} \\
}}

\newcommand{\ottep}{
\ottrulehead{\epsilon}{::=}{}\ottprodnewline
\ottfirstprodline{|}{ + }{}{}{}{}\ottprodnewline
\ottprodline{|}{ - }{}{}{}{}}

\newcommand{\ottv}{
\ottrulehead{\ottnt{v}}{::=}{\ottcom{values}}\ottprodnewline
\ottfirstprodline{|}{\ottkw{Type}}{}{}{}{}\ottprodnewline
\ottprodline{|}{ \lambda  \ottmv{x} .  \ottnt{a} }{}{}{}{}\ottprodnewline
\ottprodline{|}{ ( \ottmv{x} \!:\! \ottnt{A} )  \to   \ottnt{B} }{}{}{}{}\ottprodnewline
\ottprodline{|}{\ottkw{Unit}}{}{}{}{}\ottprodnewline
\ottprodline{|}{\ottsym{()}}{}{}{}{}\ottprodnewline
\ottprodline{|}{\ottsym{(}  \ottnt{v}  \ottsym{)}} {\textsf{S}}{}{}{}\ottprodnewline
\ottprodline{|}{ \{  \ottmv{x} \!:\! \ottnt{A} \ |\  \ottnt{B}  \} }{}{}{}{}\ottprodnewline
\ottprodline{|}{\ottsym{(}  \ottnt{a}  \ottsym{,}  \ottnt{b}  \ottsym{)}}{}{}{}{}\ottprodnewline
\ottprodline{|}{\ottkw{Bool}}{}{}{}{}\ottprodnewline
\ottprodline{|}{\ottkw{True}}{}{}{}{}\ottprodnewline
\ottprodline{|}{\ottkw{False}}{}{}{}{}\ottprodnewline
\ottprodline{|}{\ottnt{a}  \ottsym{=}  \ottnt{b}}{}{}{}{}\ottprodnewline
\ottprodline{|}{\ottkw{refl}}{}{}{}{}\ottprodnewline
\ottprodline{|}{\ottmv{K} \, \overline{a}}{}{}{}{}\ottprodnewline
\ottprodline{|}{\ottmv{T} \, \overline{a}}{}{}{}{}}

\newcommand{\ottneutral}{
\ottrulehead{\ottnt{neutral}  ,\ \ottnt{ne}}{::=}{\ottcom{neutral terms}}\ottprodnewline
\ottfirstprodline{|}{\ottmv{x}}{}{}{}{}\ottprodnewline
\ottprodline{|}{ \ottnt{ne}  \;  \ottnt{a} }{}{}{}{}\ottprodnewline
\ottprodline{|}{\ottsym{(}  \ottnt{ne}  \ottsym{)}} {\textsf{S}}{}{}{}\ottprodnewline
\ottprodline{|}{\ottkw{let} \, \ottsym{(}  \ottmv{x}  \ottsym{,}  \ottmv{y}  \ottsym{)}  \ottsym{=}  \ottnt{ne} \, \ottkw{in} \, \ottnt{a}}{}{}{}{}\ottprodnewline
\ottprodline{|}{\ottkw{if} \, \ottnt{ne} \, \ottkw{then} \, \ottnt{a} \, \ottkw{else} \, \ottnt{b}}{}{}{}{}\ottprodnewline
\ottprodline{|}{\ottkw{subst} \, \ottnt{a} \, \ottkw{by} \, \ottnt{ne}}{}{}{}{}\ottprodnewline
\ottprodline{|}{\ottkw{case} \, \ottnt{ne} \, \ottkw{of} \, \ottsym{\{} \, \ottcomp{\ottnt{pat_{\ottmv{i}}}  \to  \ottnt{a_{\ottmv{i}}}}{\ottmv{i}} \, \ottsym{\}}}{}{}{}{}}

\newcommand{\ottnf}{
\ottrulehead{\ottnt{nf}}{::=}{\ottcom{weak head normal forms}}\ottprodnewline
\ottfirstprodline{|}{\ottkw{Type}}{}{}{}{}\ottprodnewline
\ottprodline{|}{ \lambda  \ottmv{x} .  \ottnt{a} }{}{}{}{}\ottprodnewline
\ottprodline{|}{ ( \ottmv{x} \!:\! \ottnt{A} )  \to   \ottnt{B} }{}{}{}{}\ottprodnewline
\ottprodline{|}{\ottkw{Unit}}{}{}{}{}\ottprodnewline
\ottprodline{|}{\ottsym{()}}{}{}{}{}\ottprodnewline
\ottprodline{|}{\ottmv{x}}{}{}{}{}\ottprodnewline
\ottprodline{|}{ \ottnt{ne}  \;  \ottnt{b} }{}{}{}{}\ottprodnewline
\ottprodline{|}{\ottsym{(}  \ottnt{nf}  \ottsym{)}} {\textsf{S}}{}{}{}\ottprodnewline
\ottprodline{|}{ \{  \ottmv{x} \!:\! \ottnt{A} \ |\  \ottnt{B}  \} }{}{}{}{}\ottprodnewline
\ottprodline{|}{\ottsym{(}  \ottnt{a}  \ottsym{,}  \ottnt{b}  \ottsym{)}}{}{}{}{}\ottprodnewline
\ottprodline{|}{\ottkw{let} \, \ottsym{(}  \ottmv{x}  \ottsym{,}  \ottmv{y}  \ottsym{)}  \ottsym{=}  \ottnt{ne} \, \ottkw{in} \, \ottnt{a}}{}{}{}{}\ottprodnewline
\ottprodline{|}{\ottkw{Bool}}{}{}{}{}\ottprodnewline
\ottprodline{|}{\ottkw{True}}{}{}{}{}\ottprodnewline
\ottprodline{|}{\ottkw{False}}{}{}{}{}\ottprodnewline
\ottprodline{|}{\ottkw{if} \, \ottnt{ne} \, \ottkw{then} \, \ottnt{a} \, \ottkw{else} \, \ottnt{b}}{}{}{}{}\ottprodnewline
\ottprodline{|}{\ottnt{a}  \ottsym{=}  \ottnt{b}}{}{}{}{}\ottprodnewline
\ottprodline{|}{\ottkw{refl}}{}{}{}{}\ottprodnewline
\ottprodline{|}{\ottkw{subst} \, \ottnt{a} \, \ottkw{by} \, \ottnt{ne}}{}{}{}{}\ottprodnewline
\ottprodline{|}{\ottmv{K} \, \overline{a}}{}{}{}{}\ottprodnewline
\ottprodline{|}{\ottmv{T} \, \overline{a}}{}{}{}{}\ottprodnewline
\ottprodline{|}{\ottkw{case} \, \ottnt{ne} \, \ottkw{of} \, \ottsym{\{} \, \ottcomp{\ottnt{pat_{\ottmv{i}}}  \to  \ottnt{a_{\ottmv{i}}}}{\ottmv{i}} \, \ottsym{\}}}{}{}{}{}}

\newcommand{\ottargs}{
\ottrulehead{\overline{a}  ,\ \overline{b}}{::=}{}\ottprodnewline
\ottfirstprodline{|}{}{}{}{}{}\ottprodnewline
\ottprodline{|}{\ottnt{a} \, \overline{a}}{}{}{}{}\ottprodnewline
\ottprodline{|}{\ottsym{[a]}  \overline{a}}{}{}{}{}}

\newcommand{\ottpat}{
\ottrulehead{\ottnt{pat}}{::=}{}\ottprodnewline
\ottfirstprodline{|}{\ottmv{x}}{}{}{}{}\ottprodnewline
\ottprodline{|}{\ottmv{K} \, \ottnt{ps}}{}{}{}{}}

\newcommand{\ottms}{
\ottrulehead{\ottnt{ms}}{::=}{\ottcom{sequence of matches}}\ottprodnewline
\ottfirstprodline{|}{\ottkw{m} \, \ottsym{;}  \ottnt{ms}}{}{}{}{}\ottprodnewline
\ottprodline{|}{}{}{}{}{}}

\newcommand{\ottps}{
\ottrulehead{\ottnt{ps}}{::=}{\ottcom{sequence of patterns}}\ottprodnewline
\ottfirstprodline{|}{\ottnt{pat} \, \ottnt{ps}}{}{}{}{}\ottprodnewline
\ottprodline{|}{\ottnt{pat} \, \epsilon \, \ottnt{ps}}{}{}{}{}\ottprodnewline
\ottprodline{|}{}{}{}{}{}}

\newcommand{\otttelescope}{
\ottrulehead{\ottnt{telescope}  ,\ \Delta}{::=}{}\ottprodnewline
\ottfirstprodline{|}{\ottmv{x}  \ottsym{:}  \ottnt{A}  \ottsym{,}  \Delta}{}{}{}{}\ottprodnewline
\ottprodline{|}{\ottmv{x}  \ottsym{:}  \epsilon \, \ottnt{A}  \ottsym{,}  \Delta}{}{}{}{}\ottprodnewline
\ottprodline{|}{\ottmv{x}  \ottsym{=}  \ottnt{b}  \ottsym{,}  \Delta}{}{}{}{}\ottprodnewline
\ottprodline{|}{\Delta_{{\mathrm{1}}}  \ottsym{,}  \Delta_{{\mathrm{2}}}}{}{}{}{}\ottprodnewline
\ottprodline{|}{}{}{}{}{}}

\newcommand{\otttm}{
\ottrulehead{\ottnt{tm}  ,\ \ottnt{a}  ,\ \ottnt{b}  ,\ \ottnt{A}  ,\ \ottnt{B}}{::=}{\ottcom{terms and types}}\ottprodnewline
\ottfirstprodline{|}{\ottkw{Type}}{}{}{}{\ottcom{sort}}\ottprodnewline
\ottprodline{|}{\ottmv{x}}{}{}{}{\ottcom{variable}}\ottprodnewline
\ottprodline{|}{ \lambda  \ottmv{x} .  \ottnt{a} }{}{}{}{\ottcom{function definition}}\ottprodnewline
\ottprodline{|}{ \ottnt{a}  \;  \ottnt{b} }{}{}{}{\ottcom{function application}}\ottprodnewline
\ottprodline{|}{ ( \ottmv{x} \!:\! \ottnt{A} )  \to   \ottnt{B} }{}{}{}{\ottcom{dependent function type}}\ottprodnewline
\ottprodline{|}{\ottnt{a}  \ottsym{[}  \ottnt{b}  \ottsym{/}  \ottmv{x}  \ottsym{]}} {\textsf{S}}{}{}{\ottcom{substitution}}\ottprodnewline
\ottprodline{|}{\ottsym{(}  \ottnt{a}  \ottsym{:}  \ottnt{A}  \ottsym{)}}{}{}{}{\ottcom{type annotation}}\ottprodnewline
\ottprodline{|}{\ottsym{(}  \ottnt{a}  \ottsym{)}} {\textsf{S}}{}{}{\ottcom{parenthesis}}\ottprodnewline
\ottprodline{|}{\ottkw{TRUSTME}}{}{}{}{\ottcom{term that has any type}}\ottprodnewline
\ottprodline{|}{\ottkw{PRINTME}}{}{}{}{\ottcom{print the current context}}\ottprodnewline
\ottprodline{|}{\ottkw{Unit}}{}{}{}{\ottcom{unit type}}\ottprodnewline
\ottprodline{|}{\ottsym{()}}{}{}{}{\ottcom{unit term}}\ottprodnewline
\ottprodline{|}{ \{  \ottmv{x} \!:\! \ottnt{A} \ |\  \ottnt{B}  \} }{}{}{}{\ottcom{$\Sigma$-type (i.e. dependent products/dependent sums)}}\ottprodnewline
\ottprodline{|}{\ottsym{(}  \ottnt{a}  \ottsym{,}  \ottnt{b}  \ottsym{)}}{}{}{}{\ottcom{products}}\ottprodnewline
\ottprodline{|}{\ottkw{let} \, \ottsym{(}  \ottmv{x}  \ottsym{,}  \ottmv{y}  \ottsym{)}  \ottsym{=}  \ottnt{a} \, \ottkw{in} \, \ottnt{b}}{}{}{}{\ottcom{elimination form for pairs}}\ottprodnewline
\ottprodline{|}{\ottkw{Bool}}{}{}{}{\ottcom{boolean type}}\ottprodnewline
\ottprodline{|}{\ottkw{True}}{}{}{}{\ottcom{boolean value true}}\ottprodnewline
\ottprodline{|}{\ottkw{False}}{}{}{}{\ottcom{boolean value false}}\ottprodnewline
\ottprodline{|}{\ottkw{if} \, \ottnt{a} \, \ottkw{then} \, \ottnt{b_{{\mathrm{1}}}} \, \ottkw{else} \, \ottnt{b_{{\mathrm{2}}}}}{}{}{}{\ottcom{conditional}}\ottprodnewline
\ottprodline{|}{\ottkw{if} \, \ottnt{a} \, \ottkw{then} \, \ottnt{b_{{\mathrm{1}}}} \, \ottkw{else} \, \ottnt{b_{{\mathrm{2}}}}  \ottsym{[}  \ottmv{x}  \ottsym{.}  \ottnt{B}  \ottsym{]}}{}{}{}{\ottcom{conditional}}\ottprodnewline
\ottprodline{|}{\ottnt{a}  \ottsym{=}  \ottnt{b}}{}{}{}{\ottcom{equality type}}\ottprodnewline
\ottprodline{|}{\ottkw{refl}}{}{}{}{\ottcom{reflexivity proof}}\ottprodnewline
\ottprodline{|}{\ottkw{subst} \, \ottnt{a} \, \ottkw{by} \, \ottnt{b}}{}{}{}{\ottcom{equality type elimination}}\ottprodnewline
\ottprodline{|}{\ottkw{contra} \, \ottnt{a}}{}{}{}{\ottcom{false elimination}}\ottprodnewline
\ottprodline{|}{ \lambda [  \ottmv{x}  ] .  \ottnt{a} }{}{}{}{}\ottprodnewline
\ottprodline{|}{\ottnt{a}  \ottsym{[}  \ottnt{b}  \ottsym{]}}{}{}{}{}\ottprodnewline
\ottprodline{|}{ [  \ottmv{x} \!:\! \ottnt{A}  ] \rightarrow  \ottnt{B} }{}{}{}{}\ottprodnewline
\ottprodline{|}{\ottmv{K} \, \overline{a}}{}{}{}{}\ottprodnewline
\ottprodline{|}{\ottmv{T} \, \overline{a}}{}{}{}{}\ottprodnewline
\ottprodline{|}{\ottkw{case} \, \ottnt{a} \, \ottkw{of} \, \ottsym{\{} \, \ottcomp{\ottnt{pat_{\ottmv{i}}}  \to  \ottnt{a_{\ottmv{i}}}}{\ottmv{i}} \, \ottsym{\}}}{}{}{}{}\ottprodnewline
\ottprodline{|}{\ottnt{a}  \ottsym{[}  \ottnt{pat}  \ottsym{/}  \ottmv{x}  \ottsym{]}}{}{}{}{}\ottprodnewline
\ottprodline{|}{\ottkw{Nat}}{}{}{}{}\ottprodnewline
\ottprodline{|}{\ottkw{Zero}}{}{}{}{}\ottprodnewline
\ottprodline{|}{\ottkw{Succ} \, \ottnt{a}}{}{}{}{}\ottprodnewline
\ottprodline{|}{\ottkw{Void}}{}{}{}{}\ottprodnewline
\ottprodline{|}{\ottkw{ImTrue} \, \ottnt{a} \, \ottnt{b}}{}{}{}{}\ottprodnewline
\ottprodline{|}{\ottkw{SillyBool}}{}{}{}{}\ottprodnewline
\ottprodline{|}{\ottkw{let} \, \ottmv{x}  \ottsym{=}  \ottnt{a} \, \ottkw{in} \, \ottnt{b}}{}{}{}{\ottcom{name an expression}}}

\newcommand{\ottcontext}{
\ottrulehead{\ottnt{context}  ,\ \Gamma}{::=}{\ottcom{contexts}}\ottprodnewline
\ottfirstprodline{|}{ \emptyset }{}{}{}{}\ottprodnewline
\ottprodline{|}{ \Gamma ,  \ottmv{x} \! :\! \ottnt{A} }{}{}{}{}\ottprodnewline
\ottprodline{|}{ \ottmv{x} \! :\! \ottnt{A} }{}{}{}{}\ottprodnewline
\ottprodline{|}{\ottsym{(}  \Gamma  \ottsym{)}}{}{}{}{}\ottprodnewline
\ottprodline{|}{ \Gamma ,  \ottmv{x} \! :^ \epsilon \! \ottnt{A} }{}{}{}{}\ottprodnewline
\ottprodline{|}{ \ottmv{x} \! :^ \epsilon \!  \ottnt{A} }{}{}{}{}\ottprodnewline
\ottprodline{|}{ \Gamma ^ \epsilon }{}{}{}{}\ottprodnewline
\ottprodline{|}{\Gamma  \ottsym{,}  \Delta}{}{}{}{}\ottprodnewline
\ottprodline{|}{\Gamma  \ottsym{,}  \ottmv{x}  \ottsym{=}  \ottnt{a}}{}{}{}{}}

\newcommand{\ottgrammar}{\ottgrammartabular{
\ottep\ottinterrule
\ottv\ottinterrule
\ottneutral\ottinterrule
\ottnf\ottinterrule
\ottargs\ottinterrule
\ottpat\ottinterrule
\ottms\ottinterrule
\ottps\ottinterrule
\otttelescope\ottinterrule
\otttm\ottinterrule
\ottcontext\ottafterlastrule
}}

\newcommand{\ottdruleeqXXrefl}[1]{\ottdrule[#1]{%
}{
\Gamma  \vdash  \ottnt{a}  \Leftrightarrow  \ottnt{a}}{%
{\ottdrulename{eq\_refl}}{}%
}}

\newcommand{\ottdruleeqXXwhnf}[1]{\ottdrule[#1]{%
\ottpremise{ \Gamma   \vdash   \ottkw{whnf} \  \ottnt{a}  \leadsto  \ottnt{nf_{{\mathrm{1}}}} }%
\ottpremise{ \Gamma   \vdash   \ottkw{whnf} \  \ottnt{b}  \leadsto  \ottnt{nf_{{\mathrm{2}}}} }%
\ottpremise{\Gamma  \vdash  \ottnt{nf_{{\mathrm{1}}}}  \Leftrightarrow  \ottnt{nf_{{\mathrm{2}}}}}%
}{
\Gamma  \vdash  \ottnt{a}  \Leftrightarrow  \ottnt{b}}{%
{\ottdrulename{eq\_whnf}}{}%
}}

\newcommand{\ottdruleeqXXabs}[1]{\ottdrule[#1]{%
\ottpremise{\Gamma  \vdash  \ottnt{a_{{\mathrm{1}}}}  \Leftrightarrow  \ottnt{a_{{\mathrm{2}}}}}%
}{
\Gamma  \vdash   \lambda  \ottmv{x} .  \ottnt{a_{{\mathrm{1}}}}   \Leftrightarrow   \lambda  \ottmv{x} .  \ottnt{a_{{\mathrm{2}}}} }{%
{\ottdrulename{eq\_abs}}{}%
}}

\newcommand{\ottdruleeqXXpi}[1]{\ottdrule[#1]{%
\ottpremise{\Gamma  \vdash  \ottnt{A_{{\mathrm{1}}}}  \Leftrightarrow  \ottnt{A_{{\mathrm{2}}}}}%
\ottpremise{\Gamma  \vdash  \ottnt{B_{{\mathrm{1}}}}  \Leftrightarrow  \ottnt{B_{{\mathrm{2}}}}}%
}{
\Gamma  \vdash   ( \ottmv{x} \!:\! \ottnt{A_{{\mathrm{1}}}} )  \to   \ottnt{B_{{\mathrm{1}}}}   \Leftrightarrow   ( \ottmv{x} \!:\! \ottnt{A_{{\mathrm{2}}}} )  \to   \ottnt{B_{{\mathrm{2}}}} }{%
{\ottdrulename{eq\_pi}}{}%
}}

\newcommand{\ottdruleeqXXapp}[1]{\ottdrule[#1]{%
\ottpremise{\Gamma  \vdash  \ottnt{ne_{{\mathrm{1}}}}  \Leftrightarrow  \ottnt{ne_{{\mathrm{2}}}}}%
\ottpremise{\Gamma  \vdash  \ottnt{b_{{\mathrm{1}}}}  \Leftrightarrow  \ottnt{b_{{\mathrm{2}}}}}%
}{
\Gamma  \vdash   \ottnt{ne_{{\mathrm{1}}}}  \;  \ottnt{b_{{\mathrm{1}}}}   \Leftrightarrow   \ottnt{ne_{{\mathrm{2}}}}  \;  \ottnt{b_{{\mathrm{2}}}} }{%
{\ottdrulename{eq\_app}}{}%
}}

\newcommand{\ottdefnequate}[1]{\begin{ottdefnblock}[#1]{$\Gamma  \vdash  \ottnt{a}  \Leftrightarrow  \ottnt{b}$}{}
 compare $a$ and $b$ for equality \ottusedrule{\ottdruleeqXXrefl{}}
\ottusedrule{\ottdruleeqXXwhnf{}}
\ottusedrule{\ottdruleeqXXabs{}}
\ottusedrule{\ottdruleeqXXpi{}}
\ottusedrule{\ottdruleeqXXapp{}}
\end{ottdefnblock}}

\newcommand{\ottdefnsJEquate}{
\ottdefnequate{}}

\newcommand{\ottdrulewhnfXXlamXXbeta}[1]{\ottdrule[#1]{%
\ottpremise{ \Gamma   \vdash   \ottkw{whnf} \  \ottnt{a}  \leadsto  \ottsym{(}   \lambda  \ottmv{x} .  \ottnt{a'}   \ottsym{)} }%
\ottpremise{ \Gamma   \vdash   \ottkw{whnf} \  \ottsym{(}  \ottnt{a'}  \ottsym{[}  \ottnt{b}  \ottsym{/}  \ottmv{x}  \ottsym{]}  \ottsym{)}  \leadsto  \ottnt{nf} }%
}{
 \Gamma   \vdash   \ottkw{whnf} \   \ottnt{a}  \;  \ottnt{b}   \leadsto  \ottnt{nf} }{%
{\ottdrulename{whnf\_lam\_beta}}{}%
}}

\newcommand{\ottdrulewhnfXXtype}[1]{\ottdrule[#1]{%
}{
 \Gamma   \vdash   \ottkw{whnf} \  \ottkw{Type}  \leadsto  \ottkw{Type} }{%
{\ottdrulename{whnf\_type}}{}%
}}

\newcommand{\ottdrulewhnfXXlam}[1]{\ottdrule[#1]{%
}{
 \Gamma   \vdash   \ottkw{whnf} \   \lambda  \ottmv{x} .  \ottnt{a}   \leadsto   \lambda  \ottmv{x} .  \ottnt{a}  }{%
{\ottdrulename{whnf\_lam}}{}%
}}

\newcommand{\ottdrulewhnfXXpi}[1]{\ottdrule[#1]{%
}{
 \Gamma   \vdash   \ottkw{whnf} \   ( \ottmv{x} \!:\! \ottnt{A} )  \to   \ottnt{B}   \leadsto   ( \ottmv{x} \!:\! \ottnt{A} )  \to   \ottnt{B}  }{%
{\ottdrulename{whnf\_pi}}{}%
}}

\newcommand{\ottdrulewhnfXXvar}[1]{\ottdrule[#1]{%
\ottpremise{\ottmv{x}  \ottsym{=}  \ottnt{a} \, \in \, \Gamma}%
\ottpremise{ \Gamma   \vdash   \ottkw{whnf} \  \ottnt{a}  \leadsto  \ottnt{nf} }%
}{
 \Gamma   \vdash   \ottkw{whnf} \  \ottmv{x}  \leadsto  \ottnt{nf} }{%
{\ottdrulename{whnf\_var}}{}%
}}

\newcommand{\ottdrulewhnfXXapp}[1]{\ottdrule[#1]{%
\ottpremise{ \Gamma   \vdash   \ottkw{whnf} \  \ottnt{a}  \leadsto  \ottnt{ne} }%
}{
 \Gamma   \vdash   \ottkw{whnf} \   \ottnt{a}  \;  \ottnt{b}   \leadsto   \ottnt{ne}  \;  \ottnt{b}  }{%
{\ottdrulename{whnf\_app}}{}%
}}

\newcommand{\ottdrulewhnfXXannot}[1]{\ottdrule[#1]{%
\ottpremise{ \Gamma   \vdash   \ottkw{whnf} \  \ottnt{a}  \leadsto  \ottnt{nf} }%
}{
 \Gamma   \vdash   \ottkw{whnf} \  \ottsym{(}  \ottnt{a}  \ottsym{:}  \ottnt{A}  \ottsym{)}  \leadsto  \ottnt{nf} }{%
{\ottdrulename{whnf\_annot}}{}%
}}

\newcommand{\ottdrulewhnfXXletpair}[1]{\ottdrule[#1]{%
\ottpremise{ \Gamma   \vdash   \ottkw{whnf} \  \ottnt{a}  \leadsto  \ottsym{(}  \ottnt{a_{{\mathrm{1}}}}  \ottsym{,}  \ottnt{a_{{\mathrm{2}}}}  \ottsym{)} }%
\ottpremise{ \Gamma   \vdash   \ottkw{whnf} \  \ottsym{(}  \ottnt{b}  \ottsym{[}  \ottnt{a_{{\mathrm{1}}}}  \ottsym{/}  \ottmv{x}  \ottsym{]}  \ottsym{[}  \ottnt{a_{{\mathrm{2}}}}  \ottsym{/}  \ottmv{y}  \ottsym{]}  \ottsym{)}  \leadsto  \ottnt{nf} }%
}{
 \Gamma   \vdash   \ottkw{whnf} \  \ottkw{let} \, \ottsym{(}  \ottmv{x}  \ottsym{,}  \ottmv{y}  \ottsym{)}  \ottsym{=}  \ottnt{a} \, \ottkw{in} \, \ottnt{b}  \leadsto  \ottnt{nf} }{%
{\ottdrulename{whnf\_letpair}}{}%
}}

\newcommand{\ottdrulewhnfXXprodcong}[1]{\ottdrule[#1]{%
\ottpremise{ \Gamma   \vdash   \ottkw{whnf} \  \ottnt{a}  \leadsto  \ottnt{ne} }%
}{
 \Gamma   \vdash   \ottkw{whnf} \  \ottkw{let} \, \ottsym{(}  \ottmv{x}  \ottsym{,}  \ottmv{y}  \ottsym{)}  \ottsym{=}  \ottnt{a} \, \ottkw{in} \, \ottnt{b}  \leadsto  \ottkw{let} \, \ottsym{(}  \ottmv{x}  \ottsym{,}  \ottmv{y}  \ottsym{)}  \ottsym{=}  \ottnt{ne} \, \ottkw{in} \, \ottnt{b} }{%
{\ottdrulename{whnf\_prodcong}}{}%
}}

\newcommand{\ottdrulewhnfXXifXXtrue}[1]{\ottdrule[#1]{%
\ottpremise{ \Gamma   \vdash   \ottkw{whnf} \  \ottnt{a}  \leadsto  \ottkw{True} }%
\ottpremise{ \Gamma   \vdash   \ottkw{whnf} \  \ottnt{b_{{\mathrm{1}}}}  \leadsto  \ottnt{nf} }%
}{
 \Gamma   \vdash   \ottkw{whnf} \  \ottkw{if} \, \ottnt{a} \, \ottkw{then} \, \ottnt{b_{{\mathrm{1}}}} \, \ottkw{else} \, \ottnt{b_{{\mathrm{2}}}}  \leadsto  \ottnt{nf} }{%
{\ottdrulename{whnf\_if\_true}}{}%
}}

\newcommand{\ottdrulewhnfXXifXXfalse}[1]{\ottdrule[#1]{%
\ottpremise{ \Gamma   \vdash   \ottkw{whnf} \  \ottnt{a}  \leadsto  \ottkw{False} }%
\ottpremise{ \Gamma   \vdash   \ottkw{whnf} \  \ottnt{b_{{\mathrm{2}}}}  \leadsto  \ottnt{nf} }%
}{
 \Gamma   \vdash   \ottkw{whnf} \  \ottkw{if} \, \ottnt{a} \, \ottkw{then} \, \ottnt{b_{{\mathrm{1}}}} \, \ottkw{else} \, \ottnt{b_{{\mathrm{2}}}}  \leadsto  \ottnt{nf} }{%
{\ottdrulename{whnf\_if\_false}}{}%
}}

\newcommand{\ottdrulewhnfXXbool}[1]{\ottdrule[#1]{%
}{
 \Gamma   \vdash   \ottkw{whnf} \  \ottkw{Bool}  \leadsto  \ottkw{Bool} }{%
{\ottdrulename{whnf\_bool}}{}%
}}

\newcommand{\ottdrulewhnfXXtrue}[1]{\ottdrule[#1]{%
}{
 \Gamma   \vdash   \ottkw{whnf} \  \ottkw{True}  \leadsto  \ottkw{True} }{%
{\ottdrulename{whnf\_true}}{}%
}}

\newcommand{\ottdrulewhnfXXfalse}[1]{\ottdrule[#1]{%
}{
 \Gamma   \vdash   \ottkw{whnf} \  \ottkw{False}  \leadsto  \ottkw{False} }{%
{\ottdrulename{whnf\_false}}{}%
}}

\newcommand{\ottdrulewhnfXXifXXcong}[1]{\ottdrule[#1]{%
\ottpremise{ \Gamma   \vdash   \ottkw{whnf} \  \ottnt{a}  \leadsto  \ottnt{ne} }%
}{
 \Gamma   \vdash   \ottkw{whnf} \  \ottkw{if} \, \ottnt{a} \, \ottkw{then} \, \ottnt{b_{{\mathrm{1}}}} \, \ottkw{else} \, \ottnt{b_{{\mathrm{2}}}}  \leadsto  \ottkw{if} \, \ottnt{ne} \, \ottkw{then} \, \ottnt{b_{{\mathrm{1}}}} \, \ottkw{else} \, \ottnt{b_{{\mathrm{2}}}} }{%
{\ottdrulename{whnf\_if\_cong}}{}%
}}

\newcommand{\ottdrulewhnfXXtyeq}[1]{\ottdrule[#1]{%
}{
 \Gamma   \vdash   \ottkw{whnf} \  \ottsym{(}  \ottnt{a}  \ottsym{=}  \ottnt{b}  \ottsym{)}  \leadsto  \ottsym{(}  \ottnt{a}  \ottsym{=}  \ottnt{b}  \ottsym{)} }{%
{\ottdrulename{whnf\_tyeq}}{}%
}}

\newcommand{\ottdrulewhnfXXrefl}[1]{\ottdrule[#1]{%
}{
 \Gamma   \vdash   \ottkw{whnf} \  \ottkw{refl}  \leadsto  \ottkw{refl} }{%
{\ottdrulename{whnf\_refl}}{}%
}}

\newcommand{\ottdrulewhnfXXsubst}[1]{\ottdrule[#1]{%
\ottpremise{ \Gamma   \vdash   \ottkw{whnf} \  \ottnt{b}  \leadsto  \ottkw{refl} }%
\ottpremise{ \Gamma   \vdash   \ottkw{whnf} \  \ottnt{a}  \leadsto  \ottnt{nf} }%
}{
 \Gamma   \vdash   \ottkw{whnf} \  \ottsym{(}  \ottkw{subst} \, \ottnt{a} \, \ottkw{by} \, \ottnt{b}  \ottsym{)}  \leadsto  \ottnt{nf} }{%
{\ottdrulename{whnf\_subst}}{}%
}}

\newcommand{\ottdrulewhnfXXsubstXXcong}[1]{\ottdrule[#1]{%
\ottpremise{ \Gamma   \vdash   \ottkw{whnf} \  \ottnt{b}  \leadsto  \ottnt{ne} }%
}{
 \Gamma   \vdash   \ottkw{whnf} \  \ottsym{(}  \ottkw{subst} \, \ottnt{a} \, \ottkw{by} \, \ottnt{ne}  \ottsym{)}  \leadsto  \ottsym{(}  \ottkw{subst} \, \ottnt{a} \, \ottkw{by} \, \ottnt{ne}  \ottsym{)} }{%
{\ottdrulename{whnf\_subst\_cong}}{}%
}}

\newcommand{\ottdrulewhnfXXtcon}[1]{\ottdrule[#1]{%
}{
 \Gamma   \vdash   \ottkw{whnf} \  \ottmv{T} \, \overline{a}  \leadsto  \ottmv{T} \, \overline{a} }{%
{\ottdrulename{whnf\_tcon}}{}%
}}

\newcommand{\ottdrulewhnfXXdcon}[1]{\ottdrule[#1]{%
}{
 \Gamma   \vdash   \ottkw{whnf} \  \ottmv{K} \, \overline{a}  \leadsto  \ottmv{K} \, \overline{a} }{%
{\ottdrulename{whnf\_dcon}}{}%
}}

\newcommand{\ottdrulewhnfXXcase}[1]{\ottdrule[#1]{%
\ottpremise{\ottkw{match} \, \ottsym{(}  \ottmv{K} \, \overline{a}  \ottsym{)}  \ottsym{(pati}  \to  \ottsym{ai)}  \leadsto  \ottnt{b}}%
\ottpremise{ \Gamma   \vdash   \ottkw{whnf} \  \ottnt{b}  \leadsto  \ottnt{nf} }%
}{
 \Gamma   \vdash   \ottkw{whnf} \  \ottkw{case} \, \ottsym{(}  \ottmv{K} \, \overline{a}  \ottsym{)} \, \ottkw{of} \, \ottsym{\{} \, \ottcomp{\ottnt{pat_{\ottmv{i}}}  \to  \ottnt{a_{\ottmv{i}}}}{\ottmv{i}} \, \ottsym{\}}  \leadsto  \ottnt{nf} }{%
{\ottdrulename{whnf\_case}}{}%
}}

\newcommand{\ottdrulewhnfXXcaseXXcong}[1]{\ottdrule[#1]{%
\ottpremise{\ottnt{a}  \leadsto  \ottnt{ne}}%
}{
 \Gamma   \vdash   \ottkw{whnf} \  \ottkw{case} \, \ottnt{a} \, \ottkw{of} \, \ottsym{\{} \, \ottcomp{\ottnt{pat_{\ottmv{i}}}  \to  \ottnt{a_{\ottmv{i}}}}{\ottmv{i}} \, \ottsym{\}}  \leadsto  \ottkw{case} \, \ottnt{ne} \, \ottkw{of} \, \ottsym{\{} \, \ottcomp{\ottnt{pat_{\ottmv{i}}}  \to  \ottnt{a_{\ottmv{i}}}}{\ottmv{i}} \, \ottsym{\}} }{%
{\ottdrulename{whnf\_case\_cong}}{}%
}}

\newcommand{\ottdrulewhnfXXdef}[1]{\ottdrule[#1]{%
\ottpremise{\ottmv{x}  \ottsym{=}  \ottnt{a} \, \in \, \Gamma}%
\ottpremise{ \Gamma   \vdash   \ottkw{whnf} \  \ottnt{a}  \leadsto  \ottnt{v} }%
}{
 \Gamma   \vdash   \ottkw{whnf} \  \ottmv{x}  \leadsto  \ottnt{v} }{%
{\ottdrulename{whnf\_def}}{}%
}}

\newcommand{\ottdrulewhnfXXletXXbeta}[1]{\ottdrule[#1]{%
\ottpremise{ \Gamma   \vdash   \ottkw{whnf} \  \ottsym{(}  \ottnt{b}  \ottsym{[}  \ottnt{a}  \ottsym{/}  \ottmv{x}  \ottsym{]}  \ottsym{)}  \leadsto  \ottnt{v} }%
}{
 \Gamma   \vdash   \ottkw{whnf} \  \ottsym{(}  \ottkw{let} \, \ottmv{x}  \ottsym{=}  \ottnt{a} \, \ottkw{in} \, \ottnt{b}  \ottsym{)}  \leadsto  \ottnt{v} }{%
{\ottdrulename{whnf\_let\_beta}}{}%
}}

\newcommand{\ottdefnwhnf}[1]{\begin{ottdefnblock}[#1]{$ \Gamma   \vdash   \ottkw{whnf} \  \ottnt{a}  \leadsto  \ottnt{nf} $}{\ottcom{weak head normalize a to nf (if possible)}}
\ottusedrule{\ottdrulewhnfXXlamXXbeta{}}
\ottusedrule{\ottdrulewhnfXXtype{}}
\ottusedrule{\ottdrulewhnfXXlam{}}
\ottusedrule{\ottdrulewhnfXXpi{}}
\ottusedrule{\ottdrulewhnfXXvar{}}
\ottusedrule{\ottdrulewhnfXXapp{}}
\ottusedrule{\ottdrulewhnfXXannot{}}
\ottusedrule{\ottdrulewhnfXXletpair{}}
\ottusedrule{\ottdrulewhnfXXprodcong{}}
\ottusedrule{\ottdrulewhnfXXifXXtrue{}}
\ottusedrule{\ottdrulewhnfXXifXXfalse{}}
\ottusedrule{\ottdrulewhnfXXbool{}}
\ottusedrule{\ottdrulewhnfXXtrue{}}
\ottusedrule{\ottdrulewhnfXXfalse{}}
\ottusedrule{\ottdrulewhnfXXifXXcong{}}
\ottusedrule{\ottdrulewhnfXXtyeq{}}
\ottusedrule{\ottdrulewhnfXXrefl{}}
\ottusedrule{\ottdrulewhnfXXsubst{}}
\ottusedrule{\ottdrulewhnfXXsubstXXcong{}}
\ottusedrule{\ottdrulewhnfXXtcon{}}
\ottusedrule{\ottdrulewhnfXXdcon{}}
\ottusedrule{\ottdrulewhnfXXcase{}}
\ottusedrule{\ottdrulewhnfXXcaseXXcong{}}
\ottusedrule{\ottdrulewhnfXXdef{}}
\ottusedrule{\ottdrulewhnfXXletXXbeta{}}
\end{ottdefnblock}}

\newcommand{\ottdefnsJwhnf}{
\ottdefnwhnf{}}

\newcommand{\ottdruledone}[1]{\ottdrule[#1]{%
}{
\ottnt{a}  \leadsto^{\ast}  \ottnt{a}}{%
{\ottdrulename{done}}{}%
}}

\newcommand{\ottdrulestep}[1]{\ottdrule[#1]{%
\ottpremise{\ottnt{a}  \leadsto  \ottnt{b}}%
\ottpremise{\ottnt{b}  \leadsto^{\ast}  \ottnt{a'}}%
}{
\ottnt{a}  \leadsto^{\ast}  \ottnt{a'}}{%
{\ottdrulename{step}}{}%
}}

\newcommand{\ottdefnreduction}[1]{\begin{ottdefnblock}[#1]{$\ottnt{a}  \leadsto^{\ast}  \ottnt{b}$}{\ottcom{multi-step head reduction}}
\ottusedrule{\ottdruledone{}}
\ottusedrule{\ottdrulestep{}}
\end{ottdefnblock}}

\newcommand{\ottdrulesXXlambeta}[1]{\ottdrule[#1]{%
}{
 \ottsym{(}   \lambda  \ottmv{x} .  \ottnt{a}   \ottsym{)}  \;  \ottnt{b}   \leadsto  \ottnt{a}  \ottsym{[}  \ottnt{b}  \ottsym{/}  \ottmv{x}  \ottsym{]}}{%
{\ottdrulename{s\_lambeta}}{}%
}}

\newcommand{\ottdrulesXXapp}[1]{\ottdrule[#1]{%
\ottpremise{\ottnt{a}  \leadsto  \ottnt{a'}}%
}{
 \ottnt{a}  \;  \ottnt{b}   \leadsto   \ottnt{a'}  \;  \ottnt{b} }{%
{\ottdrulename{s\_app}}{}%
}}

\newcommand{\ottdrulesXXLetPairProd}[1]{\ottdrule[#1]{%
}{
\ottkw{let} \, \ottsym{(}  \ottmv{x}  \ottsym{,}  \ottmv{y}  \ottsym{)}  \ottsym{=}  \ottsym{(}  \ottnt{a_{{\mathrm{1}}}}  \ottsym{,}  \ottnt{a_{{\mathrm{2}}}}  \ottsym{)} \, \ottkw{in} \, \ottnt{b}  \leadsto  \ottnt{b}  \ottsym{[}  \ottnt{a_{{\mathrm{1}}}}  \ottsym{/}  \ottmv{x}  \ottsym{]}  \ottsym{[}  \ottnt{a_{{\mathrm{2}}}}  \ottsym{/}  \ottmv{y}  \ottsym{]}}{%
{\ottdrulename{s\_LetPairProd}}{}%
}}

\newcommand{\ottdrulesXXLetPair}[1]{\ottdrule[#1]{%
\ottpremise{\ottnt{a}  \leadsto  \ottnt{a'}}%
}{
\ottkw{let} \, \ottsym{(}  \ottmv{x}  \ottsym{,}  \ottmv{y}  \ottsym{)}  \ottsym{=}  \ottnt{a} \, \ottkw{in} \, \ottnt{b}  \leadsto  \ottkw{let} \, \ottsym{(}  \ottmv{x}  \ottsym{,}  \ottmv{y}  \ottsym{)}  \ottsym{=}  \ottnt{a'} \, \ottkw{in} \, \ottnt{b}}{%
{\ottdrulename{s\_LetPair}}{}%
}}

\newcommand{\ottdrulesXXifXXtrue}[1]{\ottdrule[#1]{%
}{
\ottkw{if} \, \ottkw{True} \, \ottkw{then} \, \ottnt{b_{{\mathrm{1}}}} \, \ottkw{else} \, \ottnt{b_{{\mathrm{2}}}}  \leadsto  \ottnt{b_{{\mathrm{1}}}}}{%
{\ottdrulename{s\_if\_true}}{}%
}}

\newcommand{\ottdrulesXXifXXfalse}[1]{\ottdrule[#1]{%
}{
\ottkw{if} \, \ottkw{False} \, \ottkw{then} \, \ottnt{b_{{\mathrm{1}}}} \, \ottkw{else} \, \ottnt{b_{{\mathrm{2}}}}  \leadsto  \ottnt{b_{{\mathrm{1}}}}}{%
{\ottdrulename{s\_if\_false}}{}%
}}

\newcommand{\ottdrulesXXif}[1]{\ottdrule[#1]{%
\ottpremise{\ottnt{a}  \leadsto  \ottnt{a'}}%
}{
\ottkw{if} \, \ottnt{a} \, \ottkw{then} \, \ottnt{b_{{\mathrm{1}}}} \, \ottkw{else} \, \ottnt{b_{{\mathrm{2}}}}  \leadsto  \ottkw{if} \, \ottnt{a'} \, \ottkw{then} \, \ottnt{b_{{\mathrm{1}}}} \, \ottkw{else} \, \ottnt{b_{{\mathrm{2}}}}}{%
{\ottdrulename{s\_if}}{}%
}}

\newcommand{\ottdrulesXXletbeta}[1]{\ottdrule[#1]{%
}{
\ottkw{let} \, \ottmv{x}  \ottsym{=}  \ottnt{a} \, \ottkw{in} \, \ottnt{b}  \leadsto  \ottnt{b}  \ottsym{[}  \ottnt{a}  \ottsym{/}  \ottmv{x}  \ottsym{]}}{%
{\ottdrulename{s\_letbeta}}{}%
}}

\newcommand{\ottdefnstep}[1]{\begin{ottdefnblock}[#1]{$\ottnt{a}  \leadsto  \ottnt{b}$}{\ottcom{single-step operational semantics, i. e. head reduction}}
\ottusedrule{\ottdrulesXXlambeta{}}
\ottusedrule{\ottdrulesXXapp{}}
\ottusedrule{\ottdrulesXXLetPairProd{}}
\ottusedrule{\ottdrulesXXLetPair{}}
\ottusedrule{\ottdrulesXXifXXtrue{}}
\ottusedrule{\ottdrulesXXifXXfalse{}}
\ottusedrule{\ottdrulesXXif{}}
\ottusedrule{\ottdrulesXXletbeta{}}
\end{ottdefnblock}}

\newcommand{\ottdefnsJOp}{
\ottdefnreduction{}\ottdefnstep{}}

\newcommand{\ottdruleeXXbeta}[1]{\ottdrule[#1]{%
}{
\Gamma  \vdash   \ottsym{(}   \lambda  \ottmv{x} .  \ottnt{a}   \ottsym{)}  \;  \ottnt{b}   \ottsym{=}  \ottnt{a}  \ottsym{[}  \ottnt{b}  \ottsym{/}  \ottmv{x}  \ottsym{]}}{%
{\ottdrulename{e\_beta}}{}%
}}

\newcommand{\ottdruleeXXrefl}[1]{\ottdrule[#1]{%
}{
\Gamma  \vdash  \ottnt{A}  \ottsym{=}  \ottnt{A}}{%
{\ottdrulename{e\_refl}}{}%
}}

\newcommand{\ottdruleeXXsym}[1]{\ottdrule[#1]{%
\ottpremise{\Gamma  \vdash  \ottnt{A}  \ottsym{=}  \ottnt{B}}%
}{
\Gamma  \vdash  \ottnt{B}  \ottsym{=}  \ottnt{A}}{%
{\ottdrulename{e\_sym}}{}%
}}

\newcommand{\ottdruleeXXtrans}[1]{\ottdrule[#1]{%
\ottpremise{\Gamma  \vdash  \ottnt{A_{{\mathrm{1}}}}  \ottsym{=}  \ottnt{A_{{\mathrm{2}}}}}%
\ottpremise{\Gamma  \vdash  \ottnt{A_{{\mathrm{2}}}}  \ottsym{=}  \ottnt{A_{{\mathrm{3}}}}}%
}{
\Gamma  \vdash  \ottnt{A_{{\mathrm{1}}}}  \ottsym{=}  \ottnt{A_{{\mathrm{3}}}}}{%
{\ottdrulename{e\_trans}}{}%
}}

\newcommand{\ottdruleeXXlift}[1]{\ottdrule[#1]{%
\ottpremise{ \Gamma ,  \ottmv{x} \! :\! \ottnt{A}   \vdash  \ottnt{b}  \ottsym{:}  \ottnt{B}}%
\ottpremise{\Gamma  \vdash  \ottnt{a_{{\mathrm{1}}}}  \ottsym{=}  \ottnt{a_{{\mathrm{2}}}}}%
}{
\Gamma  \vdash  \ottnt{b}  \ottsym{[}  \ottnt{a_{{\mathrm{1}}}}  \ottsym{/}  \ottmv{x}  \ottsym{]}  \ottsym{=}  \ottnt{b}  \ottsym{[}  \ottnt{a_{{\mathrm{2}}}}  \ottsym{/}  \ottmv{x}  \ottsym{]}}{%
{\ottdrulename{e\_lift}}{}%
}}

\newcommand{\ottdruleeXXpi}[1]{\ottdrule[#1]{%
\ottpremise{\Gamma  \vdash  \ottnt{A_{{\mathrm{1}}}}  \ottsym{=}  \ottnt{A_{{\mathrm{2}}}}}%
\ottpremise{ \Gamma ,  \ottmv{x} \! :\! \ottnt{A_{{\mathrm{1}}}}   \vdash  \ottnt{B_{{\mathrm{1}}}}  \ottsym{=}  \ottnt{B_{{\mathrm{2}}}}}%
}{
\Gamma  \vdash   ( \ottmv{x} \!:\! \ottnt{A_{{\mathrm{1}}}} )  \to   \ottnt{B_{{\mathrm{1}}}}   \ottsym{=}   ( \ottmv{x} \!:\! \ottnt{A_{{\mathrm{2}}}} )  \to   \ottnt{B_{{\mathrm{2}}}} }{%
{\ottdrulename{e\_pi}}{}%
}}

\newcommand{\ottdruleeXXlam}[1]{\ottdrule[#1]{%
\ottpremise{ \Gamma ,  \ottmv{x} \! :\! \ottnt{A_{{\mathrm{1}}}}   \vdash  \ottnt{b_{{\mathrm{1}}}}  \ottsym{=}  \ottnt{b_{{\mathrm{2}}}}}%
}{
\Gamma  \vdash   \lambda  \ottmv{x} .  \ottnt{b_{{\mathrm{1}}}}   \ottsym{=}   \lambda  \ottmv{x} .  \ottnt{b_{{\mathrm{2}}}} }{%
{\ottdrulename{e\_lam}}{}%
}}

\newcommand{\ottdruleeXXapp}[1]{\ottdrule[#1]{%
\ottpremise{\Gamma  \vdash  \ottnt{a_{{\mathrm{1}}}}  \ottsym{=}  \ottnt{a_{{\mathrm{2}}}}}%
\ottpremise{\Gamma  \vdash  \ottnt{b_{{\mathrm{1}}}}  \ottsym{=}  \ottnt{b_{{\mathrm{2}}}}}%
}{
\Gamma  \vdash   \ottnt{a_{{\mathrm{1}}}}  \;  \ottnt{b_{{\mathrm{1}}}}   \ottsym{=}   \ottnt{a_{{\mathrm{2}}}}  \;  \ottnt{b_{{\mathrm{2}}}} }{%
{\ottdrulename{e\_app}}{}%
}}

\newcommand{\ottdruleeXXannot}[1]{\ottdrule[#1]{%
\ottpremise{\Gamma  \vdash  \ottnt{a_{{\mathrm{1}}}}  \ottsym{=}  \ottnt{a_{{\mathrm{2}}}}}%
}{
\Gamma  \vdash  \ottsym{(}  \ottnt{a_{{\mathrm{1}}}}  \ottsym{:}  \ottnt{A}  \ottsym{)}  \ottsym{=}  \ottnt{a_{{\mathrm{2}}}}}{%
{\ottdrulename{e\_annot}}{}%
}}

\newcommand{\ottdruleeXXletpairprod}[1]{\ottdrule[#1]{%
}{
\Gamma  \vdash  \ottkw{let} \, \ottsym{(}  \ottmv{x}  \ottsym{,}  \ottmv{y}  \ottsym{)}  \ottsym{=}  \ottsym{(}  \ottnt{a_{{\mathrm{1}}}}  \ottsym{,}  \ottnt{a_{{\mathrm{2}}}}  \ottsym{)} \, \ottkw{in} \, \ottnt{b}  \ottsym{=}  \ottnt{b}  \ottsym{[}  \ottnt{a_{{\mathrm{1}}}}  \ottsym{/}  \ottmv{x}  \ottsym{]}  \ottsym{[}  \ottnt{a_{{\mathrm{2}}}}  \ottsym{/}  \ottmv{y}  \ottsym{]}}{%
{\ottdrulename{e\_letpairprod}}{}%
}}

\newcommand{\ottdruleeXXsigma}[1]{\ottdrule[#1]{%
\ottpremise{\Gamma  \vdash  \ottnt{A_{{\mathrm{1}}}}  \ottsym{=}  \ottnt{A_{{\mathrm{2}}}}}%
\ottpremise{ \Gamma ,  \ottmv{x} \! :\! \ottnt{A_{{\mathrm{1}}}}   \vdash  \ottnt{B_{{\mathrm{1}}}}  \ottsym{=}  \ottnt{B_{{\mathrm{2}}}}}%
}{
\Gamma  \vdash   \{  \ottmv{x} \!:\! \ottnt{A_{{\mathrm{1}}}} \ |\  \ottnt{B_{{\mathrm{1}}}}  \}   \ottsym{=}   \{  \ottmv{x} \!:\! \ottnt{A_{{\mathrm{2}}}} \ |\  \ottnt{B_{{\mathrm{2}}}}  \} }{%
{\ottdrulename{e\_sigma}}{}%
}}

\newcommand{\ottdruleeXXprod}[1]{\ottdrule[#1]{%
\ottpremise{\Gamma  \vdash  \ottnt{b_{{\mathrm{1}}}}  \ottsym{=}  \ottnt{b'_{{\mathrm{1}}}}}%
\ottpremise{\Gamma  \vdash  \ottnt{b_{{\mathrm{2}}}}  \ottsym{=}  \ottnt{b'_{{\mathrm{2}}}}}%
}{
\Gamma  \vdash  \ottsym{(}  \ottnt{b_{{\mathrm{1}}}}  \ottsym{,}  \ottnt{b_{{\mathrm{2}}}}  \ottsym{)}  \ottsym{=}  \ottsym{(}  \ottnt{b'_{{\mathrm{1}}}}  \ottsym{,}  \ottnt{b'_{{\mathrm{2}}}}  \ottsym{)}}{%
{\ottdrulename{e\_prod}}{}%
}}

\newcommand{\ottdruleeXXletpair}[1]{\ottdrule[#1]{%
\ottpremise{\Gamma  \vdash  \ottnt{a}  \ottsym{=}  \ottnt{a'}}%
\ottpremise{\Gamma  \vdash  \ottnt{b}  \ottsym{=}  \ottnt{b'}}%
}{
\Gamma  \vdash  \ottkw{let} \, \ottsym{(}  \ottmv{x}  \ottsym{,}  \ottmv{y}  \ottsym{)}  \ottsym{=}  \ottnt{a} \, \ottkw{in} \, \ottnt{b}  \ottsym{=}  \ottkw{let} \, \ottsym{(}  \ottmv{x}  \ottsym{,}  \ottmv{y}  \ottsym{)}  \ottsym{=}  \ottnt{a'} \, \ottkw{in} \, \ottnt{b'}}{%
{\ottdrulename{e\_letpair}}{}%
}}

\newcommand{\ottdruleeXXifXXtrue}[1]{\ottdrule[#1]{%
}{
\Gamma  \vdash  \ottkw{if} \, \ottkw{True} \, \ottkw{then} \, \ottnt{a} \, \ottkw{else} \, \ottnt{b}  \ottsym{=}  \ottnt{a}}{%
{\ottdrulename{e\_if\_true}}{}%
}}

\newcommand{\ottdruleeXXifXXfalse}[1]{\ottdrule[#1]{%
}{
\Gamma  \vdash  \ottkw{if} \, \ottkw{False} \, \ottkw{then} \, \ottnt{a} \, \ottkw{else} \, \ottnt{b}  \ottsym{=}  \ottnt{b}}{%
{\ottdrulename{e\_if\_false}}{}%
}}

\newcommand{\ottdruleeXXif}[1]{\ottdrule[#1]{%
\ottpremise{\Gamma  \vdash  \ottnt{a}  \ottsym{=}  \ottnt{a'}}%
\ottpremise{\Gamma  \vdash  \ottnt{b_{{\mathrm{1}}}}  \ottsym{=}  \ottnt{b'_{{\mathrm{1}}}}}%
\ottpremise{\Gamma  \vdash  \ottnt{b_{{\mathrm{2}}}}  \ottsym{=}  \ottnt{b'_{{\mathrm{2}}}}}%
}{
\Gamma  \vdash  \ottkw{if} \, \ottnt{a} \, \ottkw{then} \, \ottnt{b_{{\mathrm{1}}}} \, \ottkw{else} \, \ottnt{b_{{\mathrm{2}}}}  \ottsym{=}  \ottkw{if} \, \ottnt{a'} \, \ottkw{then} \, \ottnt{b'_{{\mathrm{1}}}} \, \ottkw{else} \, \ottnt{b'_{{\mathrm{2}}}}}{%
{\ottdrulename{e\_if}}{}%
}}

\newcommand{\ottdruleeXXifXXeta}[1]{\ottdrule[#1]{%
}{
\Gamma  \vdash  \ottkw{if} \, \ottnt{a} \, \ottkw{then} \, \ottnt{b} \, \ottkw{else} \, \ottnt{b}  \ottsym{=}  \ottnt{b}}{%
{\ottdrulename{e\_if\_eta}}{}%
}}

\newcommand{\ottdruleeXXsubstXXbeta}[1]{\ottdrule[#1]{%
}{
\Gamma  \vdash  \ottkw{subst} \, \ottnt{a} \, \ottkw{by} \, \ottkw{refl}  \ottsym{=}  \ottnt{a}}{%
{\ottdrulename{e\_subst\_beta}}{}%
}}

\newcommand{\ottdruleeXXsubst}[1]{\ottdrule[#1]{%
\ottpremise{\Gamma  \vdash  \ottnt{a}  \ottsym{=}  \ottnt{a'}}%
\ottpremise{\Gamma  \vdash  \ottnt{b}  \ottsym{=}  \ottnt{b'}}%
}{
\Gamma  \vdash  \ottkw{subst} \, \ottnt{a} \, \ottkw{by} \, \ottnt{b}  \ottsym{=}  \ottkw{subst} \, \ottnt{a'} \, \ottkw{by} \, \ottnt{b'}}{%
{\ottdrulename{e\_subst}}{}%
}}

\newcommand{\ottdruleeXXtyeq}[1]{\ottdrule[#1]{%
\ottpremise{\Gamma  \vdash  \ottnt{a_{{\mathrm{1}}}}  \ottsym{=}  \ottnt{b_{{\mathrm{1}}}}}%
\ottpremise{\Gamma  \vdash  \ottnt{a_{{\mathrm{2}}}}  \ottsym{=}  \ottnt{b_{{\mathrm{2}}}}}%
}{
\Gamma  \vdash  \ottsym{(}  \ottnt{a_{{\mathrm{1}}}}  \ottsym{=}  \ottnt{a_{{\mathrm{2}}}}  \ottsym{)}  \ottsym{=}  \ottsym{(}  \ottnt{b_{{\mathrm{1}}}}  \ottsym{=}  \ottnt{b_{{\mathrm{2}}}}  \ottsym{)}}{%
{\ottdrulename{e\_tyeq}}{}%
}}

\newcommand{\ottdruleeXXcontra}[1]{\ottdrule[#1]{%
\ottpremise{\Gamma  \vdash  \ottnt{a}  \ottsym{=}  \ottnt{a'}}%
}{
\Gamma  \vdash  \ottkw{contra} \, \ottnt{a}  \ottsym{=}  \ottkw{contra} \, \ottnt{a'}}{%
{\ottdrulename{e\_contra}}{}%
}}

\newcommand{\ottdruleeXXeapp}[1]{\ottdrule[#1]{%
\ottpremise{\Gamma  \vdash  \ottnt{a_{{\mathrm{1}}}}  \ottsym{=}  \ottnt{a_{{\mathrm{2}}}}}%
}{
\Gamma  \vdash  \ottnt{a_{{\mathrm{1}}}}  \ottsym{[}  \ottnt{b_{{\mathrm{1}}}}  \ottsym{]}  \ottsym{=}  \ottnt{a_{{\mathrm{2}}}}  \ottsym{[}  \ottnt{b_{{\mathrm{2}}}}  \ottsym{]}}{%
{\ottdrulename{e\_eapp}}{}%
}}

\newcommand{\ottdruleeXXelam}[1]{\ottdrule[#1]{%
\ottpremise{ \Gamma ,  \ottmv{x} \! :^  -  \! \ottnt{A}   \vdash  \ottnt{a_{{\mathrm{1}}}}  \ottsym{=}  \ottnt{a_{{\mathrm{2}}}}}%
}{
\Gamma  \vdash   \lambda [  \ottmv{x}  ] .  \ottnt{a_{{\mathrm{1}}}}   \ottsym{=}   \lambda [  \ottmv{x}  ] .  \ottnt{a_{{\mathrm{2}}}} }{%
{\ottdrulename{e\_elam}}{}%
}}

\newcommand{\ottdruleeXXepi}[1]{\ottdrule[#1]{%
\ottpremise{\Gamma  \vdash  \ottnt{A_{{\mathrm{1}}}}  \ottsym{=}  \ottnt{A_{{\mathrm{2}}}}}%
\ottpremise{ \Gamma ,  \ottmv{x} \! :^  +  \! \ottnt{A_{{\mathrm{1}}}}   \vdash  \ottnt{B_{{\mathrm{1}}}}  \ottsym{=}  \ottnt{B_{{\mathrm{2}}}}}%
}{
\Gamma  \vdash   [  \ottmv{x} \!:\! \ottnt{A_{{\mathrm{1}}}}  ] \rightarrow  \ottnt{B_{{\mathrm{1}}}}   \ottsym{=}   [  \ottmv{x} \!:\! \ottnt{A_{{\mathrm{2}}}}  ] \rightarrow  \ottnt{B_{{\mathrm{2}}}} }{%
{\ottdrulename{e\_epi}}{}%
}}

\newcommand{\ottdruleeXXvar}[1]{\ottdrule[#1]{%
\ottpremise{\ottmv{x}  \ottsym{=}  \ottnt{a} \, \in \, \Gamma}%
}{
\Gamma  \vdash  \ottmv{x}  \ottsym{=}  \ottnt{a}}{%
{\ottdrulename{e\_var}}{}%
}}

\newcommand{\ottdruleeXXletbeta}[1]{\ottdrule[#1]{%
}{
\Gamma  \vdash  \ottkw{let} \, \ottmv{x}  \ottsym{=}  \ottnt{a} \, \ottkw{in} \, \ottnt{b}  \ottsym{=}  \ottnt{b}  \ottsym{[}  \ottnt{a}  \ottsym{/}  \ottmv{x}  \ottsym{]}}{%
{\ottdrulename{e\_letbeta}}{}%
}}

\newcommand{\ottdruleeXXlet}[1]{\ottdrule[#1]{%
\ottpremise{\Gamma  \vdash  \ottnt{a_{{\mathrm{1}}}}  \ottsym{=}  \ottnt{a_{{\mathrm{2}}}}}%
\ottpremise{ \Gamma ,  \ottmv{x} \! :\! \ottnt{A}   \ottsym{,}  \ottmv{x}  \ottsym{=}  \ottnt{a_{{\mathrm{1}}}}  \vdash  \ottnt{b_{{\mathrm{1}}}}  \ottsym{=}  \ottnt{b_{{\mathrm{2}}}}}%
}{
\Gamma  \vdash  \ottkw{let} \, \ottmv{x}  \ottsym{=}  \ottnt{a_{{\mathrm{1}}}} \, \ottkw{in} \, \ottnt{b_{{\mathrm{1}}}}  \ottsym{=}  \ottkw{let} \, \ottmv{x}  \ottsym{=}  \ottnt{a_{{\mathrm{2}}}} \, \ottkw{in} \, \ottnt{b_{{\mathrm{2}}}}}{%
{\ottdrulename{e\_let}}{}%
}}

\newcommand{\ottdefneq}[1]{\begin{ottdefnblock}[#1]{$\Gamma  \vdash  \ottnt{A}  \ottsym{=}  \ottnt{B}$}{\ottcom{Definitional equality}}
\ottusedrule{\ottdruleeXXbeta{}}
\ottusedrule{\ottdruleeXXrefl{}}
\ottusedrule{\ottdruleeXXsym{}}
\ottusedrule{\ottdruleeXXtrans{}}
\ottusedrule{\ottdruleeXXlift{}}
\ottusedrule{\ottdruleeXXpi{}}
\ottusedrule{\ottdruleeXXlam{}}
\ottusedrule{\ottdruleeXXapp{}}
\ottusedrule{\ottdruleeXXannot{}}
\ottusedrule{\ottdruleeXXletpairprod{}}
\ottusedrule{\ottdruleeXXsigma{}}
\ottusedrule{\ottdruleeXXprod{}}
\ottusedrule{\ottdruleeXXletpair{}}
\ottusedrule{\ottdruleeXXifXXtrue{}}
\ottusedrule{\ottdruleeXXifXXfalse{}}
\ottusedrule{\ottdruleeXXif{}}
\ottusedrule{\ottdruleeXXifXXeta{}}
\ottusedrule{\ottdruleeXXsubstXXbeta{}}
\ottusedrule{\ottdruleeXXsubst{}}
\ottusedrule{\ottdruleeXXtyeq{}}
\ottusedrule{\ottdruleeXXcontra{}}
\ottusedrule{\ottdruleeXXeapp{}}
\ottusedrule{\ottdruleeXXelam{}}
\ottusedrule{\ottdruleeXXepi{}}
\ottusedrule{\ottdruleeXXvar{}}
\ottusedrule{\ottdruleeXXletbeta{}}
\ottusedrule{\ottdruleeXXlet{}}
\end{ottdefnblock}}

\newcommand{\ottdefnsJEq}{
\ottdefneq{}}

\newcommand{\ottdruletXXvar}[1]{\ottdrule[#1]{%
\ottpremise{\ottmv{x}  \ottsym{:}  \ottnt{A} \, \in \, \Gamma}%
}{
\Gamma  \vdash  \ottmv{x}  \ottsym{:}  \ottnt{A}}{%
{\ottdrulename{t\_var}}{}%
}}

\newcommand{\ottdruletXXslambda}[1]{\ottdrule[#1]{%
\ottpremise{ \Gamma ,  \ottmv{x} \! :\! \ottnt{A}   \vdash  \ottnt{a}  \ottsym{:}  \ottnt{B}}%
}{
\Gamma  \vdash   \lambda  \ottmv{x} .  \ottnt{a}   \ottsym{:}   ( \ottmv{x} \!:\! \ottnt{A} )  \to   \ottnt{B} }{%
{\ottdrulename{t\_slambda}}{}%
}}

\newcommand{\ottdruletXXlambda}[1]{\ottdrule[#1]{%
\ottpremise{ \Gamma ,  \ottmv{x} \! :\! \ottnt{A}   \vdash  \ottnt{a}  \ottsym{:}  \ottnt{B}}%
\ottpremise{\Gamma  \vdash  \ottnt{A}  \ottsym{:}  \ottkw{Type}}%
}{
\Gamma  \vdash   \lambda  \ottmv{x} .  \ottnt{a}   \ottsym{:}   ( \ottmv{x} \!:\! \ottnt{A} )  \to   \ottnt{B} }{%
{\ottdrulename{t\_lambda}}{}%
}}

\newcommand{\ottdruletXXaltlambda}[1]{\ottdrule[#1]{%
\ottpremise{ \Gamma ,  \ottmv{x} \! :\! \ottnt{A}   \vdash  \ottnt{a}  \ottsym{:}  \ottnt{B}  \ottsym{[}  \ottmv{y}  \ottsym{/}  \ottmv{x}  \ottsym{]}}%
\ottpremise{\Gamma  \vdash  \ottnt{A}  \ottsym{:}  \ottkw{Type}}%
}{
\Gamma  \vdash   \lambda  \ottmv{x} .  \ottnt{a}   \ottsym{:}   ( \ottmv{y} \!:\! \ottnt{A} )  \to   \ottnt{B} }{%
{\ottdrulename{t\_altlambda}}{}%
}}

\newcommand{\ottdruletXXpi}[1]{\ottdrule[#1]{%
\ottpremise{\Gamma  \vdash  \ottnt{A}  \ottsym{:}  \ottkw{Type}}%
\ottpremise{ \Gamma ,  \ottmv{x} \! :\! \ottnt{A}   \vdash  \ottnt{B}  \ottsym{:}  \ottkw{Type}}%
}{
\Gamma  \vdash   ( \ottmv{x} \!:\! \ottnt{A} )  \to   \ottnt{B}   \ottsym{:}  \ottkw{Type}}{%
{\ottdrulename{t\_pi}}{}%
}}

\newcommand{\ottdruletXXtype}[1]{\ottdrule[#1]{%
}{
\Gamma  \vdash  \ottkw{Type}  \ottsym{:}  \ottkw{Type}}{%
{\ottdrulename{t\_type}}{}%
}}

\newcommand{\ottdruletXXapp}[1]{\ottdrule[#1]{%
\ottpremise{\Gamma  \vdash  \ottnt{a}  \ottsym{:}   ( \ottmv{x} \!:\! \ottnt{A} )  \to   \ottnt{B} }%
\ottpremise{\Gamma  \vdash  \ottnt{b}  \ottsym{:}  \ottnt{A}}%
}{
\Gamma  \vdash   \ottnt{a}  \;  \ottnt{b}   \ottsym{:}  \ottnt{B}  \ottsym{[}  \ottnt{b}  \ottsym{/}  \ottmv{x}  \ottsym{]}}{%
{\ottdrulename{t\_app}}{}%
}}

\newcommand{\ottdruletXXconv}[1]{\ottdrule[#1]{%
\ottpremise{\Gamma  \vdash  \ottnt{a}  \ottsym{:}  \ottnt{A}}%
\ottpremise{\Gamma  \vdash  \ottnt{A}  \ottsym{=}  \ottnt{B}}%
}{
\Gamma  \vdash  \ottnt{a}  \ottsym{:}  \ottnt{B}}{%
{\ottdrulename{t\_conv}}{}%
}}

\newcommand{\ottdruletXXsigma}[1]{\ottdrule[#1]{%
\ottpremise{\Gamma  \vdash  \ottnt{A}  \ottsym{:}  \ottkw{Type}}%
\ottpremise{ \Gamma ,  \ottmv{x} \! :\! \ottnt{A}   \vdash  \ottnt{B}  \ottsym{:}  \ottkw{Type}}%
}{
\Gamma  \vdash   \{  \ottmv{x} \!:\! \ottnt{A} \ |\  \ottnt{B}  \}   \ottsym{:}  \ottkw{Type}}{%
{\ottdrulename{t\_sigma}}{}%
}}

\newcommand{\ottdruletXXpair}[1]{\ottdrule[#1]{%
\ottpremise{\Gamma  \vdash  \ottnt{a}  \ottsym{:}  \ottnt{A}}%
\ottpremise{\Gamma  \vdash  \ottnt{b}  \ottsym{:}  \ottnt{B}  \ottsym{[}  \ottnt{a}  \ottsym{/}  \ottmv{x}  \ottsym{]}}%
}{
\Gamma  \vdash  \ottsym{(}  \ottnt{a}  \ottsym{,}  \ottnt{b}  \ottsym{)}  \ottsym{:}   \{  \ottmv{x} \!:\! \ottnt{A} \ |\  \ottnt{B}  \} }{%
{\ottdrulename{t\_pair}}{}%
}}

\newcommand{\ottdruletXXletpairXXweak}[1]{\ottdrule[#1]{%
\ottpremise{\Gamma  \vdash  \ottnt{a}  \ottsym{:}   \{  \ottmv{x} \!:\! \ottnt{A_{{\mathrm{1}}}} \ |\  \ottnt{A_{{\mathrm{2}}}}  \} }%
\ottpremise{  \Gamma ,  \ottmv{x} \! :\! \ottnt{A_{{\mathrm{1}}}}  ,  \ottmv{y} \! :\! \ottnt{A_{{\mathrm{2}}}}   \vdash  \ottnt{b}  \ottsym{:}  \ottnt{B}}%
\ottpremise{\Gamma  \vdash  \ottnt{B}  \ottsym{:}  \ottkw{Type}}%
}{
\Gamma  \vdash  \ottkw{let} \, \ottsym{(}  \ottmv{x}  \ottsym{,}  \ottmv{y}  \ottsym{)}  \ottsym{=}  \ottnt{a} \, \ottkw{in} \, \ottnt{b}  \ottsym{:}  \ottnt{B}}{%
{\ottdrulename{t\_letpair\_weak}}{}%
}}

\newcommand{\ottdruletXXbool}[1]{\ottdrule[#1]{%
}{
\Gamma  \vdash  \ottkw{Bool}  \ottsym{:}  \ottkw{Type}}{%
{\ottdrulename{t\_bool}}{}%
}}

\newcommand{\ottdruletXXtrue}[1]{\ottdrule[#1]{%
}{
\Gamma  \vdash  \ottkw{True}  \ottsym{:}  \ottkw{Bool}}{%
{\ottdrulename{t\_true}}{}%
}}

\newcommand{\ottdruletXXfalse}[1]{\ottdrule[#1]{%
}{
\Gamma  \vdash  \ottkw{False}  \ottsym{:}  \ottkw{Bool}}{%
{\ottdrulename{t\_false}}{}%
}}

\newcommand{\ottdruletXXifXXsimple}[1]{\ottdrule[#1]{%
\ottpremise{\Gamma  \vdash  \ottnt{a}  \ottsym{:}  \ottkw{Bool}}%
\ottpremise{\Gamma  \vdash  \ottnt{b_{{\mathrm{1}}}}  \ottsym{:}  \ottnt{A}}%
\ottpremise{\Gamma  \vdash  \ottnt{b_{{\mathrm{2}}}}  \ottsym{:}  \ottnt{A}}%
}{
\Gamma  \vdash  \ottkw{if} \, \ottnt{a} \, \ottkw{then} \, \ottnt{b_{{\mathrm{1}}}} \, \ottkw{else} \, \ottnt{b_{{\mathrm{2}}}}  \ottsym{:}  \ottnt{A}}{%
{\ottdrulename{t\_if\_simple}}{}%
}}

\newcommand{\ottdruletXXifXXfull}[1]{\ottdrule[#1]{%
\ottpremise{\Gamma  \vdash  \ottnt{a}  \ottsym{:}  \ottkw{Bool}}%
\ottpremise{\Gamma  \vdash  \ottnt{b_{{\mathrm{1}}}}  \ottsym{:}  \ottnt{A}  \ottsym{[}  \ottkw{True}  \ottsym{/}  \ottmv{x}  \ottsym{]}}%
\ottpremise{\Gamma  \vdash  \ottnt{b_{{\mathrm{2}}}}  \ottsym{:}  \ottnt{A}  \ottsym{[}  \ottkw{False}  \ottsym{/}  \ottmv{x}  \ottsym{]}}%
\ottpremise{ \Gamma ,  \ottmv{x} \! :\! \ottkw{Bool}   \vdash  \ottnt{A}  \ottsym{:}  \ottkw{Type}}%
}{
\Gamma  \vdash  \ottkw{if} \, \ottnt{a} \, \ottkw{then} \, \ottnt{b_{{\mathrm{1}}}} \, \ottkw{else} \, \ottnt{b_{{\mathrm{2}}}}  \ottsym{:}  \ottnt{A}  \ottsym{[}  \ottnt{a}  \ottsym{/}  \ottmv{x}  \ottsym{]}}{%
{\ottdrulename{t\_if\_full}}{}%
}}

\newcommand{\ottdruletXXif}[1]{\ottdrule[#1]{%
\ottpremise{\Gamma  \vdash  \ottmv{x}  \ottsym{:}  \ottkw{Bool}}%
\ottpremise{\Gamma  \vdash  \ottnt{b_{{\mathrm{1}}}}  \ottsym{:}  \ottnt{A}  \ottsym{[}  \ottkw{True}  \ottsym{/}  \ottmv{x}  \ottsym{]}}%
\ottpremise{\Gamma  \vdash  \ottnt{b_{{\mathrm{2}}}}  \ottsym{:}  \ottnt{A}  \ottsym{[}  \ottkw{False}  \ottsym{/}  \ottmv{x}  \ottsym{]}}%
\ottpremise{\Gamma  \vdash  \ottnt{A}  \ottsym{:}  \ottkw{Type}}%
}{
\Gamma  \vdash  \ottkw{if} \, \ottmv{x} \, \ottkw{then} \, \ottnt{b_{{\mathrm{1}}}} \, \ottkw{else} \, \ottnt{b_{{\mathrm{2}}}}  \ottsym{:}  \ottnt{A}}{%
{\ottdrulename{t\_if}}{}%
}}

\newcommand{\ottdruletXXeq}[1]{\ottdrule[#1]{%
\ottpremise{\Gamma  \vdash  \ottnt{a}  \ottsym{:}  \ottnt{A}}%
\ottpremise{\Gamma  \vdash  \ottnt{b}  \ottsym{:}  \ottnt{A}}%
}{
\Gamma  \vdash  \ottnt{a}  \ottsym{=}  \ottnt{b}  \ottsym{:}  \ottkw{Type}}{%
{\ottdrulename{t\_eq}}{}%
}}

\newcommand{\ottdruletXXrefl}[1]{\ottdrule[#1]{%
\ottpremise{\Gamma  \vdash  \ottnt{a}  \ottsym{=}  \ottnt{b}}%
\ottpremise{\Gamma  \vdash  \ottnt{a}  \ottsym{=}  \ottnt{b}  \ottsym{:}  \ottkw{Type}}%
}{
\Gamma  \vdash  \ottkw{refl}  \ottsym{:}  \ottnt{a}  \ottsym{=}  \ottnt{b}}{%
{\ottdrulename{t\_refl}}{}%
}}

\newcommand{\ottdruletXXreflXXalt}[1]{\ottdrule[#1]{%
\ottpremise{\Gamma  \vdash  \ottnt{a}  \ottsym{:}  \ottnt{A}}%
}{
\Gamma  \vdash  \ottkw{refl}  \ottsym{:}  \ottnt{a}  \ottsym{=}  \ottnt{a}}{%
{\ottdrulename{t\_refl\_alt}}{}%
}}

\newcommand{\ottdruletXXsubstXXsimple}[1]{\ottdrule[#1]{%
\ottpremise{\Gamma  \vdash  \ottnt{a}  \ottsym{:}  \ottnt{A}  \ottsym{[}  \ottnt{a_{{\mathrm{1}}}}  \ottsym{/}  \ottmv{x}  \ottsym{]}}%
\ottpremise{\Gamma  \vdash  \ottnt{b}  \ottsym{:}  \ottnt{a_{{\mathrm{1}}}}  \ottsym{=}  \ottnt{a_{{\mathrm{2}}}}}%
}{
\Gamma  \vdash  \ottkw{subst} \, \ottnt{a} \, \ottkw{by} \, \ottnt{b}  \ottsym{:}  \ottnt{A}  \ottsym{[}  \ottnt{a_{{\mathrm{2}}}}  \ottsym{/}  \ottmv{x}  \ottsym{]}}{%
{\ottdrulename{t\_subst\_simple}}{}%
}}

\newcommand{\ottdruletXXsubst}[1]{\ottdrule[#1]{%
\ottpremise{\Gamma  \vdash  \ottnt{a}  \ottsym{:}  \ottnt{A}  \ottsym{[}  \ottnt{a_{{\mathrm{1}}}}  \ottsym{/}  \ottmv{x}  \ottsym{]}  \ottsym{[}  \ottkw{refl}  \ottsym{/}  \ottmv{y}  \ottsym{]}}%
\ottpremise{\Gamma  \vdash  \ottnt{b}  \ottsym{:}  \ottnt{a_{{\mathrm{1}}}}  \ottsym{=}  \ottnt{a_{{\mathrm{2}}}}}%
}{
\Gamma  \vdash  \ottkw{subst} \, \ottnt{a} \, \ottkw{by} \, \ottnt{b}  \ottsym{:}  \ottnt{A}  \ottsym{[}  \ottnt{a_{{\mathrm{2}}}}  \ottsym{/}  \ottmv{x}  \ottsym{]}  \ottsym{[}  \ottnt{b}  \ottsym{/}  \ottmv{y}  \ottsym{]}}{%
{\ottdrulename{t\_subst}}{}%
}}

\newcommand{\ottdruletXXcontra}[1]{\ottdrule[#1]{%
\ottpremise{\Gamma  \vdash  \ottnt{A}  \ottsym{:}  \ottkw{Type}}%
\ottpremise{\Gamma  \vdash  \ottnt{a}  \ottsym{:}  \ottkw{True}  \ottsym{=}  \ottkw{False}}%
}{
\Gamma  \vdash  \ottkw{contra} \, \ottnt{a}  \ottsym{:}  \ottnt{A}}{%
{\ottdrulename{t\_contra}}{}%
}}

\newcommand{\ottdruletXXevar}[1]{\ottdrule[#1]{%
\ottpremise{ \ottmv{x} :^  +  \ottnt{A}   \in   \Gamma }%
}{
\Gamma  \vdash  \ottmv{x}  \ottsym{:}  \ottnt{A}}{%
{\ottdrulename{t\_evar}}{}%
}}

\newcommand{\ottdruletXXelambda}[1]{\ottdrule[#1]{%
\ottpremise{ \Gamma ,  \ottmv{x} \! :^  -  \! \ottnt{A}   \vdash  \ottnt{a}  \ottsym{:}  \ottnt{B}}%
\ottpremise{ \Gamma ^  +    \vdash  \ottnt{A}  \ottsym{:}  \ottkw{Type}}%
}{
\Gamma  \vdash   \lambda [  \ottmv{x}  ] .  \ottnt{a}   \ottsym{:}   [  \ottmv{x} \!:\! \ottnt{A}  ] \rightarrow  \ottnt{B} }{%
{\ottdrulename{t\_elambda}}{}%
}}

\newcommand{\ottdruletXXeapp}[1]{\ottdrule[#1]{%
\ottpremise{\Gamma  \vdash  \ottnt{a}  \ottsym{:}   [  \ottmv{x} \!:\! \ottnt{A}  ] \rightarrow  \ottnt{B} }%
\ottpremise{ \Gamma ^  +    \vdash  \ottnt{b}  \ottsym{:}  \ottnt{A}}%
}{
\Gamma  \vdash  \ottnt{a}  \ottsym{[}  \ottnt{b}  \ottsym{]}  \ottsym{:}  \ottnt{B}  \ottsym{[}  \ottnt{b}  \ottsym{/}  \ottmv{x}  \ottsym{]}}{%
{\ottdrulename{t\_eapp}}{}%
}}

\newcommand{\ottdruletXXepi}[1]{\ottdrule[#1]{%
\ottpremise{\Gamma  \vdash  \ottnt{A}  \ottsym{:}  \ottkw{Type}}%
\ottpremise{ \Gamma ,  \ottmv{x} \! :^  +  \! \ottnt{A}   \vdash  \ottnt{B}  \ottsym{:}  \ottkw{Type}}%
}{
\Gamma  \vdash   [  \ottmv{x} \!:\! \ottnt{A}  ] \rightarrow  \ottnt{B}   \ottsym{:}  \ottkw{Type}}{%
{\ottdrulename{t\_epi}}{}%
}}

\newcommand{\ottdruletXXNat}[1]{\ottdrule[#1]{%
}{
\Gamma  \vdash  \ottkw{Nat}  \ottsym{:}  \ottkw{Type}}{%
{\ottdrulename{t\_Nat}}{}%
}}

\newcommand{\ottdruletXXVoid}[1]{\ottdrule[#1]{%
}{
\Gamma  \vdash  \ottkw{Void}  \ottsym{:}  \ottkw{Type}}{%
{\ottdrulename{t\_Void}}{}%
}}

\newcommand{\ottdruletXXZero}[1]{\ottdrule[#1]{%
}{
\Gamma  \vdash  \ottkw{Zero}  \ottsym{:}  \ottkw{Nat}}{%
{\ottdrulename{t\_Zero}}{}%
}}

\newcommand{\ottdruletXXSucc}[1]{\ottdrule[#1]{%
\ottpremise{\Gamma  \vdash  \ottmv{n}  \ottsym{:}  \ottkw{Nat}}%
}{
\Gamma  \vdash  \ottkw{Succ} \, \ottmv{n}  \ottsym{:}  \ottkw{Nat}}{%
{\ottdrulename{t\_Succ}}{}%
}}

\newcommand{\ottdruletXXImTrue}[1]{\ottdrule[#1]{%
\ottpremise{\Gamma  \vdash  \ottnt{a_{{\mathrm{1}}}}  \ottsym{:}  \ottkw{Bool}}%
\ottpremise{\Gamma  \vdash  \ottnt{a_{{\mathrm{2}}}}  \ottsym{:}  \ottnt{a_{{\mathrm{1}}}}  \ottsym{=}  \ottkw{True}}%
}{
\Gamma  \vdash  \ottkw{ImTrue} \, \ottnt{a_{{\mathrm{1}}}} \, \ottnt{a_{{\mathrm{2}}}}  \ottsym{:}  \ottkw{SillyBool}}{%
{\ottdrulename{t\_ImTrue}}{}%
}}

\newcommand{\ottdruletXXcaseXXsimple}[1]{\ottdrule[#1]{%
\ottpremise{\Gamma  \vdash  \ottnt{a}  \ottsym{:}  \ottmv{T} \, }%
\ottpremise{\ottcomp{\Gamma  \vdash  \ottnt{pat_{\ottmv{i}}}  \ottsym{:}  \ottmv{T} \,   \Rightarrow  \Delta_{\ottmv{i}}}{\ottmv{i}}}%
\ottpremise{\ottcomp{\Gamma  \ottsym{,}  \Delta_{\ottmv{i}}  \vdash  \ottnt{a_{\ottmv{i}}}  \ottsym{:}  \ottnt{A}}{\ottmv{i}}}%
\ottpremise{\Gamma  \vdash  \ottnt{A}  \ottsym{:}  \ottkw{Type}}%
\ottpremise{\ottkw{branches} \, \ottkw{exhaustive}}%
}{
\Gamma  \vdash  \ottkw{case} \, \ottnt{a} \, \ottkw{of} \, \ottsym{\{} \, \ottcomp{\ottnt{pat_{\ottmv{i}}}  \to  \ottnt{a_{\ottmv{i}}}}{\ottmv{i}} \, \ottsym{\}}  \ottsym{:}  \ottnt{A}}{%
{\ottdrulename{t\_case\_simple}}{}%
}}

\newcommand{\ottdruletXXletXXsimple}[1]{\ottdrule[#1]{%
\ottpremise{\Gamma  \vdash  \ottnt{a}  \ottsym{:}  \ottnt{A}}%
\ottpremise{ \Gamma ,  \ottmv{x} \! :\! \ottnt{A}   \vdash  \ottnt{b}  \ottsym{:}  \ottnt{B}}%
\ottpremise{\Gamma  \vdash  \ottnt{B}  \ottsym{:}  \ottkw{Type}}%
}{
\Gamma  \vdash  \ottkw{let} \, \ottmv{x}  \ottsym{=}  \ottnt{a} \, \ottkw{in} \, \ottnt{b}  \ottsym{:}  \ottnt{B}}{%
{\ottdrulename{t\_let\_simple}}{}%
}}

\newcommand{\ottdruletXXletXXdef}[1]{\ottdrule[#1]{%
\ottpremise{\Gamma  \vdash  \ottnt{a}  \ottsym{:}  \ottnt{A}}%
\ottpremise{ \Gamma ,  \ottmv{x} \! :\! \ottnt{A}   \ottsym{,}  \ottmv{x}  \ottsym{=}  \ottnt{a}  \vdash  \ottnt{b}  \ottsym{:}  \ottnt{B}}%
\ottpremise{\Gamma  \vdash  \ottnt{B}  \ottsym{:}  \ottkw{Type}}%
}{
\Gamma  \vdash  \ottkw{let} \, \ottmv{x}  \ottsym{=}  \ottnt{a} \, \ottkw{in} \, \ottnt{b}  \ottsym{:}  \ottnt{B}}{%
{\ottdrulename{t\_let\_def}}{}%
}}

\newcommand{\ottdefntyping}[1]{\begin{ottdefnblock}[#1]{$\Gamma  \vdash  \ottnt{a}  \ottsym{:}  \ottnt{A}$}{\ottcom{Typing}}
\ottusedrule{\ottdruletXXvar{}}
\ottusedrule{\ottdruletXXslambda{}}
\ottusedrule{\ottdruletXXlambda{}}
\ottusedrule{\ottdruletXXaltlambda{}}
\ottusedrule{\ottdruletXXpi{}}
\ottusedrule{\ottdruletXXtype{}}
\ottusedrule{\ottdruletXXapp{}}
\ottusedrule{\ottdruletXXconv{}}
\ottusedrule{\ottdruletXXsigma{}}
\ottusedrule{\ottdruletXXpair{}}
\ottusedrule{\ottdruletXXletpairXXweak{}}
\ottusedrule{\ottdruletXXbool{}}
\ottusedrule{\ottdruletXXtrue{}}
\ottusedrule{\ottdruletXXfalse{}}
\ottusedrule{\ottdruletXXifXXsimple{}}
\ottusedrule{\ottdruletXXifXXfull{}}
\ottusedrule{\ottdruletXXif{}}
\ottusedrule{\ottdruletXXeq{}}
\ottusedrule{\ottdruletXXrefl{}}
\ottusedrule{\ottdruletXXreflXXalt{}}
\ottusedrule{\ottdruletXXsubstXXsimple{}}
\ottusedrule{\ottdruletXXsubst{}}
\ottusedrule{\ottdruletXXcontra{}}
\ottusedrule{\ottdruletXXevar{}}
\ottusedrule{\ottdruletXXelambda{}}
\ottusedrule{\ottdruletXXeapp{}}
\ottusedrule{\ottdruletXXepi{}}
\ottusedrule{\ottdruletXXNat{}}
\ottusedrule{\ottdruletXXVoid{}}
\ottusedrule{\ottdruletXXZero{}}
\ottusedrule{\ottdruletXXSucc{}}
\ottusedrule{\ottdruletXXImTrue{}}
\ottusedrule{\ottdruletXXcaseXXsimple{}}
\ottusedrule{\ottdruletXXletXXsimple{}}
\ottusedrule{\ottdruletXXletXXdef{}}
\end{ottdefnblock}}

\newcommand{\ottdefnsJTyping}{
\ottdefntyping{}}

\newcommand{\ottdrulegXXnil}[1]{\ottdrule[#1]{%
}{
\vdash   \emptyset }{%
{\ottdrulename{g\_nil}}{}%
}}

\newcommand{\ottdrulegXXcons}[1]{\ottdrule[#1]{%
\ottpremise{\Gamma  \vdash  \ottnt{A}  \ottsym{:}  \ottkw{Type}}%
\ottpremise{\vdash  \Gamma}%
}{
\vdash   \Gamma ,  \ottmv{x} \! :\! \ottnt{A} }{%
{\ottdrulename{g\_cons}}{}%
}}

\newcommand{\ottdefncheckCtx}[1]{\begin{ottdefnblock}[#1]{$\vdash  \Gamma$}{\ottcom{check all types in the context}}
\ottusedrule{\ottdrulegXXnil{}}
\ottusedrule{\ottdrulegXXcons{}}
\end{ottdefnblock}}

\newcommand{\ottdruleteleXXnil}[1]{\ottdrule[#1]{%
}{
\Gamma  \vdash    \Leftarrow  }{%
{\ottdrulename{tele\_nil}}{}%
}}

\newcommand{\ottdruleteleXXsig}[1]{\ottdrule[#1]{%
\ottpremise{\Gamma  \vdash  \ottnt{a}  \ottsym{:}  \ottnt{A}}%
\ottpremise{\Gamma  \vdash  \overline{a}  \Leftarrow  \Delta  \ottsym{[}  \ottnt{a}  \ottsym{/}  \ottmv{x}  \ottsym{]}}%
}{
\Gamma  \vdash  \ottnt{a} \, \overline{a}  \Leftarrow  \ottmv{x}  \ottsym{:}  \ottnt{A}  \ottsym{,}  \Delta}{%
{\ottdrulename{tele\_sig}}{}%
}}

\newcommand{\ottdruleteleXXdef}[1]{\ottdrule[#1]{%
\ottpremise{\Gamma  \vdash  \overline{a}  \Leftarrow  \Delta  \ottsym{[}  \ottnt{a}  \ottsym{/}  \ottmv{x}  \ottsym{]}}%
}{
\Gamma  \vdash  \overline{a}  \Leftarrow  \ottmv{x}  \ottsym{=}  \ottnt{a}  \ottsym{,}  \Delta}{%
{\ottdrulename{tele\_def}}{}%
}}

\newcommand{\ottdefntcArgTele}[1]{\begin{ottdefnblock}[#1]{$\Gamma  \vdash  \overline{a}  \Leftarrow  \Delta$}{}
\ottusedrule{\ottdruleteleXXnil{}}
\ottusedrule{\ottdruleteleXXsig{}}
\ottusedrule{\ottdruleteleXXdef{}}
\end{ottdefnblock}}

\newcommand{\ottdrulepsXXNil}[1]{\ottdrule[#1]{%
}{
\Gamma  \vdash    \ottsym{:}    \Rightarrow  }{%
{\ottdrulename{ps\_Nil}}{}%
}}

\newcommand{\ottdrulepsXXCons}[1]{\ottdrule[#1]{%
\ottpremise{\Gamma  \vdash  \ottnt{pat}  \ottsym{:}  \ottnt{A}  \Rightarrow  \Delta_{{\mathrm{1}}}}%
\ottpremise{\Gamma  \vdash  \ottnt{ps}  \ottsym{:}  \Delta  \Rightarrow  \Delta_{{\mathrm{2}}}}%
}{
\Gamma  \vdash  \ottnt{pat} \, \ottnt{ps}  \ottsym{:}  \ottmv{x}  \ottsym{:}  \ottnt{A}  \ottsym{,}  \Delta  \Rightarrow  \Delta_{{\mathrm{1}}}  \ottsym{,}  \Delta_{{\mathrm{2}}}}{%
{\ottdrulename{ps\_Cons}}{}%
}}

\newcommand{\ottdefndeclarePats}[1]{\begin{ottdefnblock}[#1]{$\Gamma  \vdash  \ottnt{ps}  \ottsym{:}  \Delta_{{\mathrm{1}}}  \Rightarrow  \Delta_{{\mathrm{2}}}$}{}
\ottusedrule{\ottdrulepsXXNil{}}
\ottusedrule{\ottdrulepsXXCons{}}
\end{ottdefnblock}}

\newcommand{\ottdrulepXXvar}[1]{\ottdrule[#1]{%
}{
\Gamma  \vdash  \ottmv{x}  \ottsym{:}  \ottnt{A}  \Rightarrow  \ottmv{x}  \ottsym{:}  \ottnt{A}  \ottsym{,}  }{%
{\ottdrulename{p\_var}}{}%
}}

\newcommand{\ottdrulepXXtcon}[1]{\ottdrule[#1]{%
\ottpremise{\ottmv{K}  \ottsym{:}  \Delta_{{\mathrm{1}}}  \to  \ottmv{T} \,  \, \in \, \Gamma}%
\ottpremise{\Gamma  \vdash  \ottnt{ps}  \ottsym{:}  \Delta_{{\mathrm{1}}}  \Rightarrow  \Delta_{{\mathrm{2}}}}%
}{
\Gamma  \vdash  \ottmv{K} \, \ottnt{ps}  \ottsym{:}  \ottmv{T} \,   \Rightarrow  \Delta_{{\mathrm{2}}}}{%
{\ottdrulename{p\_tcon}}{}%
}}

\newcommand{\ottdefndeclarePat}[1]{\begin{ottdefnblock}[#1]{$\Gamma  \vdash  \ottnt{pat}  \ottsym{:}  \ottnt{A}  \Rightarrow  \Delta$}{}
\ottusedrule{\ottdrulepXXvar{}}
\ottusedrule{\ottdrulepXXtcon{}}
\end{ottdefnblock}}

\newcommand{\ottdruleiXXvar}[1]{\ottdrule[#1]{%
\ottpremise{\ottmv{x}  \ottsym{:}  \ottnt{A} \, \in \, \Gamma}%
}{
\Gamma  \vdash  \ottmv{x}  \Rightarrow  \ottnt{A}}{%
{\ottdrulename{i\_var}}{}%
}}

\newcommand{\ottdruleiXXappXXsimple}[1]{\ottdrule[#1]{%
\ottpremise{\Gamma  \vdash  \ottnt{a}  \Rightarrow   ( \ottmv{x} \!:\! \ottnt{A} )  \to   \ottnt{B} }%
\ottpremise{\Gamma  \vdash  \ottnt{b}  \Leftarrow  \ottnt{A}}%
}{
\Gamma  \vdash   \ottnt{a}  \;  \ottnt{b}   \Rightarrow  \ottnt{B}  \ottsym{[}  \ottnt{b}  \ottsym{/}  \ottmv{x}  \ottsym{]}}{%
{\ottdrulename{i\_app\_simple}}{}%
}}

\newcommand{\ottdruleiXXapp}[1]{\ottdrule[#1]{%
\ottpremise{\Gamma  \vdash  \ottnt{a}  \Rightarrow  \ottnt{A}}%
\ottpremise{ \Gamma   \vdash   \ottkw{whnf} \  \ottnt{A}  \leadsto   ( \ottmv{x} \!:\! \ottnt{A_{{\mathrm{1}}}} )  \to   \ottnt{B}  }%
\ottpremise{\Gamma  \vdash  \ottnt{b}  \Leftarrow  \ottnt{A_{{\mathrm{1}}}}}%
}{
\Gamma  \vdash   \ottnt{a}  \;  \ottnt{b}   \Rightarrow  \ottnt{B}  \ottsym{[}  \ottnt{b}  \ottsym{/}  \ottmv{x}  \ottsym{]}}{%
{\ottdrulename{i\_app}}{}%
}}

\newcommand{\ottdruleiXXpi}[1]{\ottdrule[#1]{%
\ottpremise{\Gamma  \vdash  \ottnt{A}  \Leftarrow  \ottkw{Type}}%
\ottpremise{ \Gamma ,  \ottmv{x} \! :\! \ottnt{A}   \vdash  \ottnt{B}  \Leftarrow  \ottkw{Type}}%
}{
\Gamma  \vdash   ( \ottmv{x} \!:\! \ottnt{A} )  \to   \ottnt{B}   \Rightarrow  \ottkw{Type}}{%
{\ottdrulename{i\_pi}}{}%
}}

\newcommand{\ottdruleiXXtype}[1]{\ottdrule[#1]{%
}{
\Gamma  \vdash  \ottkw{Type}  \Rightarrow  \ottkw{Type}}{%
{\ottdrulename{i\_type}}{}%
}}

\newcommand{\ottdruleiXXannot}[1]{\ottdrule[#1]{%
\ottpremise{\Gamma  \vdash  \ottnt{A}  \Leftarrow  \ottkw{Type}}%
\ottpremise{\Gamma  \vdash  \ottnt{a}  \Leftarrow  \ottnt{A}}%
}{
\Gamma  \vdash  \ottsym{(}  \ottnt{a}  \ottsym{:}  \ottnt{A}  \ottsym{)}  \Rightarrow  \ottnt{A}}{%
{\ottdrulename{i\_annot}}{}%
}}

\newcommand{\ottdruleiXXsigma}[1]{\ottdrule[#1]{%
\ottpremise{\Gamma  \vdash  \ottnt{A}  \Leftarrow  \ottkw{Type}}%
\ottpremise{ \Gamma ,  \ottmv{x} \! :\! \ottnt{A}   \vdash  \ottnt{B}  \Leftarrow  \ottkw{Type}}%
}{
\Gamma  \vdash   \{  \ottmv{x} \!:\! \ottnt{A} \ |\  \ottnt{B}  \}   \Rightarrow  \ottkw{Type}}{%
{\ottdrulename{i\_sigma}}{}%
}}

\newcommand{\ottdruleiXXbool}[1]{\ottdrule[#1]{%
}{
\Gamma  \vdash  \ottkw{Bool}  \Rightarrow  \ottkw{Type}}{%
{\ottdrulename{i\_bool}}{}%
}}

\newcommand{\ottdruleiXXtrue}[1]{\ottdrule[#1]{%
}{
\Gamma  \vdash  \ottkw{True}  \Rightarrow  \ottkw{Bool}}{%
{\ottdrulename{i\_true}}{}%
}}

\newcommand{\ottdruleiXXfalse}[1]{\ottdrule[#1]{%
}{
\Gamma  \vdash  \ottkw{False}  \Rightarrow  \ottkw{Bool}}{%
{\ottdrulename{i\_false}}{}%
}}

\newcommand{\ottdruleiXXifXXsimple}[1]{\ottdrule[#1]{%
\ottpremise{\Gamma  \vdash  \ottnt{a}  \Leftarrow  \ottkw{Bool}}%
\ottpremise{\Gamma  \vdash  \ottnt{b_{{\mathrm{1}}}}  \Rightarrow  \ottnt{A_{{\mathrm{1}}}}}%
\ottpremise{\Gamma  \vdash  \ottnt{b_{{\mathrm{2}}}}  \Rightarrow  \ottnt{A_{{\mathrm{2}}}}}%
\ottpremise{\Gamma  \vdash  \ottnt{A_{{\mathrm{1}}}}  \Leftrightarrow  \ottnt{A_{{\mathrm{2}}}}}%
}{
\Gamma  \vdash  \ottkw{if} \, \ottnt{a} \, \ottkw{then} \, \ottnt{b_{{\mathrm{1}}}} \, \ottkw{else} \, \ottnt{b_{{\mathrm{2}}}}  \Rightarrow  \ottnt{A_{{\mathrm{1}}}}}{%
{\ottdrulename{i\_if\_simple}}{}%
}}

\newcommand{\ottdruleiXXifXXmotive}[1]{\ottdrule[#1]{%
\ottpremise{\Gamma  \vdash  \ottnt{a}  \Leftarrow  \ottkw{Bool}}%
\ottpremise{\Gamma  \vdash  \ottnt{b_{{\mathrm{1}}}}  \Leftarrow  \ottnt{A}  \ottsym{[}  \ottkw{True}  \ottsym{/}  \ottmv{x}  \ottsym{]}}%
\ottpremise{\Gamma  \vdash  \ottnt{b_{{\mathrm{2}}}}  \Leftarrow  \ottnt{A}  \ottsym{[}  \ottkw{False}  \ottsym{/}  \ottmv{x}  \ottsym{]}}%
\ottpremise{ \Gamma ,  \ottmv{x} \! :\! \ottkw{Bool}   \vdash  \ottnt{A}  \Leftarrow  \ottkw{Type}}%
}{
\Gamma  \vdash  \ottkw{if} \, \ottnt{a} \, \ottkw{then} \, \ottnt{b_{{\mathrm{1}}}} \, \ottkw{else} \, \ottnt{b_{{\mathrm{2}}}}  \ottsym{[}  \ottmv{x}  \ottsym{.}  \ottnt{A}  \ottsym{]}  \Rightarrow  \ottnt{A}  \ottsym{[}  \ottnt{a}  \ottsym{/}  \ottmv{x}  \ottsym{]}}{%
{\ottdrulename{i\_if\_motive}}{}%
}}

\newcommand{\ottdruleiXXifXXalt}[1]{\ottdrule[#1]{%
\ottpremise{\Gamma  \vdash  \ottnt{a}  \Leftarrow  \ottkw{Bool}}%
\ottpremise{\Gamma  \vdash  \ottnt{b_{{\mathrm{1}}}}  \Rightarrow  \ottnt{B_{{\mathrm{1}}}}}%
\ottpremise{\Gamma  \vdash  \ottnt{b_{{\mathrm{2}}}}  \Rightarrow  \ottnt{B_{{\mathrm{2}}}}}%
}{
\Gamma  \vdash  \ottkw{if} \, \ottnt{a} \, \ottkw{then} \, \ottnt{b_{{\mathrm{1}}}} \, \ottkw{else} \, \ottnt{b_{{\mathrm{2}}}}  \Rightarrow  \ottkw{if} \, \ottnt{a} \, \ottkw{then} \, \ottnt{B_{{\mathrm{1}}}} \, \ottkw{else} \, \ottnt{B_{{\mathrm{2}}}}}{%
{\ottdrulename{i\_if\_alt}}{}%
}}

\newcommand{\ottdruleiXXeq}[1]{\ottdrule[#1]{%
\ottpremise{\Gamma  \vdash  \ottnt{a}  \Rightarrow  \ottnt{A}}%
\ottpremise{\Gamma  \vdash  \ottnt{b}  \Rightarrow  \ottnt{B}}%
\ottpremise{\Gamma  \vdash  \ottnt{A}  \Leftrightarrow  \ottnt{B}}%
}{
\Gamma  \vdash  \ottsym{(}  \ottnt{a}  \ottsym{=}  \ottnt{b}  \ottsym{)}  \Rightarrow  \ottkw{Type}}{%
{\ottdrulename{i\_eq}}{}%
}}

\newcommand{\ottdruleiXXsubst}[1]{\ottdrule[#1]{%
\ottpremise{\Gamma  \vdash  \ottnt{b}  \Rightarrow  \ottnt{B}}%
\ottpremise{ \Gamma   \vdash   \ottkw{whnf} \  \ottnt{B}  \leadsto  \ottsym{(}  \ottnt{a_{{\mathrm{1}}}}  \ottsym{=}  \ottnt{a_{{\mathrm{2}}}}  \ottsym{)} }%
\ottpremise{\Gamma  \vdash  \ottnt{a}  \Rightarrow  \ottnt{A}}%
}{
\Gamma  \vdash  \ottkw{subst} \, \ottnt{a} \, \ottkw{by} \, \ottnt{b}  \Rightarrow  \ottnt{A}}{%
{\ottdrulename{i\_subst}}{}%
}}

\newcommand{\ottdruleiXXtcon}[1]{\ottdrule[#1]{%
\ottpremise{\ottmv{T}  \ottsym{:}  \Delta  \to \, \ottkw{Type} \, \in \, \Gamma}%
\ottpremise{\Gamma  \vdash  \overline{a}  \Leftarrow  \Delta}%
}{
\Gamma  \vdash  \ottmv{T} \, \overline{a}  \Rightarrow  \ottkw{Type}}{%
{\ottdrulename{i\_tcon}}{}%
}}

\newcommand{\ottdruleiXXdconXXsimple}[1]{\ottdrule[#1]{%
\ottpremise{\ottmv{K}  \ottsym{:}  \Delta  \to  \ottmv{T} \,  \, \in \, \Gamma}%
\ottpremise{\Gamma  \vdash  \overline{a}  \Leftarrow  \Delta}%
}{
\Gamma  \vdash  \ottmv{K} \, \overline{a}  \Rightarrow  \ottmv{T} \, }{%
{\ottdrulename{i\_dcon\_simple}}{}%
}}

\newcommand{\ottdruleiXXletXXsimple}[1]{\ottdrule[#1]{%
\ottpremise{\Gamma  \vdash  \ottnt{a}  \Rightarrow  \ottnt{A}}%
\ottpremise{ \Gamma ,  \ottmv{x} \! :\! \ottnt{A}   \vdash  \ottnt{b}  \Rightarrow  \ottnt{B}}%
\ottpremise{\Gamma  \vdash  \ottnt{B}  \Leftarrow  \ottkw{Type}}%
}{
\Gamma  \vdash  \ottkw{let} \, \ottmv{x}  \ottsym{=}  \ottnt{a} \, \ottkw{in} \, \ottnt{b}  \Rightarrow  \ottnt{B}}{%
{\ottdrulename{i\_let\_simple}}{}%
}}

\newcommand{\ottdruleiXXlet}[1]{\ottdrule[#1]{%
\ottpremise{\Gamma  \vdash  \ottnt{a}  \Rightarrow  \ottnt{A}}%
\ottpremise{ \Gamma ,  \ottmv{x} \! :\! \ottnt{A}   \ottsym{,}  \ottmv{x}  \ottsym{=}  \ottnt{a}  \vdash  \ottnt{b}  \Rightarrow  \ottnt{B}}%
}{
\Gamma  \vdash  \ottkw{let} \, \ottmv{x}  \ottsym{=}  \ottnt{a} \, \ottkw{in} \, \ottnt{b}  \Rightarrow  \ottnt{B}  \ottsym{[}  \ottnt{a}  \ottsym{/}  \ottmv{x}  \ottsym{]}}{%
{\ottdrulename{i\_let}}{}%
}}

\newcommand{\ottdefninferType}[1]{\begin{ottdefnblock}[#1]{$\Gamma  \vdash  \ottnt{a}  \Rightarrow  \ottnt{A}$}{\ottcom{type synthesis (algorithmic)}}
\ottusedrule{\ottdruleiXXvar{}}
\ottusedrule{\ottdruleiXXappXXsimple{}}
\ottusedrule{\ottdruleiXXapp{}}
\ottusedrule{\ottdruleiXXpi{}}
\ottusedrule{\ottdruleiXXtype{}}
\ottusedrule{\ottdruleiXXannot{}}
\ottusedrule{\ottdruleiXXsigma{}}
\ottusedrule{\ottdruleiXXbool{}}
\ottusedrule{\ottdruleiXXtrue{}}
\ottusedrule{\ottdruleiXXfalse{}}
\ottusedrule{\ottdruleiXXifXXsimple{}}
\ottusedrule{\ottdruleiXXifXXmotive{}}
\ottusedrule{\ottdruleiXXifXXalt{}}
\ottusedrule{\ottdruleiXXeq{}}
\ottusedrule{\ottdruleiXXsubst{}}
\ottusedrule{\ottdruleiXXtcon{}}
\ottusedrule{\ottdruleiXXdconXXsimple{}}
\ottusedrule{\ottdruleiXXletXXsimple{}}
\ottusedrule{\ottdruleiXXlet{}}
\end{ottdefnblock}}

\newcommand{\ottdrulecXXlambda}[1]{\ottdrule[#1]{%
\ottpremise{ \Gamma ,  \ottmv{x} \! :\! \ottnt{A}   \vdash  \ottnt{a}  \Leftarrow  \ottnt{B}}%
}{
\Gamma  \vdash   \lambda  \ottmv{x} .  \ottnt{a}   \Leftarrow   ( \ottmv{x} \!:\! \ottnt{A} )  \to   \ottnt{B} }{%
{\ottdrulename{c\_lambda}}{}%
}}

\newcommand{\ottdrulecXXinferXXsimple}[1]{\ottdrule[#1]{%
\ottpremise{ \ottnt{a} \mbox{ is not a $\lambda$-expression } }%
\ottpremise{\Gamma  \vdash  \ottnt{a}  \Rightarrow  \ottnt{A}}%
}{
\Gamma  \vdash  \ottnt{a}  \Leftarrow  \ottnt{A}}{%
{\ottdrulename{c\_infer\_simple}}{}%
}}

\newcommand{\ottdrulecXXinfer}[1]{\ottdrule[#1]{%
\ottpremise{ \ottnt{a} \mbox{ is not a $\lambda$-expression } }%
\ottpremise{\Gamma  \vdash  \ottnt{a}  \Rightarrow  \ottnt{A}}%
\ottpremise{\Gamma  \vdash  \ottnt{A}  \Leftrightarrow  \ottnt{B}}%
}{
\Gamma  \vdash  \ottnt{a}  \Leftarrow  \ottnt{B}}{%
{\ottdrulename{c\_infer}}{}%
}}

\newcommand{\ottdrulecXXwhnf}[1]{\ottdrule[#1]{%
\ottpremise{ \ottnt{A} \mbox{ is not in weak-head normal form } }%
\ottpremise{\Gamma  \vdash  \ottnt{a}  \Leftarrow  \ottnt{nf}}%
\ottpremise{ \Gamma   \vdash   \ottkw{whnf} \  \ottnt{A}  \leadsto  \ottnt{nf} }%
}{
\Gamma  \vdash  \ottnt{a}  \Leftarrow  \ottnt{A}}{%
{\ottdrulename{c\_whnf}}{}%
}}

\newcommand{\ottdrulecXXpair}[1]{\ottdrule[#1]{%
\ottpremise{\Gamma  \vdash  \ottnt{a}  \Leftarrow  \ottnt{A}}%
\ottpremise{\Gamma  \vdash  \ottnt{b}  \Leftarrow  \ottnt{B}  \ottsym{[}  \ottnt{a}  \ottsym{/}  \ottmv{x}  \ottsym{]}}%
}{
\Gamma  \vdash  \ottsym{(}  \ottnt{a}  \ottsym{,}  \ottnt{b}  \ottsym{)}  \Leftarrow   \{  \ottmv{x} \!:\! \ottnt{A} \ |\  \ottnt{B}  \} }{%
{\ottdrulename{c\_pair}}{}%
}}

\newcommand{\ottdrulecXXletpairXXsimple}[1]{\ottdrule[#1]{%
\ottpremise{\Gamma  \vdash  \ottnt{a}  \Rightarrow   \{  \ottmv{x} \!:\! \ottnt{A_{{\mathrm{1}}}} \ |\  \ottnt{A_{{\mathrm{2}}}}  \} }%
\ottpremise{  \Gamma ,  \ottmv{x} \! :\! \ottnt{A_{{\mathrm{1}}}}  ,  \ottmv{y} \! :\! \ottnt{A_{{\mathrm{2}}}}   \vdash  \ottnt{b}  \Leftarrow  \ottnt{B}}%
}{
\Gamma  \vdash  \ottkw{let} \, \ottsym{(}  \ottmv{x}  \ottsym{,}  \ottmv{y}  \ottsym{)}  \ottsym{=}  \ottnt{a} \, \ottkw{in} \, \ottnt{b}  \Leftarrow  \ottnt{B}}{%
{\ottdrulename{c\_letpair\_simple}}{}%
}}

\newcommand{\ottdrulecXXletpair}[1]{\ottdrule[#1]{%
\ottpremise{\Gamma  \vdash  \ottmv{z}  \Rightarrow   \{  \ottmv{x} \!:\! \ottnt{A_{{\mathrm{1}}}} \ |\  \ottnt{A_{{\mathrm{2}}}}  \} }%
\ottpremise{  \Gamma ,  \ottmv{x} \! :\! \ottnt{A_{{\mathrm{1}}}}  ,  \ottmv{y} \! :\! \ottnt{A_{{\mathrm{2}}}}   \vdash  \ottnt{b}  \Leftarrow  \ottnt{B}  \ottsym{[}  \ottsym{(}  \ottmv{x}  \ottsym{,}  \ottmv{y}  \ottsym{)}  \ottsym{/}  \ottmv{z}  \ottsym{]}}%
}{
\Gamma  \vdash  \ottkw{let} \, \ottsym{(}  \ottmv{x}  \ottsym{,}  \ottmv{y}  \ottsym{)}  \ottsym{=}  \ottmv{z} \, \ottkw{in} \, \ottnt{b}  \Leftarrow  \ottnt{B}}{%
{\ottdrulename{c\_letpair}}{}%
}}

\newcommand{\ottdrulecXXifXXsimple}[1]{\ottdrule[#1]{%
\ottpremise{\Gamma  \vdash  \ottnt{a}  \Leftarrow  \ottkw{Bool}}%
\ottpremise{\Gamma  \vdash  \ottnt{b_{{\mathrm{1}}}}  \Leftarrow  \ottnt{A}}%
\ottpremise{\Gamma  \vdash  \ottnt{b_{{\mathrm{2}}}}  \Leftarrow  \ottnt{A}}%
}{
\Gamma  \vdash  \ottkw{if} \, \ottnt{a} \, \ottkw{then} \, \ottnt{b_{{\mathrm{1}}}} \, \ottkw{else} \, \ottnt{b_{{\mathrm{2}}}}  \Leftarrow  \ottnt{A}}{%
{\ottdrulename{c\_if\_simple}}{}%
}}

\newcommand{\ottdrulecXXif}[1]{\ottdrule[#1]{%
\ottpremise{\Gamma  \vdash  \ottmv{x}  \Leftarrow  \ottkw{Bool}}%
\ottpremise{\Gamma  \vdash  \ottnt{b_{{\mathrm{1}}}}  \Leftarrow  \ottnt{A}  \ottsym{[}  \ottkw{True}  \ottsym{/}  \ottmv{x}  \ottsym{]}}%
\ottpremise{\Gamma  \vdash  \ottnt{b_{{\mathrm{2}}}}  \Leftarrow  \ottnt{A}  \ottsym{[}  \ottkw{False}  \ottsym{/}  \ottmv{x}  \ottsym{]}}%
}{
\Gamma  \vdash  \ottkw{if} \, \ottmv{x} \, \ottkw{then} \, \ottnt{b_{{\mathrm{1}}}} \, \ottkw{else} \, \ottnt{b_{{\mathrm{2}}}}  \Leftarrow  \ottnt{A}}{%
{\ottdrulename{c\_if}}{}%
}}

\newcommand{\ottdrulecXXrefl}[1]{\ottdrule[#1]{%
\ottpremise{\Gamma  \vdash  \ottnt{a}  \Leftrightarrow  \ottnt{b}}%
}{
\Gamma  \vdash  \ottkw{refl}  \Leftarrow  \ottnt{a}  \ottsym{=}  \ottnt{b}}{%
{\ottdrulename{c\_refl}}{}%
}}

\newcommand{\ottdrulecXXsubstXXleftXXsimple}[1]{\ottdrule[#1]{%
\ottpremise{\Gamma  \vdash  \ottnt{b}  \Rightarrow  \ottnt{B}}%
\ottpremise{ \Gamma   \vdash   \ottkw{whnf} \  \ottnt{B}  \leadsto  \ottsym{(}  \ottmv{x}  \ottsym{=}  \ottnt{a_{{\mathrm{2}}}}  \ottsym{)} }%
\ottpremise{\Gamma  \vdash  \ottnt{a}  \Leftarrow  \ottnt{A}  \ottsym{[}  \ottnt{a_{{\mathrm{2}}}}  \ottsym{/}  \ottmv{x}  \ottsym{]}}%
}{
\Gamma  \vdash  \ottkw{subst} \, \ottnt{a} \, \ottkw{by} \, \ottnt{b}  \Leftarrow  \ottnt{A}}{%
{\ottdrulename{c\_subst\_left\_simple}}{}%
}}

\newcommand{\ottdrulecXXsubstXXrightXXsimple}[1]{\ottdrule[#1]{%
\ottpremise{\Gamma  \vdash  \ottnt{b}  \Rightarrow  \ottnt{B}}%
\ottpremise{ \Gamma   \vdash   \ottkw{whnf} \  \ottnt{B}  \leadsto  \ottsym{(}  \ottnt{a_{{\mathrm{1}}}}  \ottsym{=}  \ottmv{x}  \ottsym{)} }%
\ottpremise{\Gamma  \vdash  \ottnt{a}  \Leftarrow  \ottnt{A}  \ottsym{[}  \ottnt{a_{{\mathrm{1}}}}  \ottsym{/}  \ottmv{x}  \ottsym{]}}%
}{
\Gamma  \vdash  \ottkw{subst} \, \ottnt{a} \, \ottkw{by} \, \ottnt{b}  \Leftarrow  \ottnt{A}}{%
{\ottdrulename{c\_subst\_right\_simple}}{}%
}}

\newcommand{\ottdrulecXXsubstXXleft}[1]{\ottdrule[#1]{%
\ottpremise{\Gamma  \vdash  \ottmv{y}  \Rightarrow  \ottnt{B}}%
\ottpremise{ \Gamma   \vdash   \ottkw{whnf} \  \ottnt{B}  \leadsto  \ottsym{(}  \ottmv{x}  \ottsym{=}  \ottnt{a_{{\mathrm{2}}}}  \ottsym{)} }%
\ottpremise{\Gamma  \vdash  \ottnt{a}  \Leftarrow  \ottnt{A}  \ottsym{[}  \ottnt{a_{{\mathrm{2}}}}  \ottsym{/}  \ottmv{x}  \ottsym{]}  \ottsym{[}  \ottkw{refl}  \ottsym{/}  \ottmv{y}  \ottsym{]}}%
}{
\Gamma  \vdash  \ottkw{subst} \, \ottnt{a} \, \ottkw{by} \, \ottmv{y}  \Leftarrow  \ottnt{A}}{%
{\ottdrulename{c\_subst\_left}}{}%
}}

\newcommand{\ottdrulecXXsubstXXright}[1]{\ottdrule[#1]{%
\ottpremise{\Gamma  \vdash  \ottmv{y}  \Rightarrow  \ottnt{B}}%
\ottpremise{ \Gamma   \vdash   \ottkw{whnf} \  \ottnt{B}  \leadsto  \ottsym{(}  \ottnt{a_{{\mathrm{1}}}}  \ottsym{=}  \ottmv{x}  \ottsym{)} }%
\ottpremise{\Gamma  \vdash  \ottnt{a}  \Leftarrow  \ottnt{A}  \ottsym{[}  \ottnt{a_{{\mathrm{1}}}}  \ottsym{/}  \ottmv{x}  \ottsym{]}  \ottsym{[}  \ottkw{refl}  \ottsym{/}  \ottmv{y}  \ottsym{]}}%
}{
\Gamma  \vdash  \ottkw{subst} \, \ottnt{a} \, \ottkw{by} \, \ottmv{y}  \Leftarrow  \ottnt{A}}{%
{\ottdrulename{c\_subst\_right}}{}%
}}

\newcommand{\ottdrulecXXcontra}[1]{\ottdrule[#1]{%
\ottpremise{\Gamma  \vdash  \ottnt{a}  \ottsym{:}  \ottnt{A}}%
\ottpremise{ \Gamma   \vdash   \ottkw{whnf} \  \ottnt{A}  \leadsto  \ottsym{(}  \ottnt{a_{{\mathrm{1}}}}  \ottsym{=}  \ottnt{a_{{\mathrm{2}}}}  \ottsym{)} }%
\ottpremise{ \Gamma   \vdash   \ottkw{whnf} \  \ottnt{a_{{\mathrm{1}}}}  \leadsto  \ottkw{True} }%
\ottpremise{ \Gamma   \vdash   \ottkw{whnf} \  \ottnt{a_{{\mathrm{2}}}}  \leadsto  \ottkw{False} }%
}{
\Gamma  \vdash  \ottkw{contra} \, \ottnt{a}  \Leftarrow  \ottnt{B}}{%
{\ottdrulename{c\_contra}}{}%
}}

\newcommand{\ottdrulecXXcaseXXsimple}[1]{\ottdrule[#1]{%
\ottpremise{\Gamma  \vdash  \ottnt{a}  \Rightarrow  \ottnt{A}}%
\ottpremise{ \Gamma   \vdash   \ottkw{whnf} \  \ottnt{A}  \leadsto  \ottmv{T} \,  }%
\ottpremise{\ottcomp{\Gamma  \vdash  \ottnt{pat_{\ottmv{i}}}  \ottsym{:}  \ottmv{T} \,   \Rightarrow  \Delta_{\ottmv{i}}}{\ottmv{i}}}%
\ottpremise{\ottcomp{\vdash  \ottnt{a}  \sim  \ottnt{pat_{\ottmv{i}}}  \Rightarrow  \Delta'_{\ottmv{i}}}{\ottmv{i}}}%
\ottpremise{\ottcomp{\Gamma  \ottsym{,}  \Delta_{\ottmv{i}}  \ottsym{,}  \Delta'_{\ottmv{i}}  \vdash  \ottnt{a_{\ottmv{i}}}  \ottsym{:}  \ottnt{A}}{\ottmv{i}}}%
\ottpremise{\ottkw{branches} \, \ottkw{exhaustive}}%
}{
\Gamma  \vdash  \ottkw{case} \, \ottnt{a} \, \ottkw{of} \, \ottsym{\{} \, \ottcomp{\ottnt{pat_{\ottmv{i}}}  \to  \ottnt{a_{\ottmv{i}}}}{\ottmv{i}} \, \ottsym{\}}  \Leftarrow  \ottnt{A}}{%
{\ottdrulename{c\_case\_simple}}{}%
}}

\newcommand{\ottdrulecXXcase}[1]{\ottdrule[#1]{%
\ottpremise{\Gamma  \vdash  \ottnt{a}  \Rightarrow  \ottnt{A}}%
\ottpremise{ \Gamma   \vdash   \ottkw{whnf} \  \ottnt{A}  \leadsto  \ottmv{T} \, \overline{b} }%
\ottpremise{\ottcomp{\Gamma  \vdash  \ottnt{pat_{\ottmv{i}}}  \ottsym{:}  \ottmv{T} \, \overline{b}  \Rightarrow  \Delta_{\ottmv{i}}}{\ottmv{i}}}%
\ottpremise{\ottcomp{\vdash  \ottnt{a}  \sim  \ottnt{pat_{\ottmv{i}}}  \Rightarrow  \Delta'_{\ottmv{i}}}{\ottmv{i}}}%
\ottpremise{\ottcomp{\Gamma  \ottsym{,}  \Delta_{\ottmv{i}}  \ottsym{,}  \Delta'_{\ottmv{i}}  \vdash  \ottnt{a_{\ottmv{i}}}  \ottsym{:}  \ottnt{A}}{\ottmv{i}}}%
\ottpremise{\ottkw{branches} \, \ottkw{exhaustive}}%
}{
\Gamma  \vdash  \ottkw{case} \, \ottnt{a} \, \ottkw{of} \, \ottsym{\{} \, \ottcomp{\ottnt{pat_{\ottmv{i}}}  \to  \ottnt{a_{\ottmv{i}}}}{\ottmv{i}} \, \ottsym{\}}  \Leftarrow  \ottnt{A}}{%
{\ottdrulename{c\_case}}{}%
}}

\newcommand{\ottdrulecXXdcon}[1]{\ottdrule[#1]{%
\ottpremise{\ottmv{K}  \ottsym{:}  \Delta_{{\mathrm{1}}}  \to  \Delta_{{\mathrm{2}}}  \to  \ottmv{T} \,  \, \in \, \Gamma}%
\ottpremise{\Gamma  \vdash  \overline{a}  \Leftarrow  \Delta_{{\mathrm{2}}}  \ottsym{[}  \overline{b}  \ottsym{/}  \Delta_{{\mathrm{1}}}  \ottsym{]}}%
}{
\Gamma  \vdash  \ottmv{K} \, \overline{a}  \Leftarrow  \ottmv{T} \, \overline{b}}{%
{\ottdrulename{c\_dcon}}{}%
}}

\newcommand{\ottdrulecXXlet}[1]{\ottdrule[#1]{%
\ottpremise{\Gamma  \vdash  \ottnt{a}  \Rightarrow  \ottnt{A}}%
\ottpremise{ \Gamma ,  \ottmv{x} \! :\! \ottnt{A}   \ottsym{,}  \ottmv{x}  \ottsym{=}  \ottnt{a}  \vdash  \ottnt{b}  \Leftarrow  \ottnt{B}}%
}{
\Gamma  \vdash  \ottkw{let} \, \ottmv{x}  \ottsym{=}  \ottnt{a} \, \ottkw{in} \, \ottnt{b}  \Leftarrow  \ottnt{B}}{%
{\ottdrulename{c\_let}}{}%
}}

\newcommand{\ottdrulecXXifXXdef}[1]{\ottdrule[#1]{%
\ottpremise{\Gamma  \vdash  \ottmv{x}  \Leftarrow  \ottkw{Bool}}%
\ottpremise{\Gamma  \ottsym{,}  \ottmv{x}  \ottsym{=}  \ottkw{True}  \vdash  \ottnt{b_{{\mathrm{1}}}}  \Leftarrow  \ottnt{A}}%
\ottpremise{\Gamma  \ottsym{,}  \ottmv{x}  \ottsym{=}  \ottkw{False}  \vdash  \ottnt{b_{{\mathrm{2}}}}  \Leftarrow  \ottnt{A}}%
}{
\Gamma  \vdash  \ottkw{if} \, \ottmv{x} \, \ottkw{then} \, \ottnt{b_{{\mathrm{1}}}} \, \ottkw{else} \, \ottnt{b_{{\mathrm{2}}}}  \Leftarrow  \ottnt{A}}{%
{\ottdrulename{c\_if\_def}}{}%
}}

\newcommand{\ottdrulecXXletpairXXdef}[1]{\ottdrule[#1]{%
\ottpremise{\Gamma  \vdash  \ottmv{z}  \Rightarrow   \{  \ottmv{x} \!:\! \ottnt{A_{{\mathrm{1}}}} \ |\  \ottnt{A_{{\mathrm{2}}}}  \} }%
\ottpremise{  \Gamma ,  \ottmv{x} \! :\! \ottnt{A_{{\mathrm{1}}}}  ,  \ottmv{y} \! :\! \ottnt{B_{{\mathrm{2}}}}   \ottsym{,}  \ottmv{z}  \ottsym{=}  \ottsym{(}  \ottmv{x}  \ottsym{,}  \ottmv{y}  \ottsym{)}  \vdash  \ottnt{b}  \Leftarrow  \ottnt{B}}%
}{
\Gamma  \vdash  \ottkw{let} \, \ottsym{(}  \ottmv{x}  \ottsym{,}  \ottmv{y}  \ottsym{)}  \ottsym{=}  \ottmv{z} \, \ottkw{in} \, \ottnt{b}  \Leftarrow  \ottnt{B}}{%
{\ottdrulename{c\_letpair\_def}}{}%
}}

\newcommand{\ottdefncheckType}[1]{\begin{ottdefnblock}[#1]{$\Gamma  \vdash  \ottnt{a}  \Leftarrow  \ottnt{B}$}{\ottcom{type checking (algorithmic)}}
\ottusedrule{\ottdrulecXXlambda{}}
\ottusedrule{\ottdrulecXXinferXXsimple{}}
\ottusedrule{\ottdrulecXXinfer{}}
\ottusedrule{\ottdrulecXXwhnf{}}
\ottusedrule{\ottdrulecXXpair{}}
\ottusedrule{\ottdrulecXXletpairXXsimple{}}
\ottusedrule{\ottdrulecXXletpair{}}
\ottusedrule{\ottdrulecXXifXXsimple{}}
\ottusedrule{\ottdrulecXXif{}}
\ottusedrule{\ottdrulecXXrefl{}}
\ottusedrule{\ottdrulecXXsubstXXleftXXsimple{}}
\ottusedrule{\ottdrulecXXsubstXXrightXXsimple{}}
\ottusedrule{\ottdrulecXXsubstXXleft{}}
\ottusedrule{\ottdrulecXXsubstXXright{}}
\ottusedrule{\ottdrulecXXcontra{}}
\ottusedrule{\ottdrulecXXcaseXXsimple{}}
\ottusedrule{\ottdrulecXXcase{}}
\ottusedrule{\ottdrulecXXdcon{}}
\ottusedrule{\ottdrulecXXlet{}}
\ottusedrule{\ottdrulecXXifXXdef{}}
\ottusedrule{\ottdrulecXXletpairXXdef{}}
\end{ottdefnblock}}

\newcommand{\ottdefnsJBidirectional}{
\ottdefncheckCtx{}\ottdefntcArgTele{}\ottdefndeclarePats{}\ottdefndeclarePat{}\ottdefninferType{}\ottdefncheckType{}}

\newcommand{\ottdefnss}{
\ottdefnsJEquate
\ottdefnsJwhnf
\ottdefnsJOp
\ottdefnsJEq
\ottdefnsJTyping
\ottdefnsJBidirectional
}

\newcommand{\ottall}{\ottmetavars\\[0pt]
\ottgrammar\\[5.0mm]
\ottdefnss}

  \renewottcommands[ott]

\section{Overview and Goals}

These lecture notes describe the design of a minimal dependently-typed
language called ``\pif'' and walk through the implementation of its type
checker. They are based on lectures given at the \emph{Oregon Programming
  Languages Summer School} during July 2023 and derived from earlier lectures
from summer schools during 2022, 2014 and 2013.

\paragraph{What do I expect from you?} This discussion assumes a
familiarity with the basics of the lambda calculus, including its standard
operations (alpha-equivalence, substitution, evaluation) and the basics of
type systems (especially their specification using inference rules). For
background on this material, I recommend the textbooks~\cite{tapl, pfpl}.

Furthermore, these notes also refer to an implementation of a demo type
checker and assume basic knowledge of the Haskell programming language. This
implementation is available at \url{https://github.com/sweirich/pi-forall} on
branch \texttt{2023}. As you study these notes, I encourage you to download
this repository, read through its source code, and experiment with
it. Installation instructions are available with the repository.

\paragraph{What do these notes cover?}
These notes are broken into several sections that incrementally build up the
design and implementation of a type checker for a dependently-typed
programming language. This implementation itself is available in separate
versions, each found in separate subdirectories of the repository.

\begin{figure}[ht]
\begin{center}
\begin{tabular}{llll}
Key feature & \pif subdirectory & Section\\
\hline
Core system & \texttt{version1} & Sections~\ref{sec:simple}, \ref{sec:bidirectional}, and \ref{sec:implementation} \\
Equality    & \texttt{version2} & Sections~\ref{sec:equality} and ~\ref{sec:pattern-matching}\\
Irrelevance & \texttt{version3} & Section ~\ref{sec:irrelevance} \\
Datatypes   & \texttt{full}     & Sections~\ref{sec:examples} and ~\ref{sec:datatypes} \\
\end{tabular}
\end{center}
\caption{Connection between sections and \pif versions}
\label{fig:impls}
\end{figure}

These implementations build on each other (each is an extension of the
previous version) and are summarized in Figure~\ref{fig:impls}.  As you read
each chapter, refer to its corresponding implementation to see how the
features described in that chapter can be implemented. The directory
\texttt{main} is the source of all of these implementations and contains the
markup needed to generate the four versions.

\begin{itemize}
\item Section~\ref{sec:examples} starts with some examples of the
  \texttt{full} \pif language so that you can see how the pieces fit together.

\item Section~\ref{sec:simple} presents the mathematical specification of the
  core \pif type system including its concrete syntax (as found in \pif source
  files), abstract syntax (represented with a Haskell datatype), and core
  typing rules (written using standard mathematical notation). This initial
  type system is simple and declarative. In other words, it \emph{specifies}
  what terms should type check, but cannot be directly implemented.

\item Section~\ref{sec:bidirectional} reformulates the typing rules so that
  they are \emph{syntax-directed} and correspond to a type-checking
  algorithm. The key idea of this section is to recast the typing rules as a
  \emph{bidirectional type system}.


\item Section~\ref{sec:implementation} introduces the core \pif implementation
  and walks through the type checker found in \texttt{version1}, the Haskell
  implementation of the typing rules discussed in
  Section~\ref{sec:bidirectional}. This section also shows how the \unbound library
  can assist with the implementation of variable binding and automatically derive operations for
  capture-avoiding substitution and alpha-equivalence. Finally, the section
  describes a monadic structure for the type checker, and how it can help with
  the production of error messages, the primary purpose of a type checker.


\item Section~\ref{sec:equality} discusses the role of definitional equality
  in dependently typed languages. After motivating examples, it presents both
  a specification of when terms are equal and a semi-decidable algorithm that
  can be incorporated into the type checker.

\item An important feature of dependently-typed languages is the ability for
  run-time tests to be reflected into the type
  system. Section~\ref{sec:pattern-matching} shows how to extend \pif with a
  simple form of dependent pattern matching. We look at this feature in two
  different ways: First how to reflect the gain in information about boolean
  values in each branch of an $\mathsf{if}$ expression. Second, how to reflect the
  equality represented via propositional equality.

\item Section~\ref{sec:irrelevance} introduces the idea of tracking the
  \emph{relevance} of arguments. Relevance tracking enables parts of the
  program to be identified as ``compile-time only'' and thus erased before
  execution. It also identifies parts of terms that can be ignored when
  deciding equivalence.


\item Finally, Section~\ref{sec:datatypes} introduces arbitrary datatypes and
  generalizes dependent pattern matching. This section combines the features
  introduced in the previous sections (Sections \ref{sec:equality},
  \ref{sec:pattern-matching}, and \ref{sec:irrelevance}) into a single unified
  language feature.
\end{itemize}

\paragraph{What do these notes \textbf{not} cover?}
The goal of these notes is to provide an introductory overview of the
implementation of (dependent) type theories and type checkers. As a result,
there are many related topics that are not included here. Furthermore, because
these notes come with a reference implementation, they emphasize only a single
point in a rich design space.

For a broader view, section~\ref{sec:related-work} includes a discussion of
alternative tutorials and describes how this one differs from other
approaches.  The key differences are summarized below:

\begin{itemize}
\item For simplicity, the \pif language does not enforce termination through
  type checking. Implementing a proof system like Agda or Coq requires
  additional structure, including universe levels and bounded recursion.
\item Many implementations of dependent type theories use
  \emph{normalization-by-evaluation} to decide whether types are equivalent
  during type checking.  \pif uses an alternative approach based on
  weak-head-normalization, defined using substitution. This approach is closer
  to $\lambda$-calculus theory but can be less efficient in practice.
\item For simplicity, \pif does not attempt to infer missing arguments using
  unification or other means. As a result, example programs in \pif are
  significantly more verbose than in other languages.
\item This implementation relies on the \unbound library for variable binding,
  alpha-equivalence, and substitution. As a result, these operations can be
  automatically derived. A more common approach is to use de Bruijn indices.
\item Recent work on cubical type theory, higher-inductive types, and
  univalence is not covered here. Indeed, the rules for dependent pattern matching
  that we present are incompatible with such extensions.
\end{itemize}

\paragraph{What do I want you to get out of all of this?}

\begin{enumerate}
\item An understanding of how to translate mathematical specifications of type
  systems and logics into code, i.e., how to represent the syntax of a
  programming language and how to implement a type checker. More generally,
  this involves techniques for turning a declarative specification of a system
  of judgments into an algorithm that determines whether the judgment holds.

\item Exposure to the design choices of dependently-typed languages. In this
  respect, the goal is breadth, not depth. As a result, I will provide
  \emph{simple} solutions to some of the problems that we face and sidestep
  other problems entirely. Because solutions are chosen for simplicity,
  Section~\ref{sec:related-work} includes pointers to related work if you want to go
  deeper.

\item Experience with the Haskell programming language. I think Haskell is an
  awesome tool for this sort of work and I want to demonstrate how its
  features (monads, generic programming) are well-suited for this task.

\item A tool that you can use as a basis for experimentation. When you design
  your language, how do you know what programs you can and cannot express?
  Having an implementation lets you work out (smallish) examples and will help
  convince you (and your reviewers) that you are developing something useful.
  Please use \pif as a starting point for your ideas.

\item Templates and tools for writing about type systems. The source files for
  these lecture notes are available in the same repository as the demo
  implementation and I encourage you to check them out.\footnote{See the
    \texttt{doc} subdirectory in \url{https://github.com/sweirich/pi-forall}.}
  Building these notes requires Ott~\cite{ott}, a tool specifically tailored
  for typesetting type systems and mathematical specifications of programming
  languages.
\end{enumerate}

\section{A taste of \pif}
\label{sec:examples}

Before diving into the design of the \pif language, we will start with a few
motivating examples for dependent types. This way, you will get to learn a bit
about the syntax of \pif and anticipate the definitions that are to come.

One example that dependently-typed languages are really good at is tracking
the length of lists.

In \pif, we can define a datatype for length-indexed lists (often called
vectors) as follows.\footnote{In the \texttt{full} implementation, you can find these
definitions in the file \texttt{pi/Vec.pi}.}
\begin{piforall}
data Vec (A : Type) (n : Nat) : Type where
  Nil of  [n = Zero]
  Cons of [m:Nat] (A) (Vec A m) [n = Succ m]
\end{piforall}
The type \cd{Vec A n} has two parameters: \cd{A} the type of elements in the
list and \cd{n}, a natural number indicating the length of the list. This
datatype also has two constructors: \cd{Nil} and \cd{Cons}, corresponding to
empty and non-empty lists. The \cd{Cons} constructor has three arguments,
a number \cd{m}, the element at the head of the list (of type \cd{A}), and the tail
of the list (of type \cd{Vec A m}).
Both constructors also include \emph{constraints} that must be satisfied when
they are used. The \cd{Nil} constructor constrains the length
parameter to be \cd{Zero} and the \cd{Cons} constructor
constrains it to be one more than the length of the tail of the list.

In the definition of \cd{Cons}, the \cd{m} parameter is the length of the tail
of the list. This parameter is written in square brackets to indicate that it
should be \emph{irrelevant}: this value is for type checking only and
functions should not use it.

For example, we can use \pif to check that the list below contains exactly
three boolean values.
\begin{piforall}
v3 : Vec Bool 3
v3 = Cons [2] True (Cons [1] False (Cons [0] False Nil))
\end{piforall}
Above, \pif never infers arguments, so we must always supply the length of the
tail whenever we use \cd{Cons}.

We can also map over length-indexed lists.
\begin{piforall}
map : [A:Type] -> [B:Type] -> [n:Nat] -> (A -> B) -> Vec A n -> Vec B n
map = \ [A][B][n] f v.
  case v of
    Nil -> Nil
    Cons [m] x xs -> Cons [m] (f x) (map[A][B][m] f xs)
\end{piforall}
As \pif doesn't infer arguments, the map function needs to take \cd{A}, \cd{B}
and \cd{n} explicitly at the beginning of the function definition, and they
need to be provided as arguments to the recursive call.

The type checker helps us avoid errors in the definition of \cd{map}. Like most
functional languages, if we had forgotten to call \cd{f} and instead included
\cd{x} in the \cd{Cons}, then \cd{map} would produce a vector containing elements of
type \cd{A} instead of \cd{B}. The type checker would flag this as an error. Similarly, if
we had replaces the \cd{Cons} branch with \cd{Nil} or had forgotten to \cd{Cons} \cd{f x} to the result of the recursive call in this branch, the resulting vector would be the wrong
length. The type checker would catch this error too!

The \pif language does not include the sophisticated type inference algorithms found in many
other languages, that can figure out some of the arguments provided
to a function automatically. As a result,
when we call map, we also need to supply the appropriate parameters for
the function and the vector that we are giving to \cd{map}. For example, to map
the boolean \cd{not} function, of type \cd{Bool -> Bool}, we need to provide
these two \cd{Bool}s along with the length of the vector.

\begin{piforall}
v4 : Vec Bool 3
v4 = map [Bool][Bool][3] not v3
\end{piforall}

Because we are statically tracking the length of vectors, we can write
a safe indexing operation. We index into the vector using an index where
the type bounds its range. Below, the \cd{Fin} type has a parameter
that states what length of vector it is appropriate for.
\begin{piforall}
data Fin (n : Nat) : Type where
  Zero of [m:Nat][n = Succ m]
  Succ of [m:Nat](Fin m)[n = Succ m]
\end{piforall}

For example, the type \cd{Fin 3} has exactly three values:
\begin{piforall}
x1 : Fin 3
x1 = Succ [2] (Succ [1] (Zero [0]))

x2 : Fin 3
x2 = Succ [2] (Zero [1])

x3 : Fin 3
x3 = Zero [2]
\end{piforall}

With the \cd{Fin} type, we can safely index vectors. The following
function is guaranteed to return a result because the index is within
range.
\begin{piforall}
nth : [a:Type] -> [n:Nat] -> Vec a n -> Fin n -> a
nth = \[a][n] v f. case f of
   Zero [m] -> case v of
           Cons [m'] x xs -> x
   Succ [m] f' -> case v of
           Cons [m'] x xs -> nth [a][m] xs f'
\end{piforall}
Note how, in the case analysis above, neither \cd{case} requires a branch for
the \cd{Nil} constructor. The \pif type checker can verify that the \cd{Cons}
case is exhaustive and that the \cd{Nil} case would be inaccessible. In these
cases, the type checker notes that the patterns for the missing branches don't
make sense: constructing the \cd{Nil} pattern would require satisfying the
constraint that \cd{Zero = Succ m}, which is impossible.

You may have noticed that some of the arguments to functions and constructors in the examples above are enclosed in square brackets. These arguments are \emph{irrelevant} in \pif. What this means is that the compiler can erase these arguments before running the program and can ignore these arguments when comparing types for equality. (When arguments are marked in this way, the type checker must ensure that this erasure is sound and the argument is never actually used in a meaningful way.)
Similarly, the constraints that are part of datatype definitions, such as \cd{n = Succ m} above, are enclosed in
square brackets because there is no runtime proof of the equality stored with the data constructor.

\subsection{Homework: Check out \pif}

The files in the directory \texttt{full/pi/} demonstrate the full power of the
\pif language. Take a little time to familiarize yourself with some of these
modules (such as \cd{Fin.pi} and \cd{Vec.pi}) and compare them to similar
ones that you might have seen in Agda, Haskell, or elsewhere.

To run the type checker on these modules, make sure that you have first
compiled the implementation using \texttt{stack build} at the terminal.  Then,
to run \pif on a source file, such as \cd{Fin.pi}, you can issue the command
\texttt{stack exec -{}- pi-forall Fin.pi} in the \cd{full/pi} subdirectory. If
the file type checks, you will see the contents of the file displayed in the
terminal window. Otherwise, you will see an error message from the type checker.

For a more extensive series of examples, start with \cd{pi/Lambda0.pi} for an
interpreter for a small lambda calculus and then compare it with the
implementations in \cd{pi/Lambda1.pi} and \cd{pi/Lambda2.pi}. These two
versions index the expression type with the scope depth (a natural number) and
the expression type to eliminate run-time scope and type errors from the
execution of the interpreter.

\subsubsection{Homework: Church and Scott  encodings}

The file \texttt{full/pi/NatChurch.pi} is a start at a Church encoding of
natural numbers. Replace the \cd{TRUSTME}s in this file so that it
compiles. For example, one task in this file is to define a predecessor
function. By replacing the \cd{TRUSTME} with \cd{Refl} below, you will be able
to force the type checker to normalize both sides of the equality. The code
will only type check if you have defined \cd{pred} correctly (on this
example).

\begin{piforall}
test_pred : pred two = one
test_pred = TRUSTME -- replace with Refl
\end{piforall}

\section{A Simple Core Language}
\label{sec:simple}

Now let's turn to the design and implementation of the \pif language
itself. We'll start with a small core, and then incrementally add features
throughout the rest of this tutorial.

Consider this simple dependently-typed lambda calculus, shown below. What should it
contain? At the bare minimum we start with the following five forms:

\[
\begin{array}{rcll}
     \ottnt{a},\ottnt{b},\ottnt{A},\ottnt{B} & ::=& \ottmv{x}  &\mbox{ variables  }\\
         && \lambda  \ottmv{x} .  \ottnt{a}           &\mbox{ lambda expressions (anonymous functions)} \\
         && \ottnt{a}  \;  \ottnt{b}             &\mbox{ function applications }\\
         && ( \ottmv{x} \!:\! \ottnt{A} )  \to   \ottnt{B}      &\mbox{ dependent function type, aka $\Pi$-types }\\
         &&\ottkw{Type}           &\mbox{ the `type' of types}\\
\end{array}
\]

As in many dependently-typed languages, we have the \emph{same} syntax for
both expressions and types. For clarity, the convention is to use lowercase
letters ($a$,$b$) for expressions and uppercase letters ($A$,$B$) for types.

Note that $\lambda$ and $\Pi$ above are \emph{binding forms}. They bind the
variable $x$ in $a$ and $B$ respectively. In dependent function types
$ ( \ottmv{x} \!:\! \ottnt{A} )  \to   \ottnt{B} $, if $x$ does not appear free in $B$, it is also convention
to write the type as $\ottnt{A}  \to  \ottnt{B}$.

\subsection{When do expressions in this language type check?}

We define the type system for this language using an inductive
relation shown in Figure~\ref{fig:typing}. This relation is between an
expression $\ottnt{a}$, its type $\ottnt{A}$, and a typing context $\Gamma$.

\[ \fbox{$\Gamma  \vdash  \ottnt{a}  \ottsym{:}  \ottnt{A}$} \]

A typing context $\Gamma$ is an ordered list of assumptions about variables and
their types.

    \[ \Gamma  ::= \emptyset\ |\  \Gamma ,  \ottmv{x} \! :\! \ottnt{A}  \]

We will assume that each of the variables in this list are
distinct from each other so that there will always be at most one assumption
about any variable's type.\footnote{On paper, this assumption is reasonable
  because we always extend the context with variables that come from binders,
  and we can always rename bound variables so that they differ from other
  variables in scope. Any implementation of the type system will need to somehow
  make sure that this invariant is maintained. We will do this in Section~\ref{sec:implementation} using
  a freshness monad from the \unbound library.}

\paragraph{An initial set of typing rules: Variables and Functions}

\begin{figure}[t]
\drules[t]{$\Gamma  \vdash  \ottnt{a}  \ottsym{:}  \ottnt{A}$}{Core type system}{var,lambda,app,pi,type}
\caption{Typing rules for core system}
\label{fig:typing}
\end{figure}

Consider the first two rules shown in Figure~\ref{fig:typing}, \rref{t-var}
and \rref{t-lambda}.  The first rule states that the types of variables are
recorded in the typing context. The premise $\ottmv{x}  \ottsym{:}  \ottnt{A} \, \in \, \Gamma$ requires us to
find an association between $x$ and the type $A$ in the list $\Gamma$.

The second rule, for type checking functions, introduces a new variable into
the context when we type check the body of a $\lambda$-expression. It also
requires $\ottnt{A}$, the type of the function's argument, to be a valid type.

\paragraph{Example: Polymorphic identity functions}

Note that in \rref{t-lambda}, the parameter $x$ is allowed to appear in the
result type of the function, $\ottnt{B}$. Why is this useful? Well, it gives us
\emph{parametric polymorphism} automatically.  In Haskell, we write the
identity function as follows, annotating it with a polymorphic type.

\begin{haskell}
id :: forall a. a -> a
id = \y -> y
\end{haskell}

\noindent
Because the type of \cd{id} is generic we can apply this function to any type of
argument.

We can also write a polymorphic identity function in \pif, as follows.
\begin{piforall}
id : (x:Type) -> (y : x) -> x
id = \x. \y. y
\end{piforall}

This definition is similar to Haskell, but with a few modifications. First,
\pif{} uses a single colon for type annotations and a \cd{.} for function
binding instead of Haskell's quirky \cd{::} and \cd{->}. Second, the \pif
definition of \texttt{id} uses two lambdas: one for the ``type'' argument $x$
and one for the ``term'' argument $y$. In \pif, there is no syntactic
difference between these arguments: both are arguments to the identity
function.

Finally, the \pif{} type of \cd{id} uses the dependent function type, with
general form $ ( \ottmv{x} \!:\! \ottnt{A} )  \to   \ottnt{B} $. This form is like Haskell's usual function type
$\ottnt{A}  \to  \ottnt{B}$, except that we can name the argument $\ottmv{x}$ so that it can be
referred to in the body of the type $\ottnt{B}$. In \cd{id}, dependency allows
the type argument $\ottmv{x}$ to be used later as the type of $\ottmv{y}$.  We call
types of this form $\Pi$-types. (Often dependent function types are written as
$\Pi x\!:\!A. B$ in formalizations of dependent type theory.\footnote{The
  terminology for these types is muddled: sometimes they are called dependent
  function types and sometimes they are called dependent product types. We use
  the non $\Pi$-notation to emphasize the connection to functions.})

The fact that the type of $x$ is $\ottkw{Type}$ means that $x$ plays the role of a
type variable, such as \texttt{a} in the Haskell type. Because we don't have a
syntactic distinction between types and terms, in \pif we say that ``types''
are anything of type $\ottkw{Type}$ and ``terms'' are things of type $A$ where $A$
has type $\ottkw{Type}$.

We can use the typing rules to construct a typing derivation for the identity
function as follows.

\[
\inferrule*
{
   \inferrule*
   {
              \inferrule*{ \ottmv{y}  \ottsym{:}  \ottmv{x} \, \in \, \ottsym{(}    \ottmv{x} \! :\! \ottkw{Type}  ,  \ottmv{y} \! :\! \ottmv{x}   \ottsym{)}}{  \ottmv{x} \! :\! \ottkw{Type}  ,  \ottmv{y} \! :\! \ottmv{x}   \vdash  \ottmv{y}  \ottsym{:}  \ottmv{x}}
      \qquad \inferrule*{\ottmv{x}  \ottsym{:}  \ottkw{Type} \, \in \, \ottsym{(}   \ottmv{x} \! :\! \ottkw{Type}   \ottsym{)} }{ \ottmv{x} \! :\! \ottkw{Type}   \vdash  \ottmv{x}  \ottsym{:}  \ottkw{Type}}
   }{
      \ottmv{x} \! :\! \ottkw{Type}   \vdash   \lambda  \ottmv{y} .  \ottmv{y}   \ottsym{:}   ( \ottmv{y} \!:\! \ottmv{x} )  \to   \ottmv{x} 
   }  \qquad \inferrule*{}{ \emptyset   \vdash  \ottkw{Type}  \ottsym{:}  \ottkw{Type}}
}{
   \emptyset   \vdash   \lambda  \ottmv{x} .   \lambda  \ottmv{y} .  \ottmv{y}    \ottsym{:}   ( \ottmv{x} \!:\! \ottkw{Type} )  \to   \ottsym{(}  \ottmv{y}  \ottsym{:}  \ottmv{x}  \ottsym{)}   \to  \ottmv{x}
}
\]

This derivation uses \rref{t-lambda} to type check the two $\lambda$-expressions, before using the variable
rule to ensure that both the body of the function $\ottmv{x}$ and its type $\ottmv{y}$ are
well-typed. Finally, this rule also uses \rref{t-type} to show that $\ottkw{Type}$ itself
is a valid type, see below.

Note that the \rref{t-lambda} makes subtle use of the fact that binding variables can
be freely renamed to alpha-equivalent versions. In the conclusion of the rule, the $\ottmv{x}$ in
the body of the lambda abstraction $\ottnt{a}$, is a different binder from the $\ottmv{x}$ in the
body of the function type $\ottnt{B}$. Both of these terms can be modified so that these two
$\ottmv{x}$ do not have to match. But, above the line, there is only a single $\ottmv{x}$ that appears
free in both $\ottnt{a}$ and $\ottnt{B}$! If we wanted to be explicit about what is going on, we could
also write the rule with an explicit substitution:

\[ \drule{t-altlambda}  \]

Most type theorists do not bother with this version and prefer the one that
reuses the same binder. We will do the same here, and in other similar
rules. However, we will also have to be careful about how we implement this
rule in the implementation of our type checker!

\paragraph{More typing rules:  Types}

Observe in the typing derivation above that the rule for typing
$\lambda$-expressions has a second precondition: we need to make sure that
when we add assumptions $x:A$ to the context, that $A$ really is a type.
Without this precondition, the rules would allow us to derive this nonsensical
type for the polymorphic identity function.

\[    \emptyset   \vdash   \lambda  \ottmv{x} .   \lambda  \ottmv{y} .  \ottmv{y}    \ottsym{:}   ( \ottmv{x} \!:\! \ottkw{True} )  \to   \ottsym{(}  \ottmv{y}  \ottsym{:}  \ottmv{x}  \ottsym{)}   \to  \ottmv{x} \]

This precondition means that we need some rules that conclude that types are
actually types. For example, the type of a function is itself a type, so we
will declare it so with \rref{t-pi}. This rule also ensures that the domain
and range of the function are also types.

Likewise, for polymorphism, we need the somewhat perplexing \rref{t-type}, that
declares, by fiat, that $\ottkw{Type}$ is a valid $\ottkw{Type}$.\footnote{Note that
  this rule makes our language inconsistent as a logic, as it can encode
  a logical paradox. The \cd{full} implementation includes a demonstration of
this paradox in the file \cd{Hurkens.pi}.}

\paragraph{More typing rules: Applications}

The application rule, \rref{t-app}, requires that the type of the argument
matches the domain type of the function. However, note that because the body
of the function type $B$ could have $x$ free in it, we need also need to
substitute the argument $\ottnt{b}$ for $x$ in the result.

\paragraph{Example: applying the polymorphic identity function}

In \pif we should be able to apply the polymorphic identity function to
itself. When we do this, we need to first provide the type of \cd{id},
producing an expression of type
$ ( \ottmv{y} \!:\! \ottsym{(}   ( \ottmv{x} \!:\! \ottkw{Type} )  \to   \ottsym{(}  \ottmv{y}  \ottsym{:}  \ottmv{x}  \ottsym{)}   \to  \ottmv{x}  \ottsym{)} )  \to   \ottsym{(}   ( \ottmv{x} \!:\! \ottkw{Type} )  \to   \ottsym{(}  \ottmv{y}  \ottsym{:}  \ottmv{x}  \ottsym{)}   \to  \ottmv{x}  \ottsym{)} $.  This
function can take \cd{id} as an argument.

\begin{piforall}
    idid : (x:Type) -> (y : x) -> x
    idid = id ((x:Type) -> (y : x) -> x) id
\end{piforall}

\paragraph{Example: Church booleans}

Because we have (impredicative) polymorphism, we can \emph{encode} familiar
types, such as booleans. The idea behind this encoding, called the Church
encoding, is to represent terms by their eliminators. In other words, what is
important about the value true?  The fact that when you get two choices, you
pick the first one.  Likewise, false ``means'' that with the same two choices,
you should pick the second one.  With parametric polymorphism, we can give the
two terms the same type, which we'll call ``bool''.

\begin{piforall}
true : (A:Type) -> A -> A -> A
true = \x. \y. \z. y

false : (A:Type) -> A -> A -> A
false = \x. \y. \z. z
\end{piforall}

\noindent
Thus, a conditional expression just takes a boolean and returns it.

\begin{piforall}
cond : ((A:Type) -> A -> A -> A) -> ((x:Type) -> x -> x -> x)
cond = \b. b
\end{piforall}

\paragraph{Example: void type}

The type \cd{(x:Type) -> x}, called ``void'' is usually an \emph{empty type}
in polymorphic languages.  Observe that this function takes a type \cd{x:Type},
but never takes any expressions of type \cd{x}. Thus, there is no way to return
a value of any type without some sort of type case (not part of \pif).

However, because \pif does not enforce termination, we can define a value that
has this type.

\begin{piforall}
loop : (x:Type) -> x
loop = \x . loop x
\end{piforall}

\paragraph{Typing invariant}

Overall, a property that we want for our type system is that if a term type
checks, then its \emph{type} will also type check. More formally, to state
this property, we first need to say what it means to check all types in a
context.

\begin{definition}[Context type checks]
Define \fbox{$\vdash  \Gamma$} with the following rules. \[ \drule{g-nil} \qquad \drule{g-cons} \]
\end{definition}
\noindent
With this judgement, we can state the following property of our type system.
\begin{lemma}[Regularity]
If $\Gamma  \vdash  \ottnt{a}  \ottsym{:}  \ottnt{A}$ and $\vdash  \Gamma$, then $\Gamma  \vdash  \ottnt{A}  \ottsym{:}  \ottkw{Type}$
\end{lemma}
\noindent
Making sure that this property holds for the type system is the motivation for the
$\Gamma  \vdash  \ottnt{A}  \ottsym{:}  \ottkw{Type}$ premise in \rref{t-lambda}.

\section{From typing rules to a typing algorithm}
\label{sec:bidirectional}

The rules that we have developed so far are great for saying \emph{what} terms
should type check, but they don't say \emph{how} type checking works. We have
developed these rules without thinking about how we would implement them.

A set of typing rules, or \emph{type system}, is called \emph{syntax-directed}
if it is readily apparent how to interpret that collection of typing rules as
code. In other words, for some type systems, we can directly translate them to
some Haskell function.

For this type system, we would like to implement the following Haskell
function, which when given a term and a typing context, represented as a list
of pairs of variables and their types, produces the type of the term if it
exists.\footnote{Note: If you are looking at the \pif implementation, note
  that this is not the final type of \cd{inferType}.}

\begin{haskell}
type Ctx = [(Var,Type)]

inferType :: Term -> Ctx -> Maybe Type
\end{haskell}

Let us look at our rules in Figure~\ref{fig:typing}.  Is the definition of this
function straightforward?  For example, in the variable rule, as long as we
can look up the type of a variable in the context, we can produce its
type. That means that, assuming that there is some function \cd{lookupTy},
with type \cd{Ctx -> Var -> Maybe Type}, this rule corresponds to the
following case of the \cd{inferType} function. So far so good!

\begin{haskell}
inferType (Var x) ctx = lookupTy ctx x
\end{haskell}

\noindent
Likewise, the case of the typing function for the $\ottkw{Type}$ term is also
straightforward. When we see the term $\ottkw{Type}$, we know immediately that it
is its own type.


The only stumbling block for the algorithm is \rref{t-lambda}. To type check a
function, we need to type check its body when the context has been extended
with the type of the argument. But, like Haskell, the type of the argument
$\ottnt{A}$ is not annotated on the function in \pif. So where does it come from?

There is actually an easy fix to turn our current system into an algorithmic
one. We could just annotate $\lambda$-expressions with the types of the
abstracted variables.  But, to be ready for future extension, we will do
something else.

Look at our example \pif code above: the only types that we wrote were the
types of definitions. It is good style to do that. Furthermore, there is
enough information there for type checking---wherever we define a function, we
can look at those types to know what type its argument should have.  So, by
changing our point of view, we can get away without annotating lambdas with
those argument types.

\subsection{A bidirectional type system}

Let us redefine the system using two judgments. The first one is similar to the
judgment that we saw above, and we will call it type
\emph{inference}.\footnote{The term \emph{type inference} usually refers to
  much more sophisticated deduction of an expression's type, in the context of
  much less information encoded in the syntax of the language. We are not doing
  anything difficult here, just noting that we can read the judgment with
  $\ottnt{A}$ as an output. } This judgment will be paired with (and will depend
on) a second judgment, called type \emph{checking}, that takes advantage of
known type information, such as the annotations on top-level definitions.

We express these judgments using the notation defined in
Figure~\ref{fig:bidirectional} and implement them in Haskell using the
mutually-recursive functions \texttt{inferType} and
\texttt{checkType}. Furthermore, to keep track of which rule is in which
judgment, rules that have inference as conclusion start with
\textsc{i-} and rules that have checking as conclusion start with
\textsc{c-}.

\begin{figure}
\drules[i]{$\Gamma  \vdash  \ottnt{a}  \Rightarrow  \ottnt{A}$}{in context $\Gamma$, infer that term $a$ has type $A$}
{var,app-simple,pi,type,annot}
\drules[c]{$\Gamma  \vdash  \ottnt{a}  \Leftarrow  \ottnt{A}$}{in context $\Gamma$, check that term $a$ has type $A$}
{lambda,infer-simple}
\caption{(Simple) Bidirectional type system}
\label{fig:bidirectional}
\end{figure}

Let us compare these rules with our original typing rules. For \rref{i-var},
we need to only change the colon to an inference arrow. The context tells us
the type to infer.

On the other hand, in \rref{c-lambda} we should check $\lambda$-expressions
against a known type. If that type is provided, we can propagate it to the
body of the lambda expression. (If we assume that the type provided to
the checking judgment has already been checked to be a good type, then
we can skip checking that $A$ is a $\ottkw{Type}$.)

The rule for applications, \rref{i-app-simple}, is in inference mode. Here, we
first infer the type of the function, but once we have that type, we may use
it to check the type of the argument. This mode change means that
$\lambda-$expressions that are arguments to other functions (like
\texttt{map}) do not need any annotation.

For example, if map has type
\begin{haskell}
map : (x : Type) -> (y: Type) -> (x -> y) -> List x -> List y
\end{haskell}
then in an application, we can synthesize the type of the expression
\begin{haskell}
map Nat Nat (\x.x)
\end{haskell}
as \cd{List Nat -> List Nat}.
After the two type applications, we will end up checking the function
argument against the type \cd{Nat -> Nat}.

For types, it is apparent their type is $\ottkw{Type}$, so \rref{i-pi,i-type} just
continue to infer that.

Notice that this system is incomplete. There are inference rules for every
form of expression except for lambda. On the other hand, only lambda
expressions can be checked against types.  We can make the checking judgment
more applicable with \rref{c-infer-simple} that allows us to use inference
whenever a checking rule doesn't apply. This rule only applies when the
term is not a lambda expression.

Now, let us think about the reverse problem a bit. There are programs that the
checking system won't admit but would have been acceptable by our first
system. What do they look like?

Well, they involve applications of explicit lambda terms:

\[
\inferrule*[Right=t-app]
{
       \emptyset   \vdash   \lambda  \ottmv{x} .  \ottmv{x}   \ottsym{:}  \ottkw{Bool}  \to  \ottkw{Bool}  \qquad   \emptyset   \vdash  \ottkw{True}  \ottsym{:}  \ottkw{Bool}
}{
       \emptyset   \vdash   \ottsym{(}   \lambda  \ottmv{x} .  \ottmv{x}   \ottsym{)}  \;  \ottkw{True}   \ottsym{:}  \ottkw{Bool}
}
\]

This term doesn't type check in the bidirectional system because application
requires the function to have an inferable type, but lambdas don't.
However, there is not that much need to write such terms. We can always
replace them with something equivalent by doing a $\beta$-reduction of the
application (in this case, just replace the term with $\ottkw{True}$).

In fact, the bidirectional type system has the property that it only checks
terms in \emph{normal} form, i.e., those that do not contain any
$\beta$-reductions.

\paragraph{Type annotations}
To type check nonnormal forms in \pif, we also add typing annotations as a new
form of expression to \pif, written $\ottsym{(}  \ottnt{a}  \ottsym{:}  \ottnt{A}  \ottsym{)}$, and add \rref{i-annot} to
the type system.

Type annotations allow us to supply known type information anywhere, not just
at the top level. For example, we can construct this derivation.

\[
\inferrule*[Right=i-app]
{
       \emptyset   \vdash  \ottsym{(}   \lambda  \ottmv{x} .  \ottmv{x}   \ottsym{:}  \ottkw{Bool}  \to  \ottkw{Bool}  \ottsym{)}  \Rightarrow  \ottkw{Bool}  \to  \ottkw{Bool}  \qquad   \emptyset   \vdash  \ottkw{True}  \Leftarrow  \ottkw{Bool}
}{
       \emptyset   \vdash   \ottsym{(}   \lambda  \ottmv{x} .  \ottmv{x}   \ottsym{:}  \ottkw{Bool}  \to  \ottkw{Bool}  \ottsym{)}  \;  \ottkw{True}   \Rightarrow  \ottkw{Bool}
}
\]

The nice thing about the bidirectional system is that it reduces the number of
annotations that are necessary in programs that we want to write. As we will
see, checking mode will be even more important as we add more terms to the
language.

Furthermore, we want to convince ourselves that the bidirectional system
checks the same terms as the original type system. This means that we want to
prove a property like this one\footnote{This result requires the addition of a
  typing rule for annotated terms $\ottsym{(}  \ottnt{a}  \ottsym{:}  \ottnt{A}  \ottsym{)}$ to the original system.}:

\begin{lemma}[Correctness of the bidirectional type system]
If $\Gamma  \vdash  \ottnt{a}  \Rightarrow  \ottnt{A}$ and $\vdash  \Gamma$ then $\Gamma  \vdash  \ottnt{a}  \ottsym{:}  \ottnt{A}$.
If $\Gamma  \vdash  \ottnt{a}  \Leftarrow  \ottnt{A}$ and $\vdash  \Gamma$ and $\Gamma  \vdash  \ottnt{A}  \ottsym{:}  \ottkw{Type}$ then $\Gamma  \vdash  \ottnt{a}  \ottsym{:}  \ottnt{A}$.
\end{lemma}

On the other hand, the reverse property is not true. Even if there exists some
typing derivation $\Gamma  \vdash  \ottnt{a}  \ottsym{:}  \ottnt{A}$ for some term $\ottnt{a}$, we may not be able
to infer or check that term using the algorithm. However, all is not lost:
there will always be some term $\ottnt{a'}$ that differs from $\ottnt{a}$ only in the
addition of typing annotations that can be inferred or checked instead.

One issue with this bidirectional system is that it is not closed under
substitution.  What this means is that given some term $\ottnt{b}$ with a free
variable, $ \Gamma ,  \ottmv{x} \! :\! \ottnt{A}   \vdash  \ottnt{b}  \Leftarrow  \ottnt{B}$, and another term $\ottnt{a}$ with the same
type $\Gamma  \vdash  \ottnt{a}  \Leftarrow  \ottnt{A}$ of that variable, we \emph{do not} have a derivation
$\Gamma  \vdash  \ottnt{b}  \ottsym{[}  \ottnt{a}  \ottsym{/}  \ottmv{x}  \ottsym{]}  \Leftarrow  \ottnt{A}$. The reason is that types of variables are always
inferred, but the term $\ottnt{a}$, may need the type $\ottnt{A}$ to be supplied to
the type checker.  This issue is particularly annoying in \rref{i-app} when we
replace a variable (inference mode) with a term that was validated in checking
mode.

As a result, our type system infers types, but the types that are inferred do
not have enough annotations to be checked themselves. We can express the property
that does hold about our system, using this lemma:

\begin{lemma}[Regularity]
If $\Gamma  \vdash  \ottnt{a}  \Rightarrow  \ottnt{A}$ and $\vdash  \Gamma$ then $\Gamma  \vdash  \ottnt{A}  \ottsym{:}  \ottkw{Type}$.
\end{lemma}

This issue is not significant, and we could resolve it by adding an annotation
before substitution. However, in our implementation of \pif, we do not do so
to reduce clutter.

\subsection{Homework:  Add Booleans and
\(\Sigma\)-types (Part I)}

Some fairly standard typing rules for booleans are shown below.

\drules{$\Gamma  \vdash  \ottnt{a}  \ottsym{:}  \ottnt{A}$}{Booleans}{t-bool,t-true,t-false}
\[ \drule[width=4in]{t-if-simple} \]

Likewise, we can also extend the language with $\Sigma$-types, or
products, where the type of the second component of the product can depend on the value of the first component.

\drules{$\Gamma  \vdash  \ottnt{a}  \ottsym{:}  \ottnt{A}$}{$\Sigma$-types}{t-sigma,t-pair}
\[ \drule[width=4in]{t-letpair-weak} \]

We destruct $\Sigma$-types using pattern matching. The simple rule for pattern
matching introduces variables into the context when type checking the body of
the \textsf{let} expression.  These variables are not allowed to appear free
in the result type of the pattern match.

For your exercise, rewrite the rules above in bidirectional style. Which rules
should be inference rules? Which ones should be checking rules? If you are
familiar with other systems, how do these rules compare?

\section{Putting it all together in a Haskell
implementation}
\label{sec:implementation}

In the previous section, we defined a bidirectional type system for a small
core language. Here, we will start talking about what the implementation of this
language looks like in Haskell.

First, an overview of the main files of the implementation. There are a few
more source files in the repository (in the subdirectory \cd{src/}), but these
are the primary ones.

\begin{verbatim}
 Syntax.hs      - specification of the AST
 Parser.hs      - turn strings into AST
 PrettyPrint.hs - displays AST in a (somewhat) readable form
 Main.hs        - runs the parser and type checker

 Environment.hs - defines the type checking monad
 TypeCheck.hs   - implementation of the bidirectional type checker
\end{verbatim}

\subsection{Variable binding using the \unbound library
  {[}Syntax.hs{]}}

One difficulty with implementing any sort of lambda calculus is the treatment
of variable binding. Functions ($\lambda$-expressions) and their types
($\Pi$-types) \emph{bind} variables. In the implementation of our type
checker, we will need to be able to determine whether two terms are
\emph{alpha-equivalent}, calculate the \emph{free variables} of a term, and
perform \emph{capture-avoiding substitution.} When we work with a
$\lambda$-expression, we will want to be sure that the binding variable is
\emph{fresh}, that is, distinct from all other variables in the program or in
the typing context.

In the \pif implementation, we use the \unbound library to get all of these
operations (mostly) for free. This works because we use type constructors from
this library in the definition of the abstract syntax of \pif, and those types
specify the binding structure of \pif expressions.

First, the \unbound library defines a type for variable names, called
\cd{Name} and we use this type to define \cd{TName}, the type of term
names in our AST.
\begin{verbatim}
-- | Type used for variable names in Terms
type TName = Unbound.Name Term
\end{verbatim}
The \cd{Unbound.Name} type is indexed by \cd{Term}, the AST type that it is a
name for, whose definition we will see shortly. That way \unbound can make
sure that we only substitute the right form of syntax for the right name,
i.e., we can only replace a \cd{TName} with a \cd{Term}.  The library
includes an overloaded function \cd{subst\ x\ a\ b}, which corresponds to
the substitution we have been writing as $\ottnt{b}  \ottsym{[}  \ottnt{a}  \ottsym{/}  \ottmv{x}  \ottsym{]}$ in our inference
rules,\footnote{See Guy Steele's talk about the notation for
  substitution~\cite{steele:ppop17}. This is the most common mathematical
  notation for this operation.} i.e., substitute $\ottnt{a}$ for $\ottmv{x}$ in
$\ottnt{b}$.

In general, we want to apply a substitution to many different sorts of
syntax. For example, we may want to substitute a term $\ottnt{a}$ for a term name
$\ottmv{x}$ in all of the terms that appear in the typing context. Or there may be
other sorts of syntax that terms can be part of, and we want to be able to
substitute for term variables through that syntax too. Therefore, \unbound's
substitution function has a type with the following general pattern for
\cd{subst a x b}.
\begin{lstlisting}[language=Haskell]
 class Subst a b where
    subst  :: Name a -> a -> b -> b
\end{lstlisting}

The \cd{subst} function in this class ensures that when we see that
\cd{a} is the right sort of thing to stick in for \cd{x}. The library
can automatically generate instances of the \cd{Subst} class.

With term names, we can define the abstract syntax that corresponds to our
language above, using the following datatype.

\begin{lstlisting}[language=Haskell]
data Term

  = -- | type of types  `Type`
    TyType

  | -- | variables  `x`
    Var TName

  | -- | abstractions  `\x. a`
    Lam (Unbound.Bind TName Term)

  | -- | applications `a b`
    App Term Term

  | -- | function types   `(x : A) -> B`
    TyPi Type (Unbound.Bind TName Type)

  | -- | Annotated terms `( a : A )`
    Ann Term Type

  ...
  deriving (Show, Generic)
\end{lstlisting}

As you can see, variables are represented by names. \unbound's \cd{Bind}
type constructor declares the scope of the bound variables. Both \cd{Lam}
and \cd{TyPi} bind a single variable in a \cd{Term}.  However, in a
\cd{TyPi} term, we also want to store the type $\ottnt{A}$, the type of the bound
variable $\ottmv{x}$.

Because the syntax is all shared, a \cd{Type} is just another name for a
\cd{Term}. We will use this name in the source code just for documentation.
\begin{lstlisting}[language=Haskell]
  type Type = Term
\end{lstlisting}

Among other things, the \cd{Alpha} class enables functions for alpha
equivalence and free variable calculation, with the types shown below.
Usually \unbound can derive instances of this class for us so that we don't
have to worry about defining these functions manually.

\begin{lstlisting}[language=Haskell]
aeq :: Alpha a => a -> a -> Bool
\end{lstlisting}

For \cd{Term}, we do not use the default definitions of the \cd{Alpha}
and \cd{Subst} classes. Instead, we need to provide a bit more
information.  First, we would like the definition of alpha-equivalence of
terms to ignore type annotations and source code positions for error
messages. Therefore, our instance does so for these constructors of the
\cd{Term} data type and defers to the generic definition for the rest.

For example, if \cd{xName} and \cd{yName} are two different names, then
\unbound's implementation of $\alpha$-equivalence can equate terms that
differ only in using these names for binding.
\begin{piforall}
-- '\x -> x'
idx :: Term
idx = Lam (Unbound.bind xName (Var xName))

-- '\y -> y'
idy :: Term
idy = Lam (Unbound.bind yName (Var yName))

-- >>> Unbound.aeq idx idy
-- True
\end{piforall}

Second, the \cd{Subst} instance for \cd{Term} requires telling \unbound where
variables occur in the \cd{Term}. This short function pattern matches to find
the \cd{Var} constructor, extracts its name, and returns it to \unbound. We
don't need to provide any more code to \unbound in the implementation of
substitution; the generic implementation suffices.

\begin{verbatim}
instance Unbound.Subst Term Term where
  isvar (Var x) = Just (SubstName x)
  isvar _ = Nothing
\end{verbatim}

\noindent
For more details about \unbound, see the \texttt{unbound-generics}
documentation\footnote{\url{https://github.com/lambdageek/unbound-generics}}.

\subsection{A Type Checking monad {[}Environment.hs{]}}

Recall that our plan is to write two mutually recursive functions for
type checking of the following types:

\begin{verbatim}
inferType :: Term -> Ctx -> Maybe Type

checkType :: Term -> Type -> Ctx -> Bool
\end{verbatim}

The inference function should take a term and a context and if it type checks,
produce its type. The checking function should take a term, a context, and a
type, and determine whether the term has that type. However, note that the
types above are not very informative in the case of failure: if a term does
not type check, then the result is \cd{Nothing} or \cd{False}.

For that reason, we will do something a bit different. We will define a \emph{type
  checking monad}, called \texttt{TcMonad}, that will handle the plumbing for
the typing context, and more importantly, allow us to return more information
when a program doesn't type check. Therefore our two functions will have these
types that hide the context and rely on throwing errors to indicate type
checking failure.

\begin{verbatim}
inferType :: Term -> TcMonad Type

checkType :: Term -> Type -> TcMonad ()
\end{verbatim}

Those of you who have worked with Haskell before may be familiar with the
classes
\href{https://hackage.haskell.org/package/mtl-2.1.2/docs/Control-Monad-Reader.html}{MonadReader},
and the
\href{https://hackage.haskell.org/package/mtl-2.1.2/docs/Control-Monad-Error.html}{MonadError},
which our type checking monad will be instances of. These two classes mean
that that the \cd{TcMonad} type can provide access to the type checking
context and can throw errors. We will wrap this functionality up a bit with the
following functions. The first two encapsulate the sort of context
manipulation we need to do: look up the types of variables and add a new
assumption about the type of a variable to the context (both \cd{Sig} and
\cd{Decl} are defined in \cd{Syntax.hs}).

\begin{verbatim}
-- | Find the type of a name specified in the context
-- throwing an error if the name doesn't exist
lookupTy :: TName -> TcMonad Sig

-- | Extend the context with a new binding
extendCtx :: Decl -> TcMonad Term -> TcMonad Term

-- | Throw an error at the current location, assembling the error
-- message from a list of displayable objects.
-- Terminates type checking.
err  :: (Disp a) => [a] -> TcMonad b

-- | Print a warning, but continue the current computation.
warn :: (Disp a) => a -> TcMonad ()
\end{verbatim}

We will also need this monad to be a freshness monad, to support working with
binding structure (i.e., a source of fresh names), and throw in \cd{MonadIO} so
that we can print warning messages.

\subsection{Implementing the Type Checking Algorithm
{[}Typecheck.hs{]}}

Now that we have the type checking monad available, we can start our
implementation.

The general structure of the \cd{inferType} function starts with a pattern
match for the various syntactic forms. There are also several cases for
practical reasons (annotations, source code positions, etc.) and a few cases
for homework.  If the form of the term does not match any of the forms of the
terms that we can synthesize types for, then the type checker produces an
error.

\begin{verbatim}
inferType tm = case tm of
   (Var x) -> ...

   TyType -> ...

   (TyPi tyA bnd) -> ...

   (App t1 t2) -> ...

   (Ann tm ty) -> ...

   ...

   _ -> Env.err [DS "Must have a type for", DD tm]
\end{verbatim}

For checking, the general form of the function is also a pattern match on the
syntactic forms. The functions \cd{inferType} and \cd{checkType} are defined
in terms of each other by mutual recursion.

Mixed in here, we also have a pattern for lambda expressions in checking
mode:

\begin{verbatim}
checkType tm ty = case tm of
  (Lam bnd) -> case ty of
     TyPi tyA bnd2 = ...
        -- pass in the Pi type for the lambda expression
     _ ->
       Env.err [DS "Lambda expression has a function type, not ", DD ty]
\end{verbatim}

Again there are several cases for practical reasons, plus cases for homework.
Finally, the last case covers all other forms of checking mode, by calling
inference mode and making sure that the inferred type is equal to the checked
type. This case is the implementation of \rref{c-infer}.

\begin{verbatim}
 -- c-infer
 tm -> do
      ty' <- inferType tm
      unless (Unbound.aeq ty' ty) $
         Env.err [DS "Types don't match", DD ty, DS "and", DD ty'] -}
\end{verbatim}

The function \texttt{aeq} ensures that the two types are alpha-equivalent. If
they are not, it stops the computation and throws an error.



\subsection{Exercise:  Add Booleans and
\(\Sigma\)-types (Part II)}

The code in \texttt{version1/} includes abstract and
concrete syntax for booleans and $\Sigma$-types. The \pif file
\texttt{version1/test/Hw1.pi} contains examples of using these new forms. Your
job is to get this file to compile by filling in the missing cases in
\texttt{version1/src/TypeCheck.hs} based on the bidirectional rules that you
worked out in the previous exercise.

\section{Equality in Dependently-Typed Languages}
\label{sec:equality}

You may have noticed in the previous sections that there was something
missing. Most of the examples that we did could have also been written in
System F (or something similar)!

Next, we are going to think about how adding a notion of definitional equality
can make our language more expressive.

The addition of definitional equality follows several steps. First, we will see why we even want this feature in the language in the first place. Next, we will create a declarative specification of definitional equality and extend our typing relation with
a new rule that enables it to be used. After that, we will talk about algorithmic versions of both the equality relation and how to introduce it into the algorithmic type system.
Finally, we will cover modifications to the Haskell implementation.

\subsection{Motivating Example: Type level reduction}

In full dependently-typed languages (and in full \pif) we can see the need for
definitional equality. We want to equate types that are not merely
alpha-equivalent, so that more expressions type check.

Here's an example where we want a definition of equality
that is more expressive than alpha-equivalence. Consider an encoding for
simply-typed products:

\begin{piforall}
pair : Type -> Type -> Type
pair = \p. \q. (c: Type) -> (p -> q -> c) -> c
\end{piforall}

Unfortunately, our definition of \texttt{prod} doesn't type check:

\begin{piforall}
prod : (p:Type) -> (q:Type) -> p -> q -> pair p q
prod = \p.\q. \x.\y. \c. \f. f x y
\end{piforall}

Running this example with \texttt{version1} of the type checker produces
the following error:

\begin{verbatim}
Type Error:
pi/Lec1.pi:56:22:
  Lambda expression should have a function type, not pair p q
  When checking the term
    \p. \q. \x. \y. \c. \f. f x y
  against the signature
    prod : (p : Type) -> (q : Type) -> p -> q -> pair p q
  In the expression
    \c. \f. f x y
\end{verbatim}

The problem is that even though we want \texttt{pair\ p\ q} to be equal
to the type
\texttt{(c:\ Type)\ -\textgreater{}\ (p\ -\textgreater{}\ q\ -\textgreater{}\ c)\ -\textgreater{}\ c},
the type checker does not treat these types as equal.

Note that the type checker already records in the environment that
\texttt{pair} is defined as
\texttt{\textbackslash{}p.\textbackslash{}q.\ (c:\ Type)\ -\textgreater{}\ (p\ -\textgreater{}\ q\ -\textgreater{}\ c)\ -\textgreater{}\ c}.
We'd like the type checker to look up this definition when it sees the
variable \texttt{pair} and beta-reduce this application.

\subsection{Another example needing more expressive
equality}

As another example, in the full language, we might have a type of length
indexed vectors, where vectors containing values of type \texttt{A} with
length \texttt{n} can be given the type \cd{Vec n A}. In this language, we may
have a safe \texttt{head} operation that allows us to access the first element of the
vector, as long as it is nonzero.

\begin{piforall}
head : (A : Nat) -> (n : Nat) -> Vec (Succ n) A -> Vec n A
head = ...

\end{piforall}

However, to call this function, we need to be able to show that the length of
the argument vector is equal to \texttt{Succ\ n} for some \cd{n}. This is OK if we
know the length of the vector outright.

\begin{piforall}
v1 : Vec 1 Bool
v1 = Cons [0] True Nil
\end{piforall}

So the application \texttt{head\ Bool\ 0\ v1} will type check. (Note
that \pif cannot infer the types \texttt{A} and \texttt{n}.)

However, if we construct the vector, its length may not be a literal
natural number. Take a look at the \texttt{append} function, that
takes a vec of length \texttt{m}, vector of length \text{n},
and returns a vector of length \texttt{plus m n}:

\begin{piforall}
append : (n : Nat) -> (m : Nat) -> Vec m A -> Vec n A -> Vec (plus m n) A
append = ...
\end{piforall}

In that case, to get \texttt{head\ Bool\ 1\ (append\ v1\ v1)} to type check,
we need to show that the type \texttt{Vec\ (Succ\ 1)\ Bool} is equal to the
type \texttt{Vec\ (plus\ 1\ 1)\ Bool}. If our definition of type equality is
\emph{alpha-equivalence}, then this equality will not hold. We need to enrich
our definition of equality so that it equates more terms.

\subsection{Defining equality}

The main idea is that we will:

\begin{itemize}
\item
  establish a new judgment that defines when types are equal

\[ \fbox{$\Gamma  \vdash  \ottnt{A}  \ottsym{=}  \ottnt{B}$} \]
\item
  add the following rule to our type system so that it works ``up-to''
  our defined notion of type equivalence

\[ \drule[width=3in]{t-conv}\]

\item
  Figure out how to revise the \emph{algorithmic} version of our type
  system so that it supports the above rule, deferring to an algorithmic
  version of equality.
\end{itemize}

What is a good definition of equality? We have implicitly started with a very
simple one: identify terms up to $\alpha$-equivalence. But we can do
better.

\begin{figure}
\drules[e]{$\Gamma  \vdash  \ottnt{A}  \ottsym{=}  \ottnt{B}$}{Definitional Equality}{beta,refl,sym,trans,pi,lam,var,app,lift,annot}
\caption{Definitional equality for core \pif}
\label{fig:defeq}
\end{figure}

The rules in Figure~\ref{fig:defeq} define what it means for two terms to be
equivalent by stating the properties that we want to be true of this
relation.  \Rref{e-beta} ensures that our relation \emph{contains
  beta-equivalence}. Terms that evaluate to each other should be equal.
\Rref{e-refl,e-sym,e-trans} makes sure that this relation is an
\emph{equivalence relation}. Furthermore, we want this relation to be a
\emph{congruence relation} (i.e.~if subterms are equal, then larger terms are
equal), as specified by rules \rref{e-pi,e-lam,e-app}.  We also want to be
sure that this relation has ``functionality'' (i.e.~we can lift
equalities). We declare so with \rref{e-lift}. 
\Rref{e-var} states that a variable is equivalent to its definition in the context.
Here we add variable definitions to the context, which will later be used 
when we extend the language to support dependent pattern matching. 
Finally, we want to ignore type
annotations, so \rref{e-annot} allows the equality judgment to skip over them
on the left (and with the help of \rref{e-sym}, on the right as well).

(Note: if we add booleans and $\Sigma-$types, we will also need corresponding
$\beta$-equivalence rules and congruence rules for those constructs.)

How do we know our definition is good? We want to make sure that whatever
rules that we add to the system, we define a relation that is
\emph{consistent}. We want to equate as many terms/types as possible, but
we don't want to be be able to derive an equality between $\ottkw{Type}$ and some
function type (e.g. $ ( \ottmv{x} \!:\! \ottkw{Type} )  \to   \ottkw{Type} $), or an equality between $\ottkw{True}$ and $\ottkw{False}$. To be absolutely confident about our definition, we want a proof
that such equalties are not derivable in our system.

\subsection{Using definitional equality in the algorithm}

We would like to consider our type system as having the following rule:
\[ \drule[width=4in]{t-conv} \]
but this rule is not syntax-directed; it can be used at any place in a
derivation, so how do we know where to check types for equality? And which
types should we check?

It turns out that in our bidirectional system, there are only a few places
where type equality matters.

\begin{itemize}
\item We need to check for equality when we switch from checking mode to
  inference mode in the algorithm in rule \rref{c-infer}.  Here we need to
  ensure that the type that we infer is the same as the type that is passed to
  the checker.

\[ \drule[width=4in]{c-infer} \]

In this case, the types $\ottnt{A}$ and $\ottnt{B}$ are available to the type checker,
and our equality algorithm must decide whether they are equal. We use
$\Gamma  \vdash  \ottnt{A}  \Leftrightarrow  \ottnt{B}$ as the notation for an algorithmic version of our equality
relation---the relation we defined in Figure~\ref{fig:defeq} does not readily
lead to an algorithm. (More on this algorithmic version in
Section~\ref{sec:equate} below.)

\item In the rule for application, when we infer the type of the function we
  need to make sure that the function actually has a function type. But we
  don't really know what the domain and co-domain of the function should
  be. We would like our algorithm for type equality to be able to figure this out
  for us.

\[ \drule[width=4in]{i-app} \]

In this case, we are given a single type $\ottnt{A}$ and we need to know whether
it is equivalent to some $\Pi$-type. Because $\Pi$-types are \emph{head}
forms, we can do this via reduction. This rule evaluates the type $A$ to its
head form, using the rules shown in Figure~\ref{fig:whnf}. If that form is a
$\Pi$-type, then we can access its subcomponents.

\item Whenever we call \cd{checkType} we should call it with a term that
has already been reduced to normal form. This will allow \rref{c-lambda} to
match against a literal function type.

\[ \drule[width=4in]{c-whnf} \]

\end{itemize}

\begin{figure}
\drules[whnf]{$ \Gamma   \vdash   \ottkw{whnf} \  \ottnt{a}  \leadsto  \ottnt{nf} $}{Weak head normal form reduction}{lam-beta,annot,type,lam,pi,var,app}
\caption{Relation computing the weak head normal form of a term}
\label{fig:whnf}
\end{figure}

\paragraph{Weak head normal form}

The following judgment (shown in Figure~\ref{fig:whnf})
\[ \fbox{$ \Gamma   \vdash   \ottkw{whnf} \  \ottnt{a}  \leadsto  \ottnt{nf} $} \]
describes when a term $\ottnt{a}$ reduces to some result $\ottnt{nf}$ under context $\Gamma$ 
in \emph{weak head normal form (whnf)}.  For closed terms, these rules correspond to a
big-step evaluation relation and produce a value (i.e. abstraction or type
form).  For open terms, the rules could also produce terms that are a
sequence of applications headed by a variable.  In this case, we call the term
a \emph{neutral} term, and use the metavariable $\ottnt{ne}$ to refer to it.

\[
\begin{array}{llcl}
\textit{value}     & \ottnt{v} & ::= &  \lambda  \ottmv{x} .  \ottnt{a} \ |\ \ottkw{Type}\ |\  ( \ottmv{x} \!:\! \ottnt{A} )  \to   \ottnt{B} \ \\
\textit{neutral terms} & \ottnt{ne} & ::= & \ottmv{x}\ |\  \ottnt{ne}  \;  \ottnt{a}  \\
\textit{weak-head normal form} & \ottnt{nf} & ::= & \ottnt{v}\ |\ \ottnt{ne} \\
\end{array}
\]

For example, the term
\begin{piforall}
 (\x.x) (\x.x)
\end{piforall}
is not in whnf, because there is more reduction to go to get to the head. On
the other hand, even though there are still internal reductions possible, the terms
\begin{piforall}
  \y. (\x.x) (\x.x)
\end{piforall}
and
\begin{piforall}
 (y:Type) -> (\x.x) Bool
\end{piforall}
are in weak head normal form. Furthermore, the term \texttt{x\ y} is also in
weak head normal form, because, even though we don't know what the head form
is, we cannot reduce the term anymore.

Because evaluation may not terminate, this relation is
semi-decidable. However, it is syntax-directed, so we can readily express it as
a Haskell function.

\paragraph{Using algorithmic equality}

In the \pif implementation, the function that corresponds to $\Gamma  \vdash  \ottnt{a}  \Leftrightarrow  \ottnt{b}$
has type
\begin{verbatim}
equate :: Term -> Term -> TcMonad ()
\end{verbatim}
This function ensures that the two provided types are equal, or throws a type error if
they are not.

This function is defined in terms of a helper function that implements the
rules shown in Figure~\ref{fig:whnf}.
\begin{verbatim}
whnf :: Term -> TcMonad Term
\end{verbatim}

\noindent
In \texttt{version2} of the \href{version2/src/TypeCheck.hs}{implementation},
these functions are called in a few places:
\begin{itemize}
\item \texttt{equate} is called at the end of \texttt{checkType} to make sure
that the annotated type matches the inferred type.
\item \texttt{whnf} is called in the \texttt{App} case of \texttt{inferType} to ensure that
  the function has some sort of function type.

\item \texttt{whnf} is called at the beginning of \texttt{checkType}
  to make sure that we are using the head form of the type in checking
  mode.
\end{itemize}

\subsection{Implementing definitional equality (see Equal.hs)}
\label{sec:equate}

\begin{figure}
\drules[eq]{$\Gamma  \vdash  \ottnt{A}  \Leftrightarrow  \ottnt{B}$}{Algorithmic equality}{refl,whnf,abs,pi,app}
\caption{Algorithmic equality for core \pif}
\label{fig:algeq}
\end{figure}

The rules for $\Gamma  \vdash  \ottnt{a}  \ottsym{=}  \ottnt{b}$ \emph{specify} when terms should be equal, but
they are not an algorithm. But how do we implement its algorithmic analogue
$\Gamma  \vdash  \ottnt{a}  \Leftrightarrow  \ottnt{b}$?

There are several ways to do so.

The easiest one to explain is based on reduction---for \texttt{equate} to
reduce the two arguments to a \emph{normal form} and then compare those normal
forms for equivalence.

One way to do this is with the following algorithm:

\begin{verbatim}
 equate t1 t2 = do
    nf1 <- reduce t1  -- full reduction, not just weak head reduction
    nf2 <- reduce t2
    Unbound.aeq nf1 nf2
\end{verbatim}

However, we can do better.  Sometimes we can find out that terms are
equivalent without fully reducing them. For example, if we observe that the
two terms are already $\alpha$-equivalent, then we already know that they are
equal---no need to reduce them first. Because reduction can be expensive (or
even nonterminating) we would like to only reduce as much as we need to find
out whether the terms can be identified.

Therefore, the implementation of \cd{equate} has the following form.
\begin{verbatim}
  equate t1 t2 = do
     if (Unbound.aeq t1 t2) then return () else do
      nf1 <- whnf t1  -- reduce only to 'weak head normal form'
      nf2 <- whnf t2
      case (nf1,nf2) of
        (Lam bnd1, Lam bnd2) -> do
            -- ignore variable names, but use the same
            -- fresh name for both lambda bodies
            (_, b1, _, b2) <- Unbound.unbind2Plus bnd1 bnd2
             equate b1 b2
        (App a1 a2, App b1 b2) -> do
            equate a1 b1
            equateArg a2 b2

        ... -- all other matching head forms

        -- | head forms don't match throw an error
        (_,_) -> Env.err ...
\end{verbatim}
Above, we first check for alpha-equivalence. If they
are, we do not need to do anything else. Instead, the function returns immediately
(i.e., without throwing an error, which is what the function does when it
decides that the terms are not equal). The next step is to reduce both terms
to their weak head normal forms.  If two terms have different head forms, then
we know that they must be different terms, so we can throw an error. If they have
the same head forms, then we can call the function recursively on analogous
subcomponents until the function either terminates or finds a mismatch.

This algorithm is summarized in Figure~\ref{fig:algeq}.

Why do we use weak head reduction vs.~full reduction?

\begin{itemize}
\item We can extend this algorithm to implement deferred substitutions for
  variables. Note that when comparing terms we need to have their definitions
  available. That way we can compute that \texttt{(plus\ 3\ 1)} weak head
  normalizes to 4, by looking up the definition of \texttt{plus} when
  needed. However, we don't want to substitute all variables through
  eagerly---not only does this make extra work, but error messages can be
  extremely long.
\item Furthermore, we allow recursive definitions in \pif, so normalization
  may just fail completely if we unfold recursive definitions inside
  functions before they are applied. In contrast, the definition based on
  \texttt{whnf} only unfolds recursive definitions when they stand in the way
  of equivalence, so avoids some infinite loops in the type checker.
\end{itemize}

Note that equality is not decidable in \texttt{pi-forall}. There will always
be terms that could cause \texttt{equate} to loop forever. However, the
algorithm is sound. If it says that two terms are equal, then there is some
derivation using the rules of Figure~\ref{fig:defeq} that the terms are equal.

\section{Dependent pattern matching and propositional equality}
\label{sec:pattern-matching}

One of the powerful features of dependently-typed languages is the
idea of ``learning by testing''. In other words, the information gained
in pattern matching can be reflected into the type system, allowing it
to reason in a flow-sensitive manner. We will demonstrate how this works in
this section.

\subsection{More expressive rules for if and \(\Sigma\)-type elimination}

Consider our elimination rule for if from your homework assignment:

\[ \drule[width=4in]{t-if-simple} \]

One solution to the homework assignment is to add the following two
rules to the algorithmic system, one in inference mode and one in
checking mode.

\[ \drule[width=4in]{c-if-simple}\]
\[\drule[width=4in]{i-if-simple} \]

In both of these rules, we should check that the condition has type bool (when using the checking judgement here, the algorithm will infer the type of the condition and then compare the inferred type against bool using definitional equality.) However, there is no unique best way to handle the branches of the if expression. Some examples work better using the first rule and some examples require the second rule.
For example, we can only type check the example below using the first (checking) rule:
\[  \emptyset   \vdash   \lambda  \ottmv{y} .  \ottkw{if} \, \ottmv{y} \, \ottkw{then} \, \ottsym{(}   \lambda  \ottmv{x} .  \ottmv{x}   \ottsym{)} \, \ottkw{else} \, \ottsym{(}    \lambda  \ottmv{x} .  not \,    \;  \ottmv{x}   \ottsym{)}   \Leftarrow  \ottkw{Bool}  \to  \ottkw{Bool}  \to  \ottkw{Bool} \]
and we can only type check this next example using the second (inference) rule:
\[   \ottmv{x} \! :\! \ottkw{Bool}  ,  \ottmv{y} \! :\! \ottkw{Bool}  \to  \ottkw{Bool}   \vdash   \ottsym{(}  \ottkw{if} \, \ottmv{x} \, \ottkw{then} \, \ottmv{y} \, \ottkw{else} \, not \,   \ottsym{)}  \;  true \,    \Rightarrow  \ottkw{Bool} \]

However, note that we can strengthen the specification of the type checking
rule for \cd{if} by making the result type \texttt{A} depend on whether the
condition is true or false when type checking each branch. In this way, our
type system is flow-sensitive: we do different things while type checking
based on how control flows in a \pif program.

\[ \drule[width=4in]{t-if-full} \]

For example, here is a simple program that requires this stronger rule to type
check. The function \cd{T} is an example of a \emph{large elimination}: a function
that destructs a boolean value to produce a type. Then in the type of \cd{bar} below,
each branch can return a different type of result, according to \cd{T}.

\begin{piforall}
-- function from booleans to types
T : Bool -> Type
T = \b. if b then Unit else Bool

-- returns Unit when the argument is True
bar : (b : Bool) -> T b
bar = \b . if b then () else True
\end{piforall}

It turns out that this strong type checking rule is difficult to implement. In
fact, it is not syntax-directed because $A$ and $x$ are not fixed by the
syntax. Given $\ottnt{A}  \ottsym{[}  \ottkw{True}  \ottsym{/}  \ottmv{x}  \ottsym{]}$ and $\ottnt{A}  \ottsym{[}  \ottkw{False}  \ottsym{/}  \ottmv{x}  \ottsym{]}$ and $\ottnt{A}  \ottsym{[}  \ottnt{a}  \ottsym{/}  \ottmv{x}  \ottsym{]}$ (or
anything that they are definitionally equal to!)  how can we figure out
whether they correspond to each other? Neither inference nor checking mode
works, because we don't ever see the type $\ottnt{A}$ with just the variable
$\ottmv{x}$.

So, we will not be so ambitious in \pif. We will only allow this refinement when
the scrutinee is a variable, deferring to the weaker, non-refining typing rule
for all other cases. Specificationally, we will implement this rule:

\[ \drule[width=4in]{t-if} \]

And, in our bidirectional system, we will only allow refinement
when we are in checking mode. That way, because $\ottnt{a}$ is just $\ottmv{x}$, we can
easily identify how the type should vary in each branch.

\[ \drule[width=4in]{c-if} \]

To implement this rule, we need only remember that $x$ is $\ottkw{True}$ or
$\ottkw{False}$ when checking the individual branches of the if expression.
\footnote{
Here is an alternative version, for inference mode only, suggested
during a prior summer school:

\[ \drule[width=4in]{i-if-alt} \]

It has a nice symmetry---\textsf{if} expressions are typed by \textsf{if}
types. Note, however, that to make this rule work, we will need a stronger
definitional equivalence than we have. In particular, we will want our
definition of equivalence to support the following equality:

\[ \drule[width=4in]{e-if-eta} \]

That way, if the type of the two branches of the \textsf{if} does not actually
depend on the boolean value, we can convert the \textsf{if} expression
into a more useful type.
}

We can modify the elimination rule for $\Sigma$-types similarly. Here, we are
not learning which branch was taken, but we are still learning something. If
we get to the body of the \textsf{let} expression, we know that the scrutinee
terminated and produced some product value.

\[ \drule[width=4in]{c-letpair} \]

This modification changes our definition of $\Sigma$-types from weak $\Sigma$s
to strong $\Sigma$-types. With either typing rule, we can define the first
projection.

\begin{piforall}
fst : (A:Type) -> (B : A -> Type) -> (p : { x2 : A | B x2 }) -> A
fst = \A B p. let (x,y) = p in x
\end{piforall}

But, weak $\Sigma$-types cannot define the second projection. The
following code only type checks when the above rule is available.

\begin{piforall}
snd : (A:Type) -> (B : A -> Type) -> (p : { x2 : A | B x2 }) -> B (fst A B p)
snd = \A B p. let (x,y) = p in y
\end{piforall}

\noindent
(Try this out using \texttt{version2} of the implementation and the \texttt{Lec2.pi}
input file.)

\subsubsection{Definitions and let expressions}

Now let's consider let expressions, of the form $\ottkw{let} \, \ottmv{x}  \ottsym{=}  \ottnt{a} \, \ottkw{in} \, \ottnt{b}$. The simplest rule
that we can use for this term is the following:

\[ \drule[width=4in]{t-let-simple} \]

Note the last premise: we need to make sure that the variable $\ottmv{x}$ does not
appear in the type $\ottnt{B}$ because otherwise it would escape its scope. In the
specificational version of the type system, this version of the let rule is
derivable from function application: $\ottkw{let} \, \ottmv{x}  \ottsym{=}  \ottnt{a} \, \ottkw{in} \, \ottnt{b}$ is equivalent to
$ \ottsym{(}   \lambda  \ottmv{x} .  \ottnt{b}   \ottsym{)}  \;  \ottnt{a} $. On the other hand, the bidirectional system won't type check
this definition without an annotation, so we should add this expression form
directly.

However, we can also develop a stronger rule, which statically tracks that, because this
is a let-expression, we know that $\ottmv{x}$ is equal to its definition. This view leads to the
following rule:

\[ \drule[width=4in]{t-let-def} \]

In this rule, we've added a definition to our context, i.e. $\ottmv{x}  \ottsym{=}  \ottnt{a}$. These
definitions are used during weak head normalization. If we get to a variable
in the head position, and there is a definition for it in the context, we
should look it up and keep normalizing.

Definitions in the context act like deferred substitutions. Indeed,
\rref{t-let-def} doesn't really increase the expressiveness of the type system
compared to \rref{t-let-simple}: if we had derivation that uses this rule, we
can replace it with one that uses $\ottnt{b}  \ottsym{[}  \ottnt{a}  \ottsym{/}  \ottmv{x}  \ottsym{]}$ instead of $\ottnt{b}$ in the
body of the let expression. However, for programmers, if $\ottnt{a}$ is a large
expression and important for type checking, it is more convenient for to write
$\ottmv{x}$ instead.

How do we approach let expressions in the bidirectional system? As with if and
letpair expressions, it makes sense to have both checking and inference
rules.  In checking mode, the rule is very similar to the specification. We
extend the context appropriately and use the known type to check the body of
the expression.

\[ \drule[width=4in]{c-let} \]

However, in inference mode, we have to be careful that the variable $\ottmv{x}$
does not escape its scope. One option would be just to reject the program if
this would happen.  However, a more flexible approach is to substitute this
variable away with its definition.\footnote{This rule for let can be used to
derive the more general form of refinement rules for booleans and other datatypes. In the
general case, we can annotate an elimination form with a \emph{motive}:
\[ \drule[width=4in]{i-if-motive} \]
This annotation lets us use refinement, even when the condition is not a variable, and even in inference mode. However, we can also derive this rule via let:
\[ \ottkw{if} \, \ottnt{a} \, \ottkw{then} \, \ottnt{b_{{\mathrm{1}}}} \, \ottkw{else} \, \ottnt{b_{{\mathrm{2}}}}  \ottsym{[}  \ottmv{x}  \ottsym{.}  \ottnt{A}  \ottsym{]} = \ottkw{let} \, \ottmv{x}  \ottsym{=}  \ottnt{a} \, \ottkw{in} \, \ottsym{(}  \ottkw{if} \, \ottmv{x} \, \ottkw{then} \, \ottnt{b_{{\mathrm{1}}}} \, \ottkw{else} \, \ottnt{b_{{\mathrm{2}}}}  \ottsym{:}  \ottnt{A}  \ottsym{)} \]
}

\[ \drule[width=4in]{i-let} \]

\paragraph{Definitions and refinement}

With the addition of definitions, we can revise our refining rules for booleans and
$\Sigma$-types. Instead of substituting in the result type, we can replace those rules
with suspended definitions in the context.

\[ \drule[width=4in]{c-if-def} \]

\[ \drule[width=4in]{c-letpair-def} \]

\subsection{Propositional equality}

\begin{figure}
\drules[t]{$\Gamma  \vdash  \ottnt{a}  \ottsym{:}  \ottnt{A}$}{Typing}{refl,eq,subst,contra}
\caption{Propositional equality (Specificational rules)}
\label{fig:propeq}
\end{figure}

\begin{figure}
\drules[c]{$\Gamma  \vdash  \ottnt{a}  \Leftarrow  \ottnt{A}$}{Checking}{refl,subst-left,subst-right}
\[ \drule[width=4in]{c-contra} \]
\drules[i]{$\Gamma  \vdash  \ottnt{a}  \Rightarrow  \ottnt{A}$}{Inference}{}
\vspace{-5ex}
\[ \drule[width=4in]{i-eq} \]

\caption{Propositional equality (Bidirectional rules)}
\label{fig:propeq-alg}
\end{figure}

You started proving things right away in Coq or Agda with an equality
proposition. For example, in Coq, when you say
\begin{verbatim}
Theorem plus_O_n : forall n : nat, 0 + n  = n.
\end{verbatim}
\noindent
you are using a built-in type, \texttt{a\ =\ b} that represents the
proposition that two terms are equal.

As a step towards more general indexed datatypes, we will start by adding a propositional equality type to \pif.

The main idea of the equality type is that it converts a \emph{judgment} that
two types are equal into a \emph{type} that evaluates to a special value form,
called $\ottkw{refl}$, only when two types are equal.\footnote{Recall that all
  types are inhabited in \pif, so to know whether types are equal, we must
  make sure that we have an actual proof of equality.}  In other words, the
new value form $\ottkw{refl}$ has type $\ottnt{a}  \ottsym{=}  \ottnt{b}$ when these two types are
definitionally equal to each other, as shown in \rref{t-refl} in
Figure~\ref{fig:propeq}.

Sometimes, you might see \rref{t-refl} written as follows:
\[ \drule[width=3in]{t-refl-alt} \]
However, this rule is equivalent to the above version because types are equal up to congruence. If we know that $ \emptyset   \vdash  \ottnt{a}  \ottsym{=}  \ottnt{b}$, then we know that the type $\ottnt{a}  \ottsym{=}  \ottnt{a}$ is equal to the type $\ottnt{a}  \ottsym{=}  \ottnt{b}$. In a type system with conversion, it doesn't matter which of these equal types we choose for the conclusion of the rule.

However, in the bidirectional version of the rules, shown in
Figure~\ref{fig:propeq-alg}, it does make a difference. The algorithmic typing
rule for $\ottkw{refl}$ requires checking mode so that we know which types we
need to show equivalent.  (See \rref{c-refl}.)  Furthermore, this rule allows
$\ottnt{a}$ and $\ottnt{b}$ to differ so that there are no restrictions on its
use. Instead, to implement this rule, we need only compare $\ottnt{a}$ and $\ottnt{b}$
using our algorithmic equality, as we did in Section~\ref{sec:equality}.

An equality type is well-formed when the terms on both sides of the equality
have the same type. In other words, when it implements \emph{homogeneous}
equality, as shown in \rref{t-eq}. In the algorithm, \rref{i-eq}, we
infer the types of each side of the equality type and then check to make sure
that those types are equal.

The elimination rule for propositional equality is written $\ottkw{subst} \, \ottnt{a} \, \ottkw{by} \, \ottnt{b}$
in \pif.  This term allows us to convert the type of one expression to
another, as shown in rule \rref{t-subst-simple} below. In this rule, we can change the type
of some expression $\ottnt{a}$ by replacing an occurrence of some term $\ottnt{a_{{\mathrm{1}}}}$
with an equivalent type.

\[ \drule[width=4in]{t-subst-simple} \]

How can we implement this rule? For simplicity in these lecture notes, we will
play the same trick that we did with booleans, requiring that one of the sides
of the equality be a variable.\footnote{The \pif implementation implements a more
expressive rule that uses \emph{unification} to determine the substitution to apply
to $A$. Unification means that if we have an equality type such as
$\ottsym{(}  \ottmv{x}  \ottsym{,}  \ottkw{True}  \ottsym{)}  \ottsym{=}  \ottsym{(}  \ottsym{()}  \ottsym{,}  \ottmv{y}  \ottsym{)}$
then we can produce the substitution of $\ottsym{()}$ for $\ottmv{x}$ and $\ottkw{True}$ for $\ottmv{y}$.}

\[ \drule[width=4in]{c-subst-left-simple} \]
\[ \drule[width=4in]{c-subst-right-simple} \]

Note that this elimination form for equality is powerful. We can use it to
show that propositional equality is symmetric and transitive.

\begin{piforall}
sym : (A:Type) -> (x:A) -> (y:A) -> (x = y) -> y = x
trans : (A:Type) -> (x:A) -> (y:A) -> (z:A)
      -> (x = z) -> (z = y) -> (x = y)
\end{piforall}

Furthermore, we can also extend the elimination form for propositional
equality with dependent pattern matching as we did for booleans. This
corresponds to the elimination rule \rref{t-subst} in Figure~\ref{fig:propeq},
which observes that the only way to construct a proof of equality is with the
term $\ottkw{refl}$.\footnote{This version of subst is similar to an eliminator
  for propositional equality called \texttt{J}. However, our
  eliminator is very strong and can be used to derive ``axiom'' K or the
  uniqueness of identity proofs. The eliminator for propositional equality in
  Coq and Agda is not this expressive, leaving the type system open to
  extension with other proofs of propositional equality. }

As above, the \rref{c-subst-right}, shown in Figure~\ref{fig:propeq-alg}, (and
the corresponding left rule) only applies when $b$ is also a variable.

One last addition: \texttt{contra}. If we can somehow prove a false statement,
then we should be able to prove anything. A contradiction is a proposition
between two terms that have different head forms. For now, we will use
\rref{t-contra}, shown in Figure~\ref{fig:propeq}, that requires a proof that
$\ottkw{True}$ equals $\ottkw{False}$. The algorithmic version of this rule,
\rref{c-contra}, must use weak-head normalization multiple times: first to
find the equality type and then to show that the two sides of it have
different head forms. This rule must be in checking mode because we can use
\textsf{contra} to inhabit any type.

\subsubsection{Homework (\pif: equality)}

Complete the file \href{version2/test/Hw2.pi}{\texttt{Hw2.pi}}. This file
gives you practice with working with equality propositions in \pif.

\paragraph{Example}
To see an example of propositional equality in action, let's take a look at the function \texttt{sym} in the file \href{version2/test/Hw2.pi}{\texttt{Hw2.pi}}. 
For convenience, we reproduce the function definition here:
\begin{piforall}
sym : (A:Type) -> (x:A) -> (y:A) -> (x = y) -> y = x
sym = \ A x y pf .
  subst Refl by pf 
\end{piforall}
Intuitively, the function \texttt{sym} says: if we supply two terms  \texttt{x} and \texttt{y} (both of type \texttt{A}), along with a proof \texttt{pf} that \texttt{x = y}, then we can prove that \texttt{y = x}.  
Note that the \pif expression \texttt{subst Refl by pf} is doing a pattern-match on the term inhabiting the identity type. In Coq, \pif expressions of the form \texttt{subst a by b} would be written as:
\begin{verbatim}
match b with 
  Refl => a
\end{verbatim}

To see what is happening in the body of \texttt{sym}, let's add type annotations to each of the subterms:
\begin{piforall}
sym : (A:Type) -> (x:A) -> (y:A) -> (x = y) -> y = x
sym = \ A x y pf .
  (subst (Refl : x = x) by (pf : x = y) : y = x)
\end{piforall}

In the body of \texttt{sym}, \texttt{subst} eliminates \texttt{pf} (a proof that \texttt{x = y}) when typechecking \texttt{Refl} against the type \texttt{y = x}. 
Moreover, the \texttt{subst} keyword adds the definition \texttt{x = y} to the context, meaning that both sides of the equality \texttt{y = x} normalize to \texttt{y}; this is an example of the \rref{c-subst-left} in action. 
Note that as a result of this rule, the result type of the entire expression \texttt{subst Refl by pf} changes to \texttt{y = x}, which is the desired return type for the function \texttt{sym}.

\section{Irrelevance: the \(\forall\) of \pif}
\label{sec:irrelevance}

Now, let us talk about irrelevance. In dependently typed languages, some
arguments are ``ghost'' or ``specificational'' and only there for proofs. For
efficient executables, we don't want to have to ``run'' these arguments, nor
do we want them taking up space in data structures.

Functional languages do this all the time: they erase \emph{type annotations}
and \emph{type} arguments as part of the compilation pipeline. This erasure
makes sense because parametric polymorphic functions are not allowed to
depend on types. The behavior of map must be the same no matter whether it is
operating on a list of integers or a list of booleans.

In a dependently-typed language, we would like to erase types too. And erase proofs
that are only there to make things type check. Coq does this by making a
distinction between \texttt{Prop} and \texttt{Set}. Everything in \texttt{Set}
stays around until run time, and is guaranteed not to depend on \texttt{Prop}.

We will take another approach.

In \pif we have a new kind of quantification, called ``forall'', that marks
erasable arguments. We mark forall quantified arguments with square
brackets. For example, we can mark the type argument of the polymorphic
identity function as erasable.

\begin{piforall}
id : [x:Type] -> (y : x) -> x
id = \ [x] y. y
\end{piforall}

When we apply such functions, we will put the argument in brackets too, so
we remember that \texttt{id} is not really using that type.

\begin{piforall}
t = id [Bool] True
\end{piforall}

However, we need to make sure that irrelevant arguments really are
irrelevant. We wouldn't want to allow this definition:

\begin{piforall}
id' : [x:Type] -> [y:x] -> x
id' = \ [x][y]. y
\end{piforall}
\indent
Here \texttt{id'} claims that its second parameter \cd{y} is
erasable, but it is not. And, when we check this code with
\texttt{version3} of \pif, it is rejected with the following error
message.

\begin{verbatim}
Type Error:
Lec3.pi:16:16:
  Cannot access irrelevant variables in this context
  When checking the term
    \[x] [y]. y
  against the signature
    id' : [x : Type] -> [x] -> x
  In the expression
    y
\end{verbatim}

However, note that casting is a relevant use of propositional equality. In the
example below, \pif will prevent us from marking the argument \texttt{pf} as
irrelevant.

\begin{piforall}
proprel : [a : Type] -> (pf : a = Bool) -> (x : a) -> Bool
proprel = \ [a] pf x .
  subst x by pf
\end{piforall}

The reason for this restriction is that the language includes
non-termination. We don't know whether this proof is \texttt{Refl} or some
infinite loop. So this argument must be evaluated to be sure that the two
types are actually equal. If (somehow) we knew that the argument would always
evaluate to \texttt{Refl} we could erase it.

\subsection{How do we track irrelevance in the type system?}

We need to make sure that an irrelevant variable is not ``used'' in the
relevant parts of the body of a $\lambda$-abstraction. How can we do so?

The key idea is that we mark variable assumptions in the context with an
$\epsilon$ annotation, which is either $ + $ (relevant) or $ - $
(irrelevant).  and only relevant variables can be used in terms. (``Normal''
variables, introduced by $\lambda$-expressions, will always be marked as
relevant.) As a result, we revise the variable typing rule to require the
relevant annotation.  In the \texttt{version3} implementation, type signatures
in the environment now include an \cd{Epsilon} tag, and the function
\cd{checkStage} ensures that this tag is $ + $ when the variable is used
in the term.

\[
\drule{t-evar}
\]

An irrelevant abstraction introduces its variable into the context tagged with
the irrelevant annotation. Because of the $ - $ tag, these variables are
inaccessible in the body of the function.

\[
\drule[width=3in]{t-elambda}
\]

However, this variable should be available for use in \emph{types} and other
\emph{irrelevant} parts of the term. To enable this use, we use a special
context operation $ \Gamma ^  +  $, as in the second premise in \rref{t-elambda}, and
in the \rref{t-eapp} rule below.

\[
\drule[width=3in]{t-eapp}
\]

This operation on the context, called \emph{resurrection}, converts all
``$ - $'' tags on variables to be ``$ + $''. It represents a shift in
our perspective: because we have entered an irrelevant part of the term, the
variables that were not visible before are now available after this context
modification.

Finally, when checking irrelevant $\Pi$ types, we mark the variable with
$ + $ when we add it to the context so that it may be used in the range of
the $\Pi$ type. Otherwise, we would not be able to verify the type of the
polymorphic identity function above.
\[
\drule[width=3in]{t-epi}
\]

\subsection{Compile-time irrelevance}

What does checking irrelevance buy us? Because we know that irrelevant
arguments are not actually used by the function, we know that they can be
\emph{erased} prior to execution.

However, there is another benefit to marking some arguments as irrelevant: we
can ignore them during equivalence checking. When deciding whether two terms
are equal, we don't need to look at their irrelevant components. In other
words, we can add the following equality rule to our definitional equality in
the case of applications with irrelevant arguments.

\[
\drule{e-eapp}
\]

If you look at \texttt{version3} of the implementation, you can see where we
use this idea in the definition of equality. Note in this version that
arguments are now tagged with their relevance, and when we compare arguments
that are tagged as irrelevant, we do nothing. In fact, what we are doing is
actually \emph{defining} equality over just the computationally relevant parts
of the term instead of the entire term. And we have been doing this in a limited
form from the beginning---recall that \texttt{version1} of \pif already
ignores type annotations. Just as type annotations do not affect computation
and can be ignored, the same reasoning holds for irrelevant arguments.

Why is compile-time irrelevance important?
\begin{itemize}
\item faster comparison: don't have to look at the
whole term when comparing for equality
\item more expressive: these parts of the term don't actually have
to be equal
\end{itemize}

For example, consider the example below. The parameter $p$ takes an irrelevant
argument, so we know that it must be a constant function. Therefore, it is
sound to equate any two applications of $p$ because we know that we will
always get the same result.

\begin{piforall}
irrelevance : (p : [i : Nat] -> Type) -> p [1] = p [2]
irrelevance = \p . Refl
\end{piforall}

\section{Datatypes and Indexed Datatypes}
\label{sec:datatypes}

Finally, we will add a general implementation of datatypes (including
potentially irrelevant arguments) to \pif.  The code to look at is the
``complete'' implementation in \href{full/}{\texttt{full}}.

Unfortunately, datatypes are both:

\begin{itemize}
\item
  Really important (you see them \emph{everywhere} when working with
  languages like Haskell, Coq, Agda, Idris, etc.)
\item
  Really complicated (there are a \emph{lot} of details). In general, datatypes
  subsume booleans, $\Sigma$-types, propositional equality, and can carry
  irrelevant arguments. So they combine all of the complexity of the previous
  sections in one construct.
\end{itemize}

Unlike the previous sections, where we could walk through all of the details
of the specification of the type system, not to mention its implementation, we
won't be able to do that here. There is just too much! The goal of this part
is to give you enough information so that you can pick up the Haskell code and
understand what is going on.

Even then, realize that the implementation that I am giving you is not the
complete story. Recall that we are not considering termination. That means that
we can use datatypes merely by writing recursive functions; we do not have to
reason about whether those functions terminate. Coq, Agda, and Idris include a
lot of machinery for this termination analysis, and we won't cover any of it
here.

In the rest of this section, we will work up the general specification of
datatypes piece-by-piece, generalizing from features that we already know to
more difficult cases.  We will start with ``simple'' datatypes, and then extend
them with both parameters and indices.

\subsection{``Dirt simple'' datatypes}\label{dirt-simple-datatypes}

Our first goal is simple. What do we need to get the simplest examples of
non-recursive and regular recursive datatypes working? By this I mean
datatypes that you might see in Haskell or ML, such as \texttt{Bool},
\texttt{Void} and \texttt{Nat}. Here are some examples of what we would
like these types to look like in \pif. If you are looking at the implementation,
you will find this code in \texttt{pi/Lec4.pi}.

\subsubsection{Booleans}\label{booleans}

For example, one homework assignment was to implement booleans as a primitive
language feature. We would like to replace this primitive feature with a datatype
definition of a new type constructor (\cd{Bool}), with two constructors
\cd{True} and \cd{False}, defined by the following data type declaration.

\begin{piforall}
  data Bool : Type where
     True
     False
\end{piforall}

In the homework assignment, we used \texttt{if} as the elimination form
for boolean values. For example, with the primitive \cd{if} expression,
we could write this function.

\begin{piforall}
 not : Bool -> Bool
 not = \ b . if b then False else True
\end{piforall}

For uniformity, we want to have a common elimination form for all datatypes,
called \texttt{case}. This expression form has branches for each of the data
constructors of the data type. (In \pif, we will steal Haskell's syntax for case
expressions, including layout.) For example, we might rewrite \texttt{not}
with \cd{case} like this:

\begin{piforall}
not : Bool -> Bool
not = \ b .
  case b of
       True -> False
       False -> True
\end{piforall}

\subsubsection{Void}\label{void}

The simplest datatype of all is one that has no constructors!

\begin{piforall}
data Void : Type where {}
\end{piforall}

Because there are no constructors, the elimination form for values of
this type doesn't need any cases!

\begin{piforall}
false_elim : (A:Type) -> Void -> A
false_elim = \ A v . case v of {}
\end{piforall}

The \cd{Void} type brings up the issue of \emph{exhaustiveness} in case
analysis. Can we tell whether there are enough patterns so that all of the
cases are covered? For \cd{Void} this is simple, as we can see from its declaration
that it has no cases to cover. But, as our language becomes more sophisticated, we will need
to revisit our implementation to make sure that it can correctly identify when a case
expression is exhaustive.

\subsubsection{Nat}\label{nat}

Natural numbers include a data constructor with an argument, which we can
write in the syntax of \pif as below. (For simplicity in the parser, those
parentheses must be there.)

\begin{piforall}
data Nat : Type where
   Zero
   Succ of (Nat)
\end{piforall}

In case analysis, we can give a name to that argument in the pattern.

\begin{piforall}
is_zero : Nat -> Bool
is_zero = \ x . case x of
   Zero -> True
   Succ n -> False
\end{piforall}

However, to write \cd{plus}, we need to use recursion:

\begin{piforall}
plus : Nat -> Nat -> Nat
plus = \ x y. case x of
   Zero -> y
   Succ x' -> Succ (plus x' y)
\end{piforall}

\subsubsection{Dependently-typed data constructor args}

Now, even in our ``dirt simple'' system, we will be able to encode some new
structures that are beyond what is available using regular algebraic datatypes
in functional programming languages like Haskell and ML. These structures
won't be all that useful yet, but as we add parameters and indices to our
datatypes, they will become more so. For example, here is an example of a
datatype declaration where the data constructors have dependent types.

\begin{piforall}
data SillyBool : Type where
   ImTrue  of (b: Bool) (_ : b = True)
   ImFalse of (b: Bool) (_ : b = False)
\end{piforall}

In this example, the type of \cd{ImTrue} is
\begin{piforall}
ImTrue : (b:Bool) -> (b = True) -> SillyBool
\end{piforall}

\subsection{Implementing the type checker for ``dirt simple'' datatypes}

Datatype declarations, such as \cd{data Bool}, \cd{data Void},
or \cd{data Nat} extend the context with new type constants (i.e.
type constructors) and new data constructors. It is as if we had added a
bunch of new typing rules to the type system, such as:

\drules[t]{$\Gamma  \vdash  \ottnt{a}  \ottsym{:}  \ottnt{A}$}{Typing}{Nat,Void,Zero,Succ,ImTrue}

In the general form, a \emph{simple} data type declaration includes a
name and a list of data constructors.

\begin{piforall}
   data T : Type where
      K1                 -- no arguments
      K2 of (A)          -- single arg of type A
      K3 of (x:A)        -- also single arg of type A, called x
      K4 of (x:A)(y:B)   -- two args, the type B can mention x
      K5 of (x:A)[x = a] -- one arg, plus an equality constraint about that arg
\end{piforall}

In this setting, each data constructor takes a special sort of list of
arguments that we call a \emph{telescope}. (The word telescope for this
structure was coined by de Bruijn~\cite{debruijn} to describe the scoping
behavior of this structure. The scope of each variable overlaps all of the
subsequent ones, nesting like an expandable telescope.)

We can represent this structure in our implementation by adding a new form of
declaration to the syntax.\footnote{Some parts of this implementation have
  been elided compared to \texttt{full}. We are building up to that version
  slowly.}

\begin{verbatim}
-- | type constructor names
type TyConName = String

-- | data constructor names
type DataConName = String

-- | Declarations stored in the context
data Decl =
    -- | Declaration for the type of a term `x : A`
    TypeSig Sig
  | -- | The definition of a particular name `x = a`
    Def TName Term
    ...

  | -- | Datatype definition  `data T = ...`
    Data TyConName [ConstructorDef]

-- | A Data constructor has a name and a telescope of arguments
data ConstructorDef = ConstructorDef DataConName Telescope

-- | A telescope is a list of type declarations and definitions
newtype Telescope = Telescope [Decl]

\end{verbatim}

\noindent For example, a declaration for the \texttt{Bool} datatype would be the following.
\begin{verbatim}
boolDecl :: Decl
boolDecl = Data "Bool" [ConstructorDef "False" (Telescope []),
                        ConstructorDef "True"  (Telescope []) ]
\end{verbatim}

\subsection{Type checking uses of (simple) data constructors}


In general, when we have a datatype declaration, that means that a new data type
$\ottmv{T}$ of type $\ottkw{Type}$ will be added to the context.  Furthermore, the
context should record all of the data constructors for that type, $\ottmv{K_{\ottmv{i}}}$, as
well as the telescope, written $\Delta_{\ottmv{i}}$, for each data constructor.  For
example for natural numbers, the new type $\ottmv{T}$ is $\ottkw{Nat}$, there are two
new data constructors, $\mathsf{Zero}$ and $\mathsf{Succ}$, the former has an empty
telescope, the latter has a telescope containing the declaration of a single
relevant argument of type $\ottkw{Nat}$.  Alternatively, for the \cd{SillyBool}
example, there are again two new data constructors $\mathsf{ImTrue}$ and
$ImFalse$. Each of these constructors takes a telescope with two entries,
first the declaration of a name of type $\ottkw{Bool}$ and then a definition of
that name as either $\ottkw{True}$ or $\ottkw{False}$.

The information about datatype declarations stored in the context is used to
check terms that are the applications of data constructors. For simplicity,
\pif requires that data constructors must be fully applied to all of their
arguments.

So our typing rule for data constructor applications looks a little like
this. We have $\overline{a}$ as representing the list of arguments for the data
constructor $\ottmv{K}$.

\[ \drule{i-dcon-simple} \]

We need to check that list against the telescope for the constructor, using
the judgment $\Gamma  \vdash  \overline{a}  \Leftarrow  \Delta$.

In this judgment, each argument must have
the right type. Furthermore, because of dependency, we substitute that
argument for the variable in the rest of the telescope.

\[ \drule[width=3in]{tele-sig} \]

Furthermore, we also substitute when the telescope contains definitions.

\[ \drule[width=3in]{tele-def} \]

When we get to the end of the argument list (i.e.~there are no more arguments) we
should also get to the end of the telescope.

\[ \drule{tele-nil} \]

\begin{figure}[ht]
\begin{tabular}{lll}
$\Gamma  \vdash  \overline{a}  \Leftarrow  \Delta$   & \texttt{tcArgTele}
   & Type check a list of arguments against a telescope \\
$\Gamma  \vdash  \ottnt{pat}  \ottsym{:}  \ottnt{A}  \Rightarrow  \Delta$ & \texttt{declarePat}
   & Create telescope containing all of the \\
   && variables from the pattern \\
$\vdash  \ottnt{a}  \sim  \ottnt{b}  \Rightarrow  \Delta$  & \texttt{unify}
   & Compare two terms to create a list of definitions \\
$\Delta  \ottsym{[}  \ottnt{a}  \ottsym{/}  \ottmv{x}  \ottsym{]} $  & \texttt{doSubst}
   & Substitute through a telescope \\
$\Delta_{{\mathrm{1}}}  \ottsym{[}  \overline{a}  \ottsym{/}  \Delta_{{\mathrm{2}}}  \ottsym{]}$ & \texttt{substTele}
   & Substitute a list of args for the variables \\
   && declared in a telescope \\
\end{tabular}
\caption{Functions for checking datatypes and case expressions}
\label{fig:data}
\end{figure}

Figure ~\ref{fig:data} lists the relevant judgments for type checking the use
of datatypes and the corresponding functions in the \pif implementation.  The
$\Gamma  \vdash  \overline{a}  \Leftarrow  \Delta$ judgment, described above, is implemented by the
\cd{tcArgTele} function in \texttt{TypeCheck.hs}. This function has the
following type in the implementation: (For reasons that we explain below, we
have a special type \texttt{Arg} for the arguments to the data constructor.)

\begin{verbatim}
tcArgTele :: [Arg] -> Telescope -> TcMonad [Arg]
\end{verbatim}

The \cd{tcArgTele} also function relies on the substitution function for
telescopes, written $\Delta  \ottsym{[}  \ottnt{a}  \ottsym{/}  \ottmv{x}  \ottsym{]}$ in the rules above:

\begin{verbatim}
doSubst :: [(TName, Term)] -> [Decl] -> TcMonad [Assn]
\end{verbatim}

This substitution function propagates the substitution through
the telescope.

\subsection{Pattern matching with dirt simple datatypes}

In your homework assignment, we used a special \texttt{if} term to eliminate boolean types.
Now, we would like to replace that with the more general \texttt{case} expression, of the
form $\ottkw{case} \, \ottnt{a} \, \ottkw{of} \, \ottsym{\{} \, \ottcomp{\ottnt{pat_{\ottmv{i}}}  \to  \ottnt{a_{\ottmv{i}}}}{\ottmv{i}} \, \ottsym{\}}$
that works with any form of datatype. What should the typing rule for
that sort of expression look like? There is a lot of work to do.

\[ \drule[width=4in]{c-case-simple} \]

Mathematically, this rule first checks that the scrutinee has the type of some
datatype $\ottmv{T}$. Then, for each case of the pattern match, this rule looks at
the pattern $\ottnt{pat_{\ottmv{i}}}$ and uses the context to calculate a telescope $\Delta_{\ottmv{i}}$
containing declarations of all of the variables bound in that pattern. That
telescope is added to the context to type check each branch $\ottnt{a_{\ottmv{i}}}$ against the type of
the entire expression $\ottnt{A}$. Furthermore, when
checking the type of each branch, we can refine that type because we know that
the scrutinee is equal to the pattern. To reflect this information during type checking,
we also construct the telescope $\Delta'_{\ottmv{i}}$.

How do we implement this rule in our language? The general strategy
is as follows:

\begin{enumerate}
\def\labelenumi{\arabic{enumi}.}
\item
  Infer type of the scrutinee $\ottnt{a}$ (\texttt{inferType})
\item
  Make sure that the inferred type is some type constructor $\ottmv{T}$
  (\texttt{ensureTyCon})
\item
  For each case alternative $pi \to \ottnt{a_{\ottmv{i}}}$:
\begin{itemize}
\item
  Create the declarations $\Delta_{\ottmv{i}}$ for the variables in the pattern
  (\texttt{declarePat})

\item Create a list of definitions $\Delta'_{\ottmv{i}}$ that follow from unifying the
  scrutinee $\ottnt{a}$ with the pattern $pi$ (\texttt{unify})

\item
  Check the body of the case $\ottnt{a_{\ottmv{i}}}$ in the extended context against the
  expected type $\ottnt{A}$ (\texttt{checkType})
\end{itemize}

\item
  Make sure that the patterns in the cases are exhaustive
  (\texttt{exhausivityCheck})
\end{enumerate}

\subsection{Datatypes with parameters}

The first extension of the above scheme is for \emph{parameterized
  datatypes}. For example, in \pif we can define the \texttt{Maybe} type with
the following declaration. The type parameter \texttt{A} can
be referred to in any of the telescopes for the data constructors.

\begin{piforall}
data Maybe (A : Type) : Type where
    Nothing
     Just of (A)

\end{piforall}

Because this is a dependently-typed language, the variables in the telescope
can be referred to later in the telescope. For example, with parameters, we
can implement $\Sigma$-types as a datatype, instead of making them primitive.

\begin{piforall}
data TySigma (A: Type) (B : A -> Type) : Type
    Prod of (x:A) (B)
\end{piforall}

The general form of datatype declaration with parameters includes a
telescope for the type constructor, as well as a telescope for each of
the data constructors.

\begin{piforall}
data T D : Type where
   K1 of D1
   ...
   Kn of Dn
\end{piforall}

That means that when we check an occurrence of a type constructor, we need to
make sure that its actual arguments match up with the parameters in the
telescope. For this step, we can reuse the same argument checking judgment as
we used for data constructor applications.

\[ \drule[width=3in]{i-tcon} \]

Now, to type check data constructor applications for parameterized types, we
need to modify the typing rule for data constructors. This rule below
separates the type constructor telescope ($\Delta_{{\mathrm{1}}}$) from the data constructor
telescope $\Delta_{{\mathrm{2}}}$ when looking up the data constructor from the context. When
type checking the arguments of the data constructor $\overline{a}$, we first
substitute the arguments of the type constructor $\overline{b}$ into the data
constructor telescope.

\[ \drule[width=3in]{c-dcon} \]

For example, if we are trying to check the expression
\texttt{Just\ True}, with expected type \texttt{Maybe\ Bool}, we will
first see that \texttt{Maybe} requires the telescope
\texttt{(A\ :\ Type)}. That means we need to substitute \texttt{Bool}
for \texttt{A} in \texttt{(\_\ :\ A)}, the telescope for \texttt{Just}.
That produces the telescope \texttt{(\_\ :\ Bool)}, which we will use to
check the argument \texttt{True}.

In \texttt{TypeCheck.hs}, the function
\begin{verbatim}
substTele :: [Decl] -> [Arg] -> [Decl] -> TcMonad [Decl]
\end{verbatim}
implements this operation of substituting the actual data type arguments
for the parameters.

Note that by checking the type of data constructor applications (instead
of inferring them) we don't need to explicitly provide the parameters to
the data constructor. The type system can figure them out from the
provided type.

Also, note that checking mode also enables \emph{data constructor
overloading}. In other words, we can have multiple datatypes that use
the same data constructor. Having the type available allows us to
disambiguate.

For added flexibility, we can also add code to \emph{infer} the types of data
constructors when they are not actually parameterized and when there is no
ambiguity due to overloading. For example, if we see \cd{True}, we would like
to infer that it has type \cd{Bool} without any annotation.

\subsection{Datatypes with indices}

The final step is to index our datatypes with constraints on the
parameters. Indexed types let us express inductively defined relations,
such as \texttt{beautiful} from Software Foundations.

\begin{verbatim}
Inductive beautiful : nat -> Prop :=
  b_0 : beautiful 0
| b_3 : beautiful 3
| b_5 : beautiful 5
| b_sum : forall n m, beautiful n -> beautiful m -> beautiful (n+m).
\end{verbatim}

Even though \texttt{beautiful} has type \texttt{nat\ -\textgreater{}\ Prop},
we call the \texttt{nat} argument an \emph{index} instead of a
\emph{parameter} because it varies in the result type of each data
constructor. It is not used uniformly in each case.

In \pif, we implement indices by explicitly \emph{constraining}
parameters. These constraints are just expressed as equalities written in
square brackets. In other words, we define \texttt{beautiful} this way:

\begin{piforall}
data Beautiful (n : Nat) : Type where
  B0 of [n = 0]
  B3 of [n = 3]
  B5 of [n = 5]
  Bsum of (m1:Nat)(m2:Nat)(Beautiful m1)(Beautiful m2)[m = m1+m2]
\end{piforall}

Constraints can appear anywhere in the telescope of a data constructor.
However, they are not arbitrary equality constraints---we want to
consider them as deferred substitutions. Therefore, the term on the
left must always be a variable.

These constraints interact with the type checker in a few places:

\begin{itemize}
\item
  When we use data constructors we need to be sure that the constraints
  are satisfied, by appealing to definitional equality when we are
  checking arguments against a telescope (in \texttt{tcArgTele}).


\item
  When we substitute through telescopes (in \texttt{doSubst}), we may
  need to rewrite a constraint \texttt{x\ =\ b} if we substitute for
  \texttt{x}.

\item
  When we add the pattern variables to the context in each alternative
  of a case expression, we need to also add the constraints as
  definitions (see \texttt{declarePats}).
\end{itemize}

For example, if we check an occurrence of \texttt{B3}, i.e.~
\begin{piforall}
threeIsBeautiful : Beautiful 3
threeIsBeautiful = B3
\end{piforall}
this requires substituting \texttt{3} for \texttt{n} in the telescope
\texttt{{[}n\ =\ 3{]}}. That produces an empty telescope.

\subsubsection{Homework: finite numbers in \texttt{FinHw.pi}}

The module \texttt{FinHw.pi} declares the type of numbers that are drawn
from some bounded set. For example, the type \texttt{Fin\ 1} only
includes one number (called Zero), \texttt{Fin\ 2} includes two numbers,
etc. More generally, \texttt{Fin\ n} is the type of all natural numbers
smaller than \texttt{n}, i.e.~of all valid indices for lists of size
\texttt{n}.

\begin{piforall}
data Fin (n : Nat) : Type where
   Zero of (m:Nat)[n = Succ m]
   Succ of (m:Nat)[n = Succ m] (Fin m)
\end{piforall}

The file \texttt{FinHw.pi} includes a number of definitions
that use these types. Two of these definitions are marked with
\texttt{TRUSTME}. Replace these expressions with appropriate definitions.

\subsection{Irrelevance and datatypes}

Datatypes may wish to mark some of their arguments as irrelevant, so that
these components of the data structure can be erased at run time. For example,
the \cd{Vec} type in Section~\ref{sec:examples} marked the length of the tail
of the vector as irrelevant in the telescope for the \cd{Cons} constructor.

The mechanisms that we have set up so far extend naturally to irrelevant data
constructor arguments. Note however: we can only erase \emph{data} constructor
arguments; \pif does not allow marking arguments to \emph{type} constructors
as irrelevant. This is not a significant limitation as we almost always want
these arguments to actually be relevant---we don't want to equate \cd{Vec 0
  Int} with \cd{Vec 1 Bool}. However, these type parameters are never relevant
when they are used with data constructors---we don't even include them in the
abstract syntax.

\subsubsection{Homework: Erasure and Indexed datatypes: finite numbers in \texttt{FinHw.pi} }

Now take your code in \texttt{FinHw.pi} and see if you can mark some of the
components of the \texttt{Fin} datatype as erasable. Where do we need to pass
the length of a list around at run time?

\section{Where to go next?}
\label{sec:related-work}

\subsection{Other tutorials on the implementation of dependent type systems}

\begin{itemize}

\item A. Löh, C. McBride, W. Swierstra, 
  \emph{A tutorial implementation of a dependently typed lambda calculus}    \cite{loeh:tutorial}.

  This tutorial implements the ``LambdaPi'' language and is accompanied by a
  Haskell implementation available
  online~\footnote{\url{https://www.andres-loeh.de/LambdaPi/}}. The tutorial
  starts with an implementation of the simply-typed lambda calculus, using a
  locally nameless representation and bidirectional type checking, and then
  extends to core type system with type-in-type and uses
  normalization-by-evaluation for equivalence checking. It also adds natural
  numbers and vectors along with their eliminators, but does not include a
  general form of (inductive) datatypes.

\item Lennart Augustsson. \emph{Simple, Easier!} \cite{augustsson:tutorial}

  The goal of Augustsson's tutorial, inspired by an earlier version of
  LambdaPi, is to be the simplest implementation possible. Therefore, he uses
  a string representation of variables and binding and weak-head normalization
  for the implementation of equality. His version is a Haskell implementation of
  Barendregt's lambda cube~\cite{barendregt:lambda-calculi-with-types}.

\item Coquand, Kinoshita, Nordstrom, Takeyama. \emph{A simple type-theoretic
    language: Mini-TT} \cite{coquand:tutorial}

  Coquand et al. describe another Haskell implementation of dependent types,
  called ``Mini-TT'', which shows how to implement (in about 400 lines of code)
  a small dependently-typed language that includes $\Sigma$-types, void, unit,
  and sums.  The language does not include propositional equality or indexed
  datatypes, nor does it enforce termination.  The implementation uses strings
  to represent variable binding, which are converted to de Bruijn indices
  during normalization-by-evaluation.

\item Andrej Bauer, \emph{How to implement dependent type theory} \cite{bauer:tutorial}

  Bauer explores several different implementations of dependent type theory in
  a series of blog posts using the OCaml language. The first version uses a
  named representation (i.e., strings) and generates fresh names during
  substitution and alpha-equality. This version uses full normalization to
  implement definitional equality. Then, in the second version, Bauer revises
  the implementation to use normalization by evaluation. In the third version,
  he revises again to use de Bruijn indices, explicit substitutions, and
  switches back to weak-head normalization.

\item Tiark Rompf, \emph{Implementing Dependent Types} \cite{rompf:tutorial}.

  In a recent blog post tutorial, Rompf uses Javascript to implement a core
  dependent type theory using higher-order abstract syntax.

\item Andras Kovacs, \emph{Elaboration Zoo} \cite{kovacs:tutorial}

  Kovacs' GitHub repository includes several different implementations of
  evaluation and type checking for a minimal dependently typed
  languages. These implementations are accompanied by video recordings of a
  seminar that goes over this code. Notably, this tutorial includes implicit
  argument synthesis via (higher-order) unification.

\end{itemize}

\subsection{Related work for \pif}

\paragraph{Section~\ref{sec:simple}:  Core Dependent Types}

Because they lack a rule for conversion, the typing rules for the core
dependent type system presented in Section~\ref{sec:simple} do not correspond
to any existing type system. However, jumping ahead, the core language with
the inclusion of conversion (\rref{t-conv}) has been well-studied.

As a foundation of logic, the inclusion of rule \rref{t-type} makes the system
inconsistent. Indeed, Martin-L\"of's original formulation of type
theory~\cite{martin-lof71} included this rule until the resulting paradox
was demonstrated by Girard. This inconsistency was resolved in later versions
of the systems~\cite{martin-lof73} and in the Calculus of
Constructions~\cite{Coc} by stratifying $\ottkw{Type}$: first into two levels and
then into an infinite hierarchy of universes~\cite{luo-ecc}.

The inconsistency of this system as a logic does not prevent its use as a
programming language. Cardelli~\cite{cardelli:1986} showed that this
language was type sound and explored applications for programming.

\paragraph{Section~\ref{sec:bidirectional}:  Bidirectional type checking}

The expressiveness of dependently-typed programming languages put them out of
reach of the Hindley Milner type inference algorithm found in Haskell and
ML. As a result, type checkers and elaborators for these languages must
abandon complete type inference and require users to annotate their code.

Bidirectional type systems provide a platform for propagating type annotations
through judgments, and notably, support a syntax-directed specification of
where to apply the conversion rule during type checking. As a result, such
type systems are frequently used as the basis for type checking, with
elaboration for implicit arguments layered on top.

Pierce and Turner~\cite{pierce:lti} popularized the use of bidirectional type
systems, in the context of polymorphic languages with subtyping.  The Haskell
language uses bidirectional type checking layered on top of Hindley-Milner
unification in order to support type inference for higher-rank
polymorphism~\cite{practical-type-inference}. This type system extension
allows functions to take polymorphic functions as arguments. Notably, the
paper includes a well-documented reference implementation of the type system
in the appendix.

David Christiansen's tutorial~\cite{christiansen:tutorial-bidirectional}
introduces the idea of bidirectional type checking as an independent
topic. For another description of the role of bidirectional type systems in
the context of dependent types, see L\"oh, McBride and Swierstra's
tutorial~\cite{loeh:tutorial}. Unlike \pif, this tutorial separates the syntax
of terms based whether they are inferable or checkable.

Finally, Dunfield and Krishnaswami's extensive survey
paper~\cite{dunfield:bidirectional-survey} catalogs and organizes recent
results in this area.

\paragraph{Section~\ref{sec:equality}: Definitional equality}

Many implementations, such as Coq and Agda, use a procedure called
\emph{normalization-by-evaluation} to decide whether two terms are
equivalent. In general, this algorithm is favored due to its speed (it can
avoid interpretation overhead by relying on an evaluator) and because it can
be extended with typed equalities that derive from $\eta$-equivalence.  In
Coq, this procedure is implemented by compilation to OCaml bytecode, through
the procedure described in this paper~\cite{gregoire:strong-reduction}. More
detailed information about NbE is available in Abel's habilitation
thesis~\cite{abel_2013}. A simple implementation of NbE as described in this
document is available~\footnote{\url{https://github.com/jozefg/nbe-for-mltt}}.

\paragraph{Sections~\ref{sec:pattern-matching} and \ref{sec:datatypes}: Dependently-typed pattern matching}

The implementation of indexed datatypes in \pif is inspired by the compilation
of \emph{Generalized Algebraic Datatypes} in
Haskell~\cite{vytiniotis2011outsidein,vytiniotis2006simple}. The
distinguishing feature of this formalization is that internally, datatype
indices are represented by constrained parameters and so are an extension of
the parameterized recursive sums of products found in functional programming
languages.

In contrast, the theory of inductive datatypes found in Coq and Agda must
ensure that they cannot be used to implement nonterminating functions. This
means that there are several restrictions that must be applied: inductive
datatypes must be strictly positive, their eliminators must be defined
separately, and they must satisfy constraints based on universe-levels.  For
more information about how inductive definitions work in these settings, see
the
Coq~\footnote{\url{https://coq.inria.fr/refman/language/core/inductive.html#theory-of-inductive-definitions}}
or
Agda~\footnote{\url{https://agda.readthedocs.io/en/v2.6.2.2/language/data-types.html}}
manuals.

\paragraph{Section~\ref{sec:irrelevance}: Compile-time and run-time irrelevance}

The implementation of irrelevance in \pif is inspired by the Dependent
Dependency Calculus (DDC)~\cite{choudhury:ddc}, a pure type system
parameterized by a dependency lattice. The selection of the lattice determines
the precise form of irrelevance described by the type system: does the system
include run-time irrelevance only, compile-time irrelevance only, both sorts
of irrelevance combined together (as in \pif), or both sorts and allow them to
be distinguished.

\paragraph{Compile-time irrelevance only}

Pfenning~\cite{pfenning:2001} used a modal type system to identify arguments
that can be ignored when checking equivalence. This system, extended by Abel
and Scherer~\cite{abel-scherer}, became the basis for Agda's implementation of
compile-time irrelevance.

\paragraph{Run-time irrelevance only}

The Coq proof assistance distinguishes the \cd{Prop} universe from the
\cd{Type} universe for several reasons. However, one feature of this
distinction is that all terms in the \cd{Prop} universe can be erased prior to
execution, leading to faster evaluation of extracted code.

More recently, \emph{quantitative type theory}~\cite{atkey} unifies run-time
irrelevance and linear type systems by restricting the number of times that
variables may be used in a given context. This resulting system has been
adopted for the implementations of irrelevance in the Agda and
Idris~\cite{idris2} languages.

\paragraph{Both run-time and compile-time irrelevance, but no distinction}

The Implicit Calculus of Constructions~\cite{miquel} ensures that some
arguments are treated irrelevantly (at both run time and compile time) by simply
not including them in the syntax of terms. This work was extended to decidable
type checking through the use of an erasure operation---terms may include
irrelevant arguments, but these arguments must be erased prior to run time and
equality checking\cite{icc-star}. This approach lead to Erasure Pure Type
Systems (EPTS)~\cite{mishra-linger:epts}, and the treatment of irrelevance in
Dependent Haskell~\cite{weirich:systemd}.

\section*{Acknowledgments}
Thanks to ``Team Weirich'' from OPLSS 2023 (Ernest Ng, David L\"{a}wen, Litao
Zhou, and Katherine Philip) and OPLSS 2022 (Kartik Singhal, Jonathan Chan,
Shengyi Jiang, and Thomas Binetruy), the Penn PLClub, and the attendees of
OPLSS 2022, OPLSS 2014 and OPLSS 2013, for feedback and suggestions about
these lecture notes. Some parts of the implementation This work was supported
by the National Science Foundation under grants NSF 2006535, NSF 1703835 and
NSF 0910786.

\phantomsection
\addcontentsline{toc}{section}{References}
\bibliographystyle{alphaurl}
\bibliography{weirich}



\end{document}